\documentclass[cmp]{svjour}

\usepackage{amsmath,a4,amssymb,epsfig} 
\usepackage[hypertex]{hyperref}

\allowdisplaybreaks[4]

\makeatletter
\@addtoreset{equation}{section}

\date{}

\makeatother

\newtheorem{lem}{Lemma} 
\newtheorem{dfn}[lem]{Definition}
\newtheorem{def-lem}[lem]{Definition/Lemma} 
\newtheorem{thm}[lem]{Theorem}
\newtheorem{cor}[lem]{Corollary}
\newtheorem{prp}[lem]{Proposition}

\begin{document}

\title{Power-counting theorem for non-local matrix models
    and renormalisation}

\author{Harald Grosse\inst{1} \and 
        Raimar Wulkenhaar\inst{2}}

\institute{Institut f\"ur Theoretische Physik, Universit\"at Wien,
Boltzmanngasse 5, A-1090 Wien, Austria,
\email{harald.grosse@univie.ac.at}
\and 
Max-Planck-Institut f\"ur Mathematik in den
  Naturwissenschaften,
Inselstra\ss{}e 22-26, D-04103 Leipzig, Germany, 
\email{raimar.wulkenhaar@mis.mpg.de}}

\maketitle

\begin{abstract}
  Solving the exact renormalisation group equation \`a la
  Wilson-Polchinski perturbatively, we derive a power-counting theorem
  for general matrix models with arbitrarily non-local propagators.
  The power-counting degree is determined by two scaling dimensions of
  the cut-off propagator and various topological data of ribbon
  graphs. As a necessary condition for the renormalisability of a
  model, the two scaling dimensions have to be large enough relative
  to the dimension of the underlying space. In order to have a
  renormalisable model one needs additional locality
  properties---typically arising from orthogonal polynomials---which
  relate the relevant and marginal interaction coefficients to a
  finite number of base couplings. The main application of our
  power-counting theorem is the renormalisation of field theories 
  on noncommutative $\mathbb{R}^D$ in matrix formulation.
\end{abstract}

\section{Introduction}

Noncommutative quantum field theories show in most cases a phenomenon
called \emph{UV/IR-mixing} \cite{Minwalla:1999px} which seems to
prevent the perturbative renormalisation. There is an enormous number
of articles on this problem, most of them performing one-loop
calculations extra\-polated to higher order. A systematic analysis of
noncommutative (massive) field theories at \emph{any loop order} was
performed by Chepelev and Roiban
\cite{Chepelev:1999tt,Chepelev:2000hm}. They calculated the
integral of an arbitrary Feynman graph using the parametric integral
representation and expressed the result in terms of determinants
involving the incidence matrix and the intersection matrix. They
succeeded to evaluate the leading contribution to the determinants in
terms of topological properties of \emph{ribbon graphs} wrapped
around Riemann surfaces. In this
way a power-counting theorem was established which led to the
identification of two power-counting non-renormalisable classes of
ribbon graphs. The \emph{Rings}-type class consists of graphs with
classically divergent subgraphs wrapped around the same handle of the
Riemann surface. The \emph{Com}-type class consists of planar graphs
with external legs ending at several disconnected boundary components
with the momentum flow into a boundary component being identically
zero\footnote{These graphs were called ``Swiss Cheese'' in
  \cite{Becchi:2002kj}}.

Except for models with enough symmetry, noncommutative field theories
are not renormalisable by standard techniques. One may speculate that
the reason is the too na\"{\i}ve way of performing the various limits.
Namely, a field theory is a dynamical system with infinitely many
degrees of freedom defined by a certain limiting procedure of a system
with finitely many degrees of freedom. One may perform the limits
formally to the path integral and evaluate it by Feynman graphs which
are often meaningless. It is the art of renormalisation to give a
meaning to these graphs. This approach works well in the commutative
case, but in the noncommutative situation it seems to be not 
successful.

A procedure which deals more carefully with the limits is the
renormalisation group approach due to Wilson \cite{Wilson:1973jj},
which was adapted by Polchinski to a very efficient renormalisability
proof of commutative $\phi^4$-theory in four dimensions
\cite{Polchinski:1983gv}.  There are already some attempts
\cite{Griguolo:2001ez} to use Polchinski's method to renormalise
noncommutative field theories.  We are, however, not convinced that
the claimed results (UV-renormalisability) are so easy to obtain. The
main argument in \cite{Griguolo:2001ez} is that the Polchinski
equation is a one-loop equation so that the authors simply compute an
integral having exactly one loop. It is, however, not true that
nothing new happens at higher loop order. For instance, all one-loop
graphs can be drawn on a genus-zero Riemann surface. The entire
complexity of Riemann surfaces of higher genus as discussed by
Chepelev and Roiban \cite{Chepelev:1999tt,Chepelev:2000hm} shows up at
higher loop order and is completely ignored by the authors of
\cite{Griguolo:2001ez}.  As we show in this paper, the same discussion
of Riemann surfaces is necessary in the renormalisation group
approach, too.

The mentioned complexity is due to the phase factors described by the
intersection matrix which result in convergent but not absolutely
convergent momentum integrals.  As such it is very difficult to access
this complexity by Polchinski's procedure \cite{Polchinski:1983gv}
which is based on taking \emph{norms} of the contributions.
Moreover, Chepelev and Roiban established the link between
power-counting and the topology of the Riemann surface via the
parametric integral representation, which is based on Gau\ss{}ian
integrations. These are not available for Polchinski's method where we
deal with cut-off integrals. In conclusion, we believe it is extremely
difficult (if not impossible) to use the exact renormalisation group
equation for noncommutative field theories in momentum space. The best
one can hope is to restrict oneself to limiting cases where e.g.\ the
non-planar graphs are suppressed \cite{Becchi:2002kj,Becchi:2003dg}.
Even this restricted model has rich topological features.

Fortunately, there exists a base $f_{mn}$ for the algebra under
consideration\footnote{For another matrix realisation of the
  noncommutative $\mathbb{R}^D$ and its treatment by renormalisation
  group methods, see \cite{Nicholson:2003wp}.} where the
$\star$-product is reduced to an ordinary product of (infinite)
matrices, $f_{mn} \star f_{kl}= \delta_{nk} f_{ml}$, see
\cite{Gracia-Bondia:1987kw}. The interaction $\int d^Dx(\phi\star
\phi\star\phi\star \phi)$ can then be written as $\mathrm{tr}(\phi^4)$
where $\phi$ is now an infinite matrix (with entries of rapid decay).
The price for the simplification of the interaction is that the
kinetic matrix, or rather its inverse, the propagator, becomes very
complicated. However, in Polchinski's approach the propagator is
anyway made complicated when multiplying it with the smooth cut-off
function $K[\Lambda]$.  The parameter $\Lambda$ is an energy scale
which varies between the renormalisation scale $\Lambda_R$ and the
initial scale $\Lambda_0 \gg \Lambda_R$. Introducing in the bilinear
(kinetic) part of the action the cut-off function and replacing the
$\phi^4$-interaction by a $\Lambda$-dependent effective action
$L[\phi,\Lambda]$, the philosophy is to determine $L[\phi,\Lambda]$
such that the generating functional $Z[J,\Lambda]$ is actually
independent of $\Lambda$.

In this paper we provide the prerequisites to investigate the
renormalisation of general non-local matrix models. We prove a
power-counting theorem for the effective action $L[\phi,\Lambda]$ by
solving (better: estimating) the Polchinski equation perturbatively.
Our derivation and solution of the matrix Polchinski equation combines
the original ideas of \cite{Polchinski:1983gv} with some of the
improvements made in \cite{Keller:1992ej}. In particular, we follow
\cite{Keller:1992ej} to obtain $\Lambda_0$-independent estimations for
the interaction coefficients.  The Polchinski equation for matrix
models can be visualised by ribbon graphs.  The power-counting degree
of divergence of a ribbon graph depends on the topological data of the
graph and on two scaling dimensions of the cut-off propagator. In this
way, suitable scaling dimensions provide a simple criterion to decide
whether a non-local matrix model has the chance of resulting in a
renormalisable or not. However, having the right scaling dimensions is
not sufficient for the renormalisability of a model, because a
divergent interaction is parametrised by an infinite number of matrix
indices. Thus, a renormalisable model needs further
structures\footnote{In the first version of this paper we had proposed
  a ``reduction-of-couplings'' mechanism \cite{Zimmermann:1984sx} to
  get a finite number of relevant/marginal base couplings. Meanwhile
  it turned out \cite{Grosse:2003nw,Grosse:2004yu} that the models of
  interest provide automatically such structures in form of orthogonal
  polynomials.  } which relate these infinitely many interaction
coefficients to a finite number of base couplings.

Nevertheless, the dimension criterion proven in this paper is of great
value. For instance, it discards immediately the standard
$\phi^4$-models on noncommutative $\mathbb{R}^D$, $D=2,4$, in the
matrix base. These models have the wrong scaling dimension, which is
nothing but the manifestation of the old UV/IR-mixing problem
\cite{Minwalla:1999px}. Looking closer at the origin of the wrong
scaling dimensions it is not difficult to find a deformation of the
free action which has the chance to be a renormalisable model.  In
the matrix base of the noncommutative $\mathbb{R}^D$, the Laplace
operator becomes a tri-diagonal band matrix. The main diagonal behaves
nicely, but the two adjacent diagonals are ``too big'' and
compensate the desired behaviour of the main diagonal. Making the
adjacent diagonals ``smaller'' one preserves the properties of the
main diagonal and obtains the good scaling dimensions required for a
renormalisable model. The deformation of the adjacent diagonals
corresponds to the inclusion of a harmonic oscillator potential in the
free field action. 

We treat in \cite{Grosse:2003nw} the $\phi^4$-model on noncommutative
$\mathbb{R}^2$ within the Wilson-Polchinski approach in more detail.
We prove that this model is renormalisable when adding the harmonic
oscillator potential. Remarkably, the model remains renormalisable
when scaling the oscillator potential in a certain way to zero 
with the removal $\Lambda_0 \to \infty$ of the cut-off. 

We prove in \cite{Grosse:2004yu} that the $\phi^4$-model on
noncommutative $\mathbb{R}^4$ is renormalisable to all orders by
imposing normalisation conditions for the physical mass, the field
amplitude, the frequency of the harmonic oscillator potential and the
coupling constant. In particular, the harmonic oscillator potential
cannot be removed from the model. It gives the explicit solution of
the UV/IR-duality which suggests that noncommutativity relevant at
short distances goes hand in hand with a different physics at very
large distances.  The oscillator potential makes the $\phi^4$-action
covariant with respect to a duality transformation
\cite{Langmann:2002cc} between positions and momenta.

We stress that the power-counting theorem proven in this paper was
indispensable to have from the start the right $\phi^4$-model for the
renormalisation proof \cite{Grosse:2003nw,Grosse:2004yu}. Many
noncommutative field theories have a matrix formulation, too. We think
of fuzzy spaces and $q$-deformed models. Our general power-counting
theorem can play an important r\^ole in the renormalisation proof of
these examples.

\section{The exact renormalisation group equation}
\label{secPol}

We consider a $\phi^4$-matrix model with a general (non-diagonal) kinetic
term, 
\begin{align}
  S[\phi] &= \mathcal{V}_D \Big( \sum_{m,n,k,l} 
  \frac{1}{2} G_{mn;kl} \phi_{mn} \phi_{kl} 
+ \sum_{m,n,k,l} \frac{\lambda}{4!}
\phi_{mn} \phi_{nk} \phi_{kl} \phi_{lm} \Big)\;,
\label{actionG}
\end{align}
where $m,n,k,l \in \mathbb{N}^q$. For the noncommutative
$\mathbb{R}^D$, $D$ even, we have $q=\frac{D}{2}$. 
The factor $\mathcal{V}_D$ is the volume of an elementary cell. The
choice of $\phi^4$ is no restriction but for us the most natural one
because we are interested in four-dimensional models. Standard matrix
models are given by 
\begin{align}
q=1 \;,  \qquad 
G_{mn;kl} & =\frac{1}{\mu_0^2} \delta_{ml}\delta_{nk}\;.
\end{align}
For reviews on matrix models and their applications we refer to
\cite{Dijkgraaf:1991qh,DiFrancesco:1993nw}. The idea to apply
renormalisation group techniques to matrix models is also not new
\cite{Brezin:1992yc}. The difference of our approach is that we will
not demand that the action can be written as the trace of a polynomial
in the field, that is, we allow for matrix-valued 
kinetic terms. The only restriction we are imposing is
\begin{align}
  G_{mn;kl} &= 0 \qquad\text{unless } m+k=n+l\;.
\label{restindex}  
\end{align}
The restriction (\ref{restindex}) is due to the fact that the action
comes from a trace. It is verified for the noncommutative
$\mathbb{R}^D$ due to the $\big(O(2)\big)^{\frac{D}{2}}$-symmetry of
both the interaction and the kinetic term. The kinetic matrix
$G_{mn;kl}$ contains the entire information about the differential
calculus, including the underlying (Riemannian) geometry, and the
masses of the model. More important than the kinetic matrix $G$ will
be its inverse, the propagator $\Delta$ defined by
\begin{align}
  \sum_{k,l} G_{mn;kl} \Delta_{lk;sr} = \sum_{k,l} \Delta_{nm;lk}
  G_{kl;rs} = \delta_{mr} \delta_{ns}\;.
\label{GD}
\end{align}
Due to (\ref{restindex}) we have the same index restrictions
for the propagator:
\begin{align}
  \Delta_{nm;lk} &= 0 \qquad\text{unless } m+k=n+l\;. 
\label{restindexD}
\end{align}
Let us introduce a notion of locality:
\begin{dfn}
  A matrix model is called \underline{local} if
  $\Delta_{nm;lk}=\Delta(m,n) \delta_{ml} \delta_{nk}$ for some
  function $\Delta(m,n)$, otherwise \underline{non-local}.
\label{def-local}
\end{dfn}

We add sources $J$ to the action (\ref{actionG}) and define a 
(Euclidean) quantum field theory by the generating functional
(partition function)
\begin{align}
  Z[J]&= \int \Big(\prod_{a,b} d
  \phi_{ab}\Big) \,\exp\big(-S[\phi]- \mathcal{V}_D 
\sum_{m,n} \phi_{mn} J_{nm}\big)\;.
\label{pathintm}
\end{align}
According to Polchinski's derivation of the exact renormalisation
group equation we now consider a (at first sight) different problem
than (\ref{pathintm}). Via a cut-off function $K[m,\Lambda]$, which is
smooth in $\Lambda$ and satisfies $K[m,\infty]=1$, we modify the
weight of a matrix index $m$ as a function of a certain \emph{scale}
$\Lambda$:
\begin{align}
  Z[J,\Lambda]&= \int \Big(\prod_{a,b} d
  \phi_{ab}\Big) \,\exp\big(-S[\phi,J,\Lambda]\big)\;, 
\label{ZJL}
  \\*
  S[\phi,J,\Lambda] &= \mathcal{V}_D \Big( \sum_{m,n,k,l} \frac{1}{2}
  \phi_{mn} G^K_{mn;kl} (\Lambda)\, \phi_{kl} +
  L[\phi,\Lambda] \nonumber
  \\*
  & +\sum_{m,n,k,l} \phi_{mn} F_{mn;kl}[\Lambda] J_{kl} +
  \sum_{m,n,k,l} \frac{1}{2} J_{mn} E_{mn;kl}[\Lambda] J_{kl} 
+ C[\Lambda] \Big)\;,
\\*
G^K_{mn;kl}(\Lambda) &:= \Big(\prod_{i \in m,n,k,l} K[i,\Lambda]^{-1} 
\Big) G_{mn;kl} \;, 
\label{pathintp}
\end{align}
with $L[0,\Lambda]=0$. Accordingly, we define
\begin{align}
\Delta^K_{nm;lk}(\Lambda) =\Big(\prod_{i \in m,n,k,l}
K[i,\Lambda]\Big)  \Delta_{nm;lk}\;.
\label{GKDK1}  
\end{align}
For indices $m=(m^1,\dots,m^{\frac{D}{2}})\in
\mathbb{N}^{\frac{D}{2}}$ we would write the cut-off function as a
product $K[m,\Lambda]=\prod_{i=1}^{\frac{D}{2}}
K\Big(\frac{m^i}{(\mathcal{V}_D)^{\frac{2}{D}} \Lambda^2}\Big)$ where
$K(x)$ is a smooth function on $\mathbb{R}^+$ with $K(x)=1$ for $0\leq
x \leq 1$ and $K(x)=\epsilon$ for $x\geq 2$.  In the limit $\epsilon
\to 0$, the partition function (\ref{ZJL}) vanishes unless
$\phi_{mn}=0$ for $\max_i(m^i,n^i) \geq 2
(\mathcal{V}_D)^{\frac{2}{D}} \Lambda^2$, thus implementing a cut-off
of the measure $\prod_{a,b} d\phi_{ab}$ in (\ref{ZJL}). All other
formulae involve positive powers of
$K\Big(\frac{m^i}{(\mathcal{V}_D)^{\frac{2}{D}} \Lambda^2}\Big)$ which
multiply through the cut-off propagator (\ref{GKDK1}) the appearing
matrix indices. In the limit $\epsilon \to 0$, $K[m,\Lambda]$ has
finite support in $m$ so that all infinite-sized matrices are reduced
to finite ones.

The function $C[\Lambda]$ is the vacuum energy
and the matrices $E$ and $F$, which are not necessary in the
commutative case, must be introduced because the propagator $\Delta$
is non-local. It is, in general, not possible to separate the support of
the sources $J$ from the support of the $\Lambda$-variation of $K$.
Due to $K[m,\infty]=1$ we formally obtain (\ref{pathintm}) for
$\Lambda\to \infty$ in (\ref{pathintp}) if we set
\begin{align}
  L[\phi,\infty] & =
\sum_{m,n,k,l} \frac{\lambda}{4!}
\phi_{mn} \phi_{nk} \phi_{kl} \phi_{lm}\;, \hspace*{-10em}
\nonumber
\\*
C[\infty]&=0\;, & E_{mn;kl}[\infty] &=0\;, &
  F_{mn;kl}[\infty] &=\delta_{ml}\delta_{nk}\;.
\label{initinfty}
\end{align}
However, we shall expect divergences in the partition function which
require a renormalisation, i.e.\ additional (divergent) counterterms
in $L[\phi,\infty]$. In the Feynman graph solution of the partition
function one carefully adapts these counterterms so that all
divergences disappear. If such an adaptation is possible with a
\emph{finite number} of \emph{local} counterterms, the model is
considered as perturbatively renormalisable.

Following Polchinski \cite{Polchinski:1983gv} we proceed differently
to prove renormalisability. We first ask ourselves how to
choose $L,C,E,F$ in order to make $Z[J,\Lambda]$ \emph{independent} of
$\Lambda$. After straightforward calculation one finds the answer
\begin{align}
  \frac{\partial}{\partial \Lambda} Z[J,\Lambda] &= 0\qquad \text{iff}
  \\
  \Lambda \frac{\partial L[\phi,\Lambda]}{\partial \Lambda} &=
  \sum_{m,n,k,l} \frac{1}{2} \Lambda \frac{\partial
    \Delta^K_{nm;lk}(\Lambda)}{\partial \Lambda} \bigg( \frac{\partial
    L[\phi,\Lambda]}{\partial \phi_{mn}}\frac{\partial
    L[\phi,\Lambda]}{\partial \phi_{kl}} - \frac{1}{\mathcal{V}_D}
  \Big[\frac{\partial^2 L[\phi,\Lambda]}{\partial \phi_{mn}\,\partial
    \phi_{kl}}\Big]_\phi \bigg) \;,
\label{polL}
\\
\Lambda\frac{\partial F_{mn;kl}[\Lambda]}{\partial \Lambda} &=
-\sum_{m',n',k',l'} G^K_{mn;m'n'} (\Lambda) \, \Lambda \frac{\partial
  \Delta^K_{n'm';l'k'}(\Lambda)}{\partial \Lambda} F_{k'l';kl}[\Lambda]\;,
\label{polF}
\\
\Lambda \frac{\partial E_{mn;kl}[\Lambda]}{\partial \Lambda} &=
-\sum_{m',n',k',l'} F^T_{mn;m'n'} [\Lambda] \, \Lambda \frac{\partial
  \Delta^K_{n'm';l'k'}(\Lambda)}{\partial \Lambda}
F_{k'l';kl}[\Lambda]\;,
\label{polE}
\\
\Lambda \frac{\partial C[\Lambda]}{\partial \Lambda} &=
\frac{1}{\mathcal{V}_D} \sum_{m,n} \Lambda \frac{\partial}{\partial \Lambda}
\ln \big( K[m,\Lambda] K[n,\Lambda] \big) \nonumber
\\*
&-\frac{1}{2\mathcal{V}_D} \sum_{m,n,k,l} \Lambda \frac{\partial
    \Delta^K_{nm;lk}(\Lambda)}{\partial \Lambda} 
 \frac{\partial^2 L[\phi,\Lambda]}{\partial \phi_{mn}\,\partial
    \phi_{kl}}\Big|_{\phi=0}\;,
\label{polC}
\end{align}
where $\big[f[\phi]\big]_\phi:=f[\phi]-f[0]$.
Na\"{\i}vely we would integrate (\ref{polL})--(\ref{polC}) for the
initial conditions (\ref{initinfty}). Technically, this would be
achieved by imposing the conditions (\ref{initinfty}) not at
$\Lambda=\infty$ but at some finite scale $\Lambda=\Lambda_0$,
followed by taking the limit $\Lambda_0\to \infty$. This is easily
done for (\ref{polF})--(\ref{polC}):
\begin{align}
  F_{mn;kl}[\Lambda] &= \sum_{m',n'} G^K_{mn;m'n'}(\Lambda)
  \Delta^K_{n'm';kl}(\Lambda_0) \;,
  \\
  E_{mn;kl}[\Lambda] &= \sum_{m',n',k',l'} \Delta^K_{mn;m'n'}
  (\Lambda_0) \, \Big(G^K_{n'm';l'k'}(\Lambda) -
  G^K_{n'm';l'k'}(\Lambda_0) \Big) \Delta^K_{k'l';kl}(\Lambda_0)\;,
  \\
  C[\Lambda] &= \frac{2}{\mathcal{V}_D} \ln\Big( \prod_{m}
  K[m,\Lambda] K^{-1}[m,\Lambda_0] \Big)
\nonumber
\\*
&+ \frac{1}{2\mathcal{V}_D} \int_{\Lambda}^{\Lambda_0} \!\!d\Lambda'\;
\sum_{m,n,k,l} \frac{\partial
    \Delta^K_{nm;lk}(\Lambda')}{\partial \Lambda'} 
 \frac{\partial^2 L[\phi,\Lambda']}{\partial \phi_{mn}\,\partial
    \phi_{kl}}\Big|_{\phi=0}\;.
\end{align}
At $\Lambda=\Lambda_0$ the functions $F,E,C$ become independent of
$\Lambda_0$ and satisfy, in particular, (\ref{initinfty}) in the limit
$\Lambda_0\to \infty$.

The partition function $Z[J,\Lambda]$ is 
evaluated by Feynman graphs with vertices given by the
Taylor expansion coefficients
\begin{align}
  L_{m_1n_1;\dots;m_Nn_N}[\Lambda] := \frac{1}{N!}
  \Big(\frac{\partial^N L[\phi,\Lambda]}{\partial \phi_{m_1n_1}\,
    \partial \phi_{m_2n_2}\, \dots \partial \phi_{m_Nn_N}}
  \Big)_{\phi=0}
\label{LTaylor}
\end{align}
connected with each other by internal lines $\Delta^K(\Lambda)$ and to
sources $J$ by external lines $\Delta^K(\Lambda_0)$. As $K[m,\Lambda]$
has finite support in $m$ for finite $\Lambda$, the summation
variables in the above Feynman graphs are via the propagator
$\Delta^K(\Lambda)$ restricted to a finite set. Thus, loop summations
are finite, provided that the interaction coefficients
$L_{m_1n_1;\dots;m_Nn_N}[\Lambda]$ are bounded. In other words, for
the renormalisation of a non-local matrix model it is necessary to
prove that the differential equation (\ref{polL}) admits a regular
solution. As pointed out in the Introduction, to obtain a physically
reasonable quantum field theory one has additionally to prove that
there is a regular solution of (\ref{polL}) which depends on a
\emph{finite number of initial conditions} only. This requirement is
difficult to fulfil because there is, a priori, an infinite number of
degrees of freedom given by the Taylor expansion coefficients
(\ref{LTaylor}). This is the reason for the fact that renormalisable
(four-dimensional) quantum field theories are rare.

We are going to integrate (\ref{polL}) between a certain
renormalisation scale $\Lambda_R$ and the initial scale $\Lambda_0$.
We assume that $L_{m_1n_1;\dots;m_Nn_N}$ can be decomposed into parts
$L_{m_1n_1;\dots;m_Nn_N}^{(i)}$ which for $\Lambda_R\leq \Lambda \leq
\Lambda_0$ scale homogeneously:
\begin{align}
\Big|  \Lambda \frac{\partial
    L_{m_1n_1;\dots;m_Nn_N}^{(i)}[\Lambda] }{ \partial
    \Lambda}\Big| \leq \Lambda^{r_{i}}\; P^{q_i}\Big[\ln
    \frac{\Lambda}{\Lambda_R} \Big]\;.
\label{difftoy}
\end{align}
Here, $P^q[X] \geq 0$ stands for some polynomial of degree $q$ in $X
\geq 0$. Clearly, $P^q[X]$, for $X \geq 0$, can be further bound by a
polynomial with non-negative coefficients. As usual we define
\begin{dfn}
  Homogeneous parts $L_{m_1n_1;\dots;m_Nn_N}^{(i)}[\Lambda]$ in
  (\ref{difftoy}) with $r_i>0$ are called \underline{relevant}, with $r_i<0$
  \underline{irrelevant} and with $r_i=0$ \underline{marginal}.
\end{dfn}
There are two possibilities for the integration, either from $\Lambda_0$
down to $\Lambda$ or from $\Lambda_R$ up to $\Lambda$, corresponding
to the identities
\begin{subequations}
\begin{align}
&L_{m_1n_1;\dots;m_Nn_N}^{(i)}[\Lambda] 
\nonumber
\\*
&= 
L_{m_1n_1;\dots;m_Nn_N}^{(i)}[\Lambda_0] - \int_{\Lambda}^{\Lambda_0} 
\frac{d\Lambda'}{\Lambda'} \Big( \Lambda'\frac{\partial}{\partial
  \Lambda'}L_{m_1n_1;\dots;m_Nn_N}^{(i)}[\Lambda'] \Big) 
\label{int-down}
\\
&= L_{m_1n_1;\dots;m_Nn_N}^{(i)}[\Lambda_R] + \int_{\Lambda_R}^{\Lambda} 
\frac{d\Lambda'}{\Lambda'} \Big( \Lambda'\frac{\partial}{\partial
  \Lambda'}L_{m_1n_1;\dots;m_Nn_N}^{(i)}[\Lambda'] \Big) \;.
\label{int-up}
\end{align}
\end{subequations}
One has 
\begin{align}
\int dx\,x^{r-1} \Big(\ln \frac{x}{x_R}\Big)^q  = \left\{
  \begin{array}{cc}
\displaystyle \frac{(-1)^q q!}{r^{q+1}} x^r 
\sum_{j=0}^q \frac{\big(-r \ln \frac{x}{x_R} \big)^j}{j!} +  \text{const}
\qquad & \text{for } r \neq 0\;, \\[1ex]
\displaystyle \frac{1}{q{+}1} \Big(\ln \frac{x}{x_R}\Big)^{q+1} 
+  \text{const} \qquad  & \text{for } r=0 \;.
\end{array}\right.
\label{intlog}
\end{align}
At the end we are interested in the limit $\Lambda_0\to \infty$. This
requires that positive powers of $\Lambda_0$ must be prevented in the
estimations. For $r_i<0$ we we can safely take the direction
(\ref{int-down}) of integration and then, because all coefficients are
positive, the limit $\Lambda_0 \to \infty$ in the integral of
(\ref{int-down}). Thus,
\begin{align}
  \big|L_{m_1n_1;\dots;m_Nn_N}^{(i)}[\Lambda] \big| &\leq
  \big|L_{m_1n_1;\dots;m_Nn_N}^{(i)}[\Lambda_0] \big| 
+ \int_{\Lambda}^{\infty} 
\frac{d\Lambda'}{\Lambda'} \Big| \Lambda'\frac{\partial}{\partial
  \Lambda'}L_{m_1n_1;\dots;m_Nn_N}^{(i)}[\Lambda'] \Big|
\nonumber
\\*
& \leq  \big|L_{m_1n_1;\dots;m_Nn_N}^{(i)}[\Lambda_0] \big| 
+ \Lambda^{-|r_i|}
P^{q_i}\Big[\ln\frac{\Lambda}{\Lambda_R}\Big]\;,\qquad \text{for }
  r_i<0\;.
\label{intdown}
\end{align}
Here, $P^{q_i}$ is a new polynomial of degree $q_i$ with non-negative
coefficients. Now, the limit $\Lambda_0 \to \infty$ carried out later
requires that $\big|L_{m_1n_1;\dots;m_Nn_N}^{(i)}[\Lambda_0] \big|$ in
(\ref{intdown}) is bounded, i.e.\ 
$\big|L_{m_1n_1;\dots;m_Nn_N}^{(i)}[\Lambda_0] \big|<
\frac{C}{\Lambda_0^{s_i}}$, with $s_i>0$. As the resulting estimation
(\ref{intdown}) is further iterated, $s_i$ must be sufficiently large.
We do not investigate this question in detail and simply note that it
is safe to require
\begin{align}
  \big|L_{m_1n_1;\dots;m_Nn_N}^{(i)}[\Lambda_0] \big|<
  \Lambda_0^{-|r_i|}
  P^{q_i}\Big[\ln\frac{\Lambda_0}{\Lambda_R}\Big]\;,\qquad \text{for }
  r_i<0\;,
\label{intdown0}
\end{align}
for the boundary condition. 

In the other case $r_i\geq 0$, the integration direction
(\ref{int-down}) will produce divergences in $\Lambda_0 \to \infty$.
Thus, we have to choose the other direction (\ref{int-up}). The
integration (\ref{intlog}) produces alternating signs, but these can
be ignored in the maximisation. The only contribution from the lower
bound $\Lambda_R$ in the integral of (\ref{int-up}) is the term with
$j=0$ in (\ref{intlog}). There, we can obviously ignore it in the
difference $\Lambda^r-\Lambda_R^r$. We thus obtain from (\ref{intlog})
the estimation
\begin{align}
  \big|L_{m_1n_1;\dots;m_Nn_N}^{(i)}[\Lambda] \big| \leq
  \big|L_{m_1n_1;\dots;m_Nn_N}^{(i)}[\Lambda_R] \big| + \left\{
    \begin{array}{ll} 
      \Lambda^{r_i}P^{q_i}\Big[\ln\frac{\Lambda}{\Lambda_R}\Big] 
\qquad & \text{for } r_i > 0
      \\
     P^{q_i+1}\Big[\ln\frac{\Lambda}{\Lambda_R}\Big]  
\qquad & \text{for } r_i = 0\;.
\end{array}\right.
\label{intup}
\end{align}
The reduction from $P\big[\ln \frac{\Lambda_0}{\Lambda_R}\big]$ in
Polchinski's original work \cite{Polchinski:1983gv} to $P\big[\ln
\frac{\Lambda}{\Lambda_R}\big]$ is due to \cite{Keller:1992ej}.
We can summarise these considerations as follows:
\begin{def-lem}
\label{def-lem1}
Let $\big|\Lambda \frac{\partial}{\partial \Lambda}
L_{m_1n_1;\dots;m_Nn_N}^{(i)}[\Lambda]\big|$ be bounded by
(\ref{difftoy}),
\begin{align}
\Big|  \Lambda \frac{\partial
   L_{m_1n_1;\dots;m_Nn_N}^{(i)}[\Lambda]}{ \partial
    \Lambda}\Big| \leq \Lambda^{r_{i}} P^{q_i}\Big[ \ln
   \frac{\Lambda}{\Lambda_R}\Big] \;.
\label{difftoy1}
\end{align}
The integration of (\ref{difftoy1}) is for irrelevant interactions
performed from $\Lambda_0$ down to $\Lambda$ starting from an initial
condition bounded by $\big|L_{m_1n_1;\dots;m_Nn_N}^{(i)}[\Lambda_0]
\big|  <  \Lambda_0^{-|r_i|}
P^{q_i}\big[\ln\frac{\Lambda_0}{\Lambda_R}\big]$. For relevant and
marginal interactions we have to integrate (\ref{difftoy1}) from
$\Lambda_R$ up to $\Lambda$, starting from an initial condition
$L_{m_1n_1;\dots;m_Nn_N}^{(i)}[\Lambda_R]< \infty$.
Under these conventions we have
\begin{align}
  \big|L_{m_1n_1;\dots;m_Nn_N}^{(i)}[\Lambda]\big| \leq
  \Lambda^{r_{i}} P^{q_i+1}\Big[ \ln \frac{\Lambda}{\Lambda_R}\Big]\;.
\label{intresult}
\end{align}
\end{def-lem}
Let us give a few comments:
\begin{itemize}
\item The stability (\ref{difftoy1}) versus (\ref{intresult}) of the
  estimation will be very useful in the iteration process.
  
\item Integrations according to the direction (\ref{int-up}), which
  entail an initial condition
  $L_{m_1n_1;\dots;m_Nn_N}^{(i)}[\Lambda_R]$, are expensive for
  renormalisation, because each such condition (even the choice
  $L_{m_1n_1;\dots;m_Nn_N}^{(i)}[\Lambda_R]=0$) corresponds to a
  normalisation experiment. In order to have a meaningful theory,
  there has to be only a finite number of required normalisation
  experiments.  Initial data at $\Lambda_0$ do not correspond to
  normalisation conditions, because the interaction at $\Lambda_0 \to
  \infty$ is experimentally not accessible. Moreover, unless
  artificially kept alive\footnote{An example of an irrelevant
    coupling which remains present for $\Lambda_0 \to \infty$ is the
    initial $\phi^4$-interaction in two-dimensional models
    \cite{Grosse:2003nw}.}, an irrelevant coupling scales away for
  $\Lambda_0 \to \infty$ via its own dynamics.  The property
  $\lim_{\Lambda_0 \to \infty}
  L_{m_1n_1;\dots;m_Nn_N}^{(i)}[\Lambda_0]=0$ for an irrelevant
  coupling is, therefore, a result and no condition.
  
\item There might be cases where the direction (\ref{int-up}) for
  $r_i<0$ gives convergence for $\Lambda_0\to \infty$ nevertheless.
  This corresponds to the over-subtractions 
  in the BPHZ renormalisation scheme. We shall not exploit this
  possibility.

\end{itemize}

Unless there are further correlations between functions with different
indices, specifying $L_{m_1n_1;\dots;m_Nn_N}^{(i)}[\Lambda_R]$ means
to impose an \emph{infinite number} of normalisation conditions
(because of $m_i,n_i \in \mathbb{N}^{D/2}$).  Hence, a non-local
matrix model with relevant and/or marginal interactions can only be
renormalisable if some additional structures exist which relate all
divergent functions to a finite number of relevant/marginal base
interactions.  Such a distinguished property depends crucially on the
model.  Presumably, the class of models where such a reduction is
possible is rather small. It cannot be the purpose of this paper to
analyse these reductions. Instead, our strategy is to find the general
power-counting behaviour of a non-local matrix model which limits the
class of divergent functions among which the reduction has to be
studied in detail. For example, we will find that under very general
conditions on the propagator all non-planar graphs (as defined below)
are irrelevant. Such a result is already an enormous gain\footnote{We
  recall \cite{Minwalla:1999px} that non-planar graphs produce the
  trouble in noncommutative quantum field theories in momentum space.}
for the detailed investigation of a model.

Thus, our strategy is to integrate the Polchinski equation
(\ref{polL}) perturbatively between two scales $\Lambda_R$ and
$\Lambda_0$ for a self-determined choice of the boundary condition
according to Definition/Lemma~\ref{def-lem1}. The resulting
normalisation condition for relevant and marginal interactions will
not be the correct choice for a renormalisable model. Nevertheless,
the resulting estimation (\ref{intresult}) is compatible with a more
careful treatment. Taking the example \cite{Grosse:2004yu} of the
$\phi^4$-model on noncommutative $\mathbb{R}^4$, we would replace
\begin{itemize}
\item   
almost all of the relevant functions with bound
$\frac{\Lambda^2}{\mu^2} P^q[\ln\frac{\Lambda}{\Lambda_R}]$ in
(\ref{intresult}) by irrelevant functions with bound
$\big(\max(m_1,n_1,\dots,m_N,n_N)\big)^2
\frac{\mu^2}{\Lambda^2}P^q[\ln\frac{\Lambda}{\Lambda_R}]$, and

\item
almost all marginal functions with bound
$P^q[\ln\frac{\Lambda}{\Lambda_R}]$ in (\ref{intresult}) by irrelevant
functions with bound $\max(m_1,n_1,\dots,m_N,n_N)
\frac{\mu^2}{\Lambda^2}P^q[\ln\frac{\Lambda}{\Lambda_R}]$,
\end{itemize}
for some reference scale $\mu$.

\section{Ribbon graphs and their topologies}

We can symbolise the expansion coefficients $L_{m_1n_1;\dots;m_Nn_N}$
as
\begin{align}
\parbox{26mm}{\begin{picture}(20,21) 
 \put(0,0){\epsfig{scale=.9,file=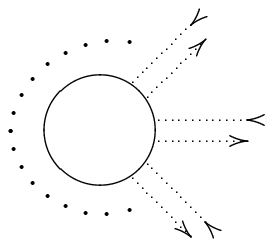,bb=69 621 141 684}}
 \put(22,13.5){\mbox{\scriptsize$n_1$}}
 \put(20,8){\mbox{\scriptsize$m_1$}}
 \put(21,3){\mbox{\scriptsize$n_2$}}
 \put(13,1){\mbox{\scriptsize$m_2$}}
 \put(19,17){\mbox{\scriptsize$m_N$}}
 \put(15,23){\mbox{\scriptsize$n_N$}}
\end{picture}}
+ (N!-1) \text{ permutations of } \{m_in_i\}\;.
\end{align}
The big circle stands for a possibly very complex interior and the
outer (dotted) double lines stand for the valences produced by
differentiation (\ref{LTaylor}) with respect to the $N$ fields
$\phi_{m_in_i}$. The arrows are merely added for bookkeeping purposes
in the proof of the power-counting theorem. Since we work with real
fields, i.e.\ $\phi_{mn}=\overline{\phi_{nm}}$, the expansion
coefficients $L_{m_1n_1;\dots;m_Nn_N}$ have to be unoriented. The
situation is different for complex fields where $\phi \neq \phi^*$
leads to an orientation of the lines. In this case we would draw both
arrows at the double line either incoming or outgoing.

The graphical interpretation of the Polchinski equation (\ref{polL})
is found when differentiating it with respect to the fields
$\phi_{m_in_i}$:
\begin{align}
\Lambda \frac{\partial}{\partial \Lambda}
\parbox{26mm}{\begin{picture}(20,21) 
 \put(0,0){\epsfig{scale=.9,file=l1a,bb=69 621 141 684}}
 \put(22,13.5){\mbox{\scriptsize$n_1$}}
 \put(20,8){\mbox{\scriptsize$m_1$}}
 \put(21,3){\mbox{\scriptsize$n_2$}}
 \put(13,1){\mbox{\scriptsize$m_2$}}
 \put(19,17){\mbox{\scriptsize$m_N$}}
 \put(15,23){\mbox{\scriptsize$n_N$}}
\end{picture}}
&=\frac{1}{2}\sum_{m,n,k,l} \sum_{N_1=1}^{N-1}
\parbox{52mm}{\begin{picture}(50,25) 
 \put(0,0){\epsfig{scale=.9,file=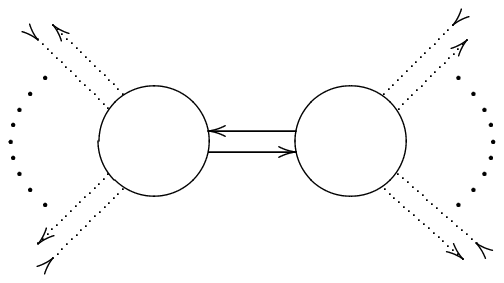,bb=69 615 214 684}}
 \put(0,4){\mbox{\scriptsize$m_1$}}
 \put(6,0){\mbox{\scriptsize$n_1$}}
 \put(0,21){\mbox{\scriptsize$n_{N_1}$}}
 \put(7,23){\mbox{\scriptsize$m_{N_1}$}}
 \put(47,20){\mbox{\scriptsize$m_{N_1+1}$}}
 \put(37,26){\mbox{\scriptsize$n_{N_1+1}$}}
 \put(48,3){\mbox{\scriptsize$n_N$}}
 \put(39,2){\mbox{\scriptsize$m_N$}}
 \put(27,14.5){\mbox{\scriptsize$k$}}
 \put(28,8){\mbox{\scriptsize$l$}}
 \put(22,15){\mbox{\scriptsize$n$}}
 \put(22,9){\mbox{\scriptsize$m$}}
\end{picture}}
\nonumber
\\*
& - ~~ \frac{1}{2\mathcal{V}_D}
\sum_{m,n,k,l} \parbox{35mm}{\begin{picture}(30,35) 
 \put(0,0){\epsfig{scale=.9,file=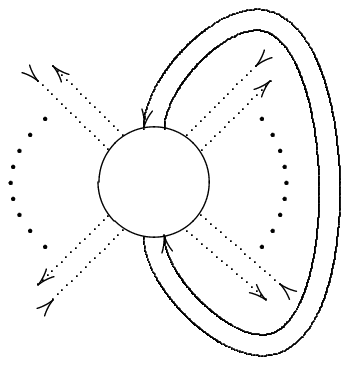,bb=69 585 168 687}}
 \put(0,9){\mbox{\scriptsize$m_1$}}
 \put(6,5){\mbox{\scriptsize$n_1$}}
 \put(-1,26){\mbox{\scriptsize$n_{i-1}$}}
 \put(7.5,29){\mbox{\scriptsize$m_{i-1}$}}
 \put(27,25){\mbox{\scriptsize$m_i$}}
 \put(22.5,30){\mbox{\scriptsize$n_i$}}
 \put(27,9){\mbox{\scriptsize$n_N$}}
 \put(20,7){\mbox{\scriptsize$m_N$}}
 \put(12,25.5){\mbox{\scriptsize$n$}}
 \put(18,26){\mbox{\scriptsize$m$}}
 \put(12,10.5){\mbox{\scriptsize$k$}}
 \put(18,10.5){\mbox{\scriptsize$l$}}
\end{picture}}
\label{polLgraph}
\end{align}
Combinatorical factors are not shown and a symmetrisation in all
indices $m_in_i$ has to be performed. On the rhs of (\ref{polLgraph})
the two valences $mn$ and $kl$ of the subgraphs are connected to the 
ends of a \emph{ribbon} which symbolises the differentiated propagator
$\parbox{10mm}{\begin{picture}(10,8)
   \put(0,3){\epsfig{scale=.9,file=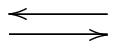,bb=71 667 101 675}}
 \put(2,6.5){\mbox{\scriptsize$n$}}
 \put(7,1){\mbox{\scriptsize$l$}}
 \put(0,1){\mbox{\scriptsize$m$}}
 \put(8,6.5){\mbox{\scriptsize$k$}}
\end{picture}}= \Lambda \frac{\partial}{\partial \Lambda} \Delta^K_{nm;lk}$. 
For local matrix models in the sense of Definition~\ref{def-local} we
can regard the ribbon as a product of single lines with interaction
given by $\Delta(m,n)$. For non-local matrix models there is an
exchange of indices within the entire ribbon.

We can regard (\ref{polL}) as a formal construction scheme for
$L[\phi,\Lambda]$ if we introduce a grading
$L[\phi,\Lambda]=\sum_{V=1}^\infty \lambda^V L^{(V)}[\phi,\Lambda]$ and
additionally impose a cut-off in $N$ for $V=1$, i.e\ 
\begin{align}
L^{(1)}_{m_1n_1;\dots;m_Nn_N}[\Lambda]=0 \qquad \text{for } N>N_0\;.
\end{align}
In order to obtain a $\phi^4$-model we choose $N_0=4$ and the grading
as the degree $V$ in the coupling constant $\lambda$. We conclude from
(\ref{polL}) that $L^{(1)}_{m_1n_1;\dots;m_4n_4}$ is independent of
$\Lambda$ so that it is identified with the original
$(\lambda/4!)\phi^4$-interaction in (\ref{actionG}):
\begin{align}
L^{(1)}_{m_1 n_1;m_2 n_2;m_3 n_3;m_4 n_4}[\Lambda] \hspace*{-3em} &
\nonumber
\\*
= \frac{1}{4! \,6}
\Big( &\delta_{n_1 m_2} \delta_{n_2
    m_3} \delta_{n_3 m_4} \delta_{n_4 m_1} + \delta_{n_1 m_3}
  \delta_{n_3 m_4} \delta_{n_4 m_2} \delta_{n_2 m_1} \nonumber
  \\
  +& \delta_{n_1 m_4} \delta_{n_4 m_2} \delta_{n_2 m_3} \delta_{n_3
    m_1} + \delta_{n_1 m_4} \delta_{n_4 m_3} \delta_{n_3 m_2}
  \delta_{n_2 m_1} \nonumber
  \\
  +& \delta_{n_1 m_3} \delta_{n_3 m_2} \delta_{n_2 m_4} \delta_{n_4
    m_1} + \delta_{n_1 m_2} \delta_{n_2 m_4} \delta_{n_4 m_3}
  \delta_{n_3 m_1} \Big)\;.
\label{L4start}
\end{align}
To the first term on the rhs of (\ref{L4start}) we associate the graph
\begin{align}
  \delta_{n_1 m_2} \delta_{n_2 m_3} \delta_{n_3 m_4} \delta_{n_4 m_1}
  =\quad \parbox{40mm}{\begin{picture}(22,22)
      \put(0,0){\epsfig{scale=.9,file=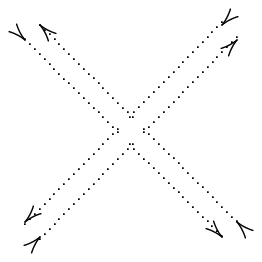,bb=71 621 134 684}}
      \put(-1,6){\mbox{\scriptsize$m_1$}}
      \put(3,0){\mbox{\scriptsize$n_1$}}
      \put(13.5,2){\mbox{\scriptsize$m_2$}}
      \put(21,4){\mbox{\scriptsize$n_2$}}
      \put(16,22){\mbox{\scriptsize$n_3$}}
      \put(20,17){\mbox{\scriptsize$m_3$}}
      \put(4.5,21){\mbox{\scriptsize$m_4$}}
      \put(-2,18){\mbox{\scriptsize$n_4$}}
\end{picture}}
\label{L4graph}
\end{align}
The graphs for the other five terms are obtained by permutation of
indices. 

As mentioned before, a complex $\phi^4$-model would be given by
oriented propagators $\parbox{10mm}{\begin{picture}(10,4)
   \put(0,0){\epsfig{scale=.9,file=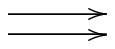,bb=71 667 101 675}}
\end{picture}}$ and examples for vertices are
\begin{align}
\phi\phi^*\phi\phi^* \sim 
\parbox{20mm}{\begin{picture}(16,16)
   \put(0,0){\epsfig{scale=.9,file=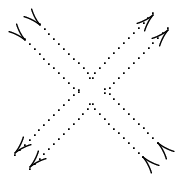,bb=71 638 117 684}}
\end{picture}} \quad 
\phi\phi\phi\phi \sim 
\parbox{20mm}{\begin{picture}(16,16)
   \put(0,0){\epsfig{scale=.9,file=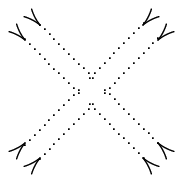,bb=71 644 111 684}}
\end{picture}} \quad 
\phi^*\phi^*\phi^*\phi^* \sim 
\parbox{20mm}{\begin{picture}(16,16)
   \put(0,0){\epsfig{scale=.9,file=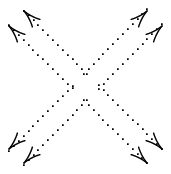,bb=71 638 117 684}}
\end{picture}} 
\end{align}
The consequence is that many graphs of the real $\phi^4$-model are now
excluded. We can thus obtain the complex $\phi^4$-model from the real
one by deleting the impossible graphs.

The iteration of (\ref{polLgraph}) with starting point (\ref{L4graph})
leads to \emph{ribbon graphs}. The first examples of the iteration are 
 \begin{align}
\parbox{30mm}{\begin{picture}(22,27)
      \put(0,0){\epsfig{scale=.9,file=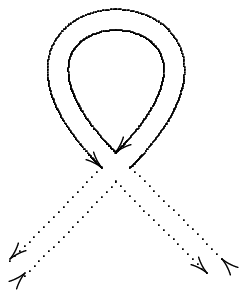,bb=71 609 134 687}}
      \put(-1,6){\mbox{\scriptsize$m_1$}}
      \put(3,0){\mbox{\scriptsize$n_1$}}
      \put(14,2){\mbox{\scriptsize$m_2$}}
      \put(21,4){\mbox{\scriptsize$n_2$}}
      \put(10,16){\mbox{\scriptsize$l$}}
\end{picture}}
\parbox{40mm}{\begin{picture}(30,22)
    \put(0,0){\epsfig{scale=.9,file=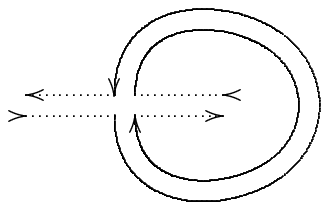,bb=71 626 158 684}}
    \put(0,6.5){\mbox{\scriptsize$n_1$}}
    \put(3,12.5){\mbox{\scriptsize$m_1$}}
    \put(15,6.5){\mbox{\scriptsize$m_2$}}
    \put(17,12.5){\mbox{\scriptsize$n_2$}}
\end{picture}}
 \parbox{40mm}{
\begin{picture}(30,20)
\put(1,0){\epsfig{scale=.9,file=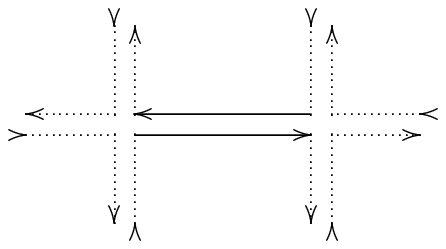,bb=71 625 187 684}}
\put(6,19){\mbox{\scriptsize$n_1$}}
\put(12.5,17){\mbox{\scriptsize$m_1$}}
\put(0,7.5){\mbox{\scriptsize$n_2$}}
\put(2,13){\mbox{\scriptsize$m_2$}}
\put(5,4){\mbox{\scriptsize$m_3$}}
\put(12.5,1){\mbox{\scriptsize$n_3$}}
\put(25.5,4){\mbox{\scriptsize$m_4$}}
\put(32.5,1){\mbox{\scriptsize$n_4$}}
\put(34,7.5){\mbox{\scriptsize$m_5$}}
\put(37,13){\mbox{\scriptsize$n_5$}}
\put(25.5,19){\mbox{\scriptsize$n_6$}}
\put(32.5,17){\mbox{\scriptsize$m_6$}}
\end{picture}}
\label{examplesofgraphs}
\end{align}
We can obviously build very complicated ribbon graphs with crossings
of lines which cannot be drawn any more in a plane.  A general ribbon
graph can, however, be drawn on a \emph{Riemann surface} of some
\emph{genus} $g$. In fact, a ribbon graph \emph{defines} the
Riemann surfaces topologically through the \emph{Euler characteristic}
$\chi$. We have to regard here the external lines of the ribbon graph
as amputated (or closed), which means to directly connect the single
lines $m_i$ with $n_i$ for each external leg $m_in_i$. A few examples
may help to understand this procedure:
\begin{align}
\parbox{40mm}{\begin{picture}(40,35) 
      \put(0,0){\epsfig{scale=.9,file=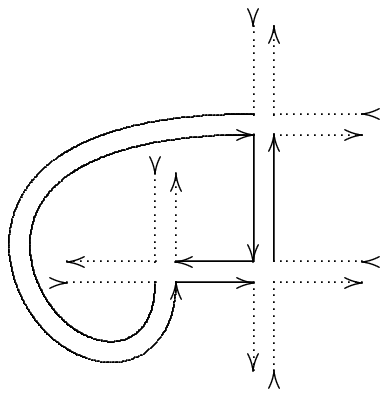,bb=74 583 178 684}}
      \put(7,7.5){\mbox{\scriptsize$n_1$}}
      \put(9,12.5){\mbox{\scriptsize$m_1$}}
      \put(30,7){\mbox{\scriptsize$m_3$}}
      \put(32,12.5){\mbox{\scriptsize$n_3$}}
      \put(30,22){\mbox{\scriptsize$m_4$}}
      \put(32,27.5){\mbox{\scriptsize$n_4$}}
      \put(21,4){\mbox{\scriptsize$m_2$}}
      \put(28,2){\mbox{\scriptsize$n_2$}}
      \put(21.5,34){\mbox{\scriptsize$n_5$}}
      \put(28,32){\mbox{\scriptsize$m_5$}}
      \put(11.5,19){\mbox{\scriptsize$n_6$}}
      \put(18,17){\mbox{\scriptsize$m_6$}}
    \end{picture}}
  \quad \Rightarrow \quad 
\parbox{35mm}{\begin{picture}(30,30) 
  \put(0,0){\epsfig{scale=.9,file=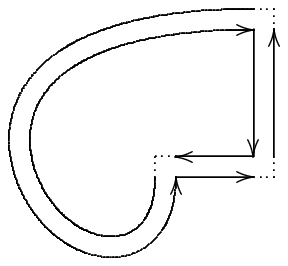,bb=74 610 153 684}}
\end{picture}}
\qquad 
\begin{array}{r@{\,}l@{\qquad}r@{\,}ll}
\tilde{L} &= 2  & B &= 2 \\
I &= 3  & N &= 6 \\
V &= 3  & V^e &= 3 \\
g &= 0 & \iota &= 0 
\end{array}
\raisetag{2ex}
\label{gt1}
\\[2ex]
\parbox{45mm}{\begin{picture}(30,25)
    \put(0,0){\epsfig{scale=.9,file=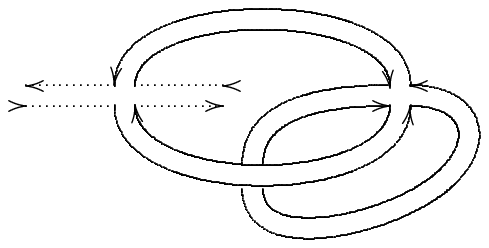,bb=71 612 204 681}}
    \put(0,11.5){\mbox{\scriptsize$n_1$}}
    \put(3,17.5){\mbox{\scriptsize$m_1$}}
    \put(15,11.5){\mbox{\scriptsize$m_2$}}
    \put(17,17.5){\mbox{\scriptsize$n_2$}}
\end{picture}}
 \quad \Rightarrow \quad 
\parbox{35mm}{\begin{picture}(30,25) 
  \put(0,0){\epsfig{scale=.9,file=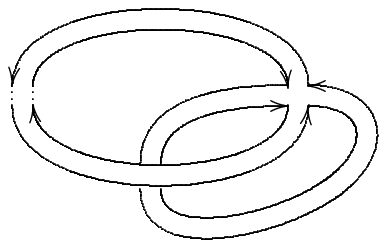,bb=71 612 178 681}}
\end{picture}}
\qquad 
\begin{array}{r@{\,}l@{\qquad}r@{\,}ll}
\tilde{L} &= 1  & B &= 1 \\
I &= 3  & N &= 2 \\
V &= 2  & V^e &= 1 \\
g &= 1 & \iota &= 1 
\end{array}
\label{gt3}
\\
\parbox{40mm}{\begin{picture}(30,25)
    \put(0,0){\epsfig{scale=.9,file=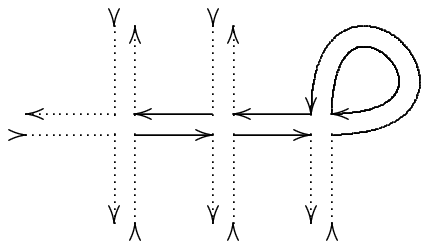,bb=71 625 187 684}}
\put(15.5,19){\mbox{\scriptsize$n_1$}}
\put(22.5,16){\mbox{\scriptsize$m_1$}}
\put(6,19){\mbox{\scriptsize$n_2$}}
\put(12.5,16){\mbox{\scriptsize$m_2$}}
\put(0,7.5){\mbox{\scriptsize$n_3$}}
\put(2,13){\mbox{\scriptsize$m_3$}}
\put(5,4){\mbox{\scriptsize$m_4$}}
\put(12.5,1){\mbox{\scriptsize$n_4$}}
\put(15,4){\mbox{\scriptsize$m_5$}}
\put(22,1){\mbox{\scriptsize$n_5$}}
\put(25,4){\mbox{\scriptsize$m_6$}}
\put(32,1){\mbox{\scriptsize$n_6$}}
\end{picture}}
 \quad \Rightarrow \quad 
\parbox{35mm}{\begin{picture}(30,10) 
  \put(0,0){\epsfig{scale=.9,file=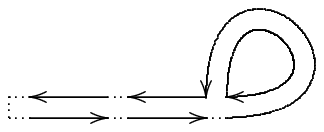,bb=71 651 161 684}}
\end{picture}}
\qquad 
\begin{array}{r@{\,}l@{\qquad}r@{\,}ll}
\tilde{L} &= 2  & B &= 1 \\
I &= 3  & N &= 6 \\
V &= 3  & V^e &= 3 \\
g &= 0 & \iota &= 0 
\end{array}
\label{gt5}
\end{align}
The genus is computed from the number $\tilde{L}$ of single-line loops
of the closed graph, the number $I$ of internal (double) lines and the
number $V$ of vertices of the graph according to
\begin{align}
\chi &= 2-2g =  \tilde{L} - I + V\;.
\label{Euler}
\end{align}
There can be several possibilities to draw the graph and its Riemann
surface, but $\tilde{L},I,V$ and thus $g$ remain
unchanged. Indeed, the Polchinski equation (\ref{polL}) interpreted as
in (\ref{polLgraph}) tells us which external legs of the vertices are
connected. It is completely irrelevant how the ribbons are drawn
between these legs. In particular, there is no distinction between
overcrossings and undercrossings. 

There are two types of loops in (amputated) ribbon graphs: 
\begin{itemize} 
\item Some of them carry at
least one external leg. They are called \emph{boundary components} (or
holes of the Riemann surface). Their
number is $B$.

\item Some of them do not carry any external leg. They are called
  \emph{inner loops}. Their number is $\tilde{L}_0=\tilde{L}-B$.
\end{itemize}
Boundary components consist of a concatenation of \emph{trajectories}
from an incoming index $n_i$ to an outgoing index $m_j$. In the
example (\ref{gt1}) the inner boundary component consists of the
single trajectory $\overrightarrow{n_1 m_6}$ whereas the outer
boundary component is made of two trajectories
$\overrightarrow{n_3m_4}$ and $\overrightarrow{n_5m_2}$. We let
$\mathfrak{o}[n_j]$ be the outgoing index to $n_j$ and
$\mathfrak{i}[m_j]$ be the incoming index to $m_j$.

We have to introduce a few additional notations for ribbon graphs.  An
\emph{external vertex} is a vertex which has at least one external
leg. We denote by $V^e$ the total number of external vertices. For the
arrangement of external legs at an external vertex there are the
following possibilities:
\begin{align}
  \parbox{30mm}{\begin{picture}(30,20)
      \put(0,0){\epsfig{scale=.9,file=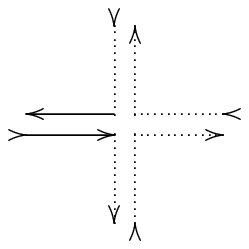,bb=71 625 130 684}}
      \put(7,19){\mbox{\scriptsize$n$}}
      \put(6,4){\mbox{\scriptsize$m$}}
\end{picture}}
\parbox{30mm}{\begin{picture}(30,20) 
    \put(0,0){\epsfig{scale=.9,file=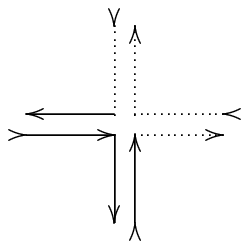,bb=71 625 130 684}}
    \put(7,19){\mbox{\scriptsize$n$}}
    \put(16,7.5){\mbox{\scriptsize$m$}}
\end{picture}}
\parbox{30mm}{\begin{picture}(30,20) 
    \put(0,0){\epsfig{scale=.9,file=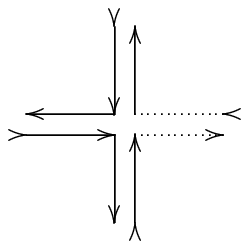,bb=71 625 130 684}}
    \put(15,7.5){\mbox{\scriptsize$m$}}
    \put(18,13){\mbox{\scriptsize$n$}}
\end{picture}}
\parbox{25mm}{\begin{picture}(25,20) 
    \put(0,0){\epsfig{scale=.9,file=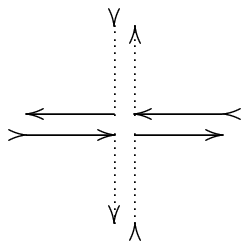,bb=71 625 130 684}}
    \put(6,19){\mbox{\scriptsize$n_1$}}
    \put(12,17){\mbox{\scriptsize$m_1$}}
    \put(5,4){\mbox{\scriptsize$m_2$}}
    \put(12,2){\mbox{\scriptsize$n_2$}}
\end{picture}}
\label{vertexe}
\end{align}
We call the first three types of external vertices \emph{simple
  vertices}. They provide one starting point and one end point of
trajectories through a ribbon graph. The fourth vertex in
(\ref{vertexe}) is called \emph{composed vertex}. It has 
two starting points and two end points of trajectories. 

A composed vertex can be decomposed by pulling the two propagators
with attached external lines apart:
\begin{align}
  \parbox{35mm}{\begin{picture}(30,20)
      \put(0,0){\epsfig{scale=.9,file=ae4,bb=71 625 130 684}}
      \put(6,19){\mbox{\scriptsize$n_1$}}
      \put(12,17){\mbox{\scriptsize$m_1$}}
      \put(5,4){\mbox{\scriptsize$m_2$}}
      \put(12,2){\mbox{\scriptsize$n_2$}}
\end{picture}}
\mapsto \qquad \parbox{35mm}{\begin{picture}(30,20)
    \put(0,0){\epsfig{scale=.9,file=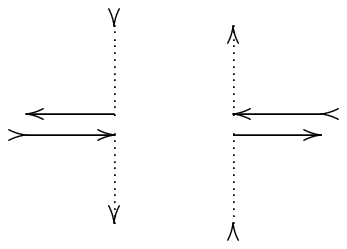,bb=71 625 158 684}}
    \put(6,19){\mbox{\scriptsize$n_1$}}
    \put(22,17){\mbox{\scriptsize$m_1$}}
    \put(5,4){\mbox{\scriptsize$m_2$}}
    \put(22,2){\mbox{\scriptsize$n_2$}}
\end{picture}}
\end{align}
In this way a given graph with composed vertices is 
decomposed into $S$ \emph{segments}. The external vertices of the segments
are either true external vertices or the halves of a composed vertex.
If composed vertices occur in loops, their decomposition does not
always increase the number of segments. We need the following
\begin{dfn}
\label{defiota}
The \underline{segmentation index} $\iota$ of a graph is the maximal
number of decompositions of composed vertices which keep the graph
connected.
\end{dfn}
It follows immediately that if $V^c$ is the number of composed
vertices of a graph and $S$ the number of segments obtained by
decomposing all composed vertices we have
\begin{align}
  \iota = V^c-S+1\;.
\label{iota}
\end{align}

In order to evaluate $L_{m_1n_1;\dots;m_Nn_N}[\Lambda]$ by connection
and contraction of subgraphs according to (\ref{polLgraph}) we need
estimations for \emph{index summations} of ribbon graphs. Namely, our
strategy is to apply the summations in (\ref{polLgraph}) either to the
propagator or the subgraph only and to maximise the other object over
the summation indices. We agree to fix all starting points of
trajectories and sum over the end points of trajectories. However, due
to (\ref{restindexD}) and (\ref{L4start}) not all summations are
independent: The sum of outgoing indices equals for each segment the
sum of incoming indices. Since there are $V^e+V^c$ (end points of)
trajectories in a ribbon graph, there are
\begin{align}
s \leq V^e+V^c-S = V^e+\iota-1
\label{numsum}
\end{align}
independent index summations. The inequality (\ref{numsum}) also holds
for the restriction to each segment if $V^e$ includes the number of
halves of composed vertices belonging to the segment. 
We let $\mathcal{E}^{s}$ be the set of $s$ end points of trajectories in a
graph over which we are going to sum, keeping the starting points of
these trajectories fixed. We define
\begin{align}
  \sum_{\mathcal{E}^{s}} \equiv \sum_{m_1} \dots
  \sum_{m_{s}}\Big|_{\mathfrak{i}[m_j]=\text{const}} 
  \qquad \text{if } \mathcal{E}^{s} =\{m_1,m_2,\dots,m_{s}\}\;.
\end{align}
Taking the example of the graph (\ref{gt1}), we can due to
$V^e+\iota=3$ apply up to two index summations, i.e.\ a summation over
at most two of the end points of trajectories $m_2,m_4,m_6$, where the
corresponding incoming indices $\mathfrak{i}[m_2]=n_5$,
$\mathfrak{i}[m_4]=n_3$ and $\mathfrak{i}[m_6]=n_1$ are kept fixed.
For the example of the graph (\ref{gt3}) we can due to $V^e+\iota=2$
apply at most one index summation, either over $m_1$ for fixed
$\mathfrak{i}[m_1]=n_2$ or over $\mathfrak{i}[m_2]=n_1$.
For $\mathcal{E}^1 = \{m_2\}$ we would consider
\begin{align}
\sum_{m_2} \left(~\parbox{42mm}{\begin{picture}(30,25)
    \put(0,0){\epsfig{scale=.9,file=gt3,bb=71 612 204 681}}
    \put(0,11.5){\mbox{\scriptsize$n_1$}}
    \put(3,17.5){\mbox{\scriptsize$m_1$}}
    \put(15,11.5){\mbox{\scriptsize$m_2$}}
    \put(17,17.5){\mbox{\scriptsize$n_2$}}
\end{picture}}\right)_{n_1=\text{const}}
\label{indsumexample}
\end{align}
Note that for given $n_2$ the other outgoing index is determined to
$m_1=n_1+n_2-m_2$ through index conservation at propagators
(\ref{restindexD}) and vertices (\ref{L4graph}). It is part of the
proof to show that the index summation (\ref{indsumexample}) is
bounded independently of the incoming indices $n_1,n_2$.

\section{Formulation of the power-counting theorem}

We first have to transform the Polchinski equation (\ref{polL}) into a
dimensionless form. It is important here that in the class of models
we consider there is always a dimensionful parameter, 
\begin{align}
\mu = \big(\mathcal{V}_D\big)^{-\frac{1}{D}}\;,
\end{align}
which instead of $\Lambda$ can be used to absorb the mass
dimensions. The effective action $L[\phi,\Lambda]$ has total mass dimension
$D$, a field $\phi$ has dimension $\frac{D-2}{2}$ and the dimension of
the coupling constant for the $\lambda \phi^4$ interaction is
$4-D$. We thus decompose $L[\phi,\Lambda]$ according to the number of
fields and the order in the coupling constant:
\begin{align}
L[\phi,\Lambda] = \sum_{V=1}^\infty 
\sum_{N=2}^{2V+2} \frac{1}{N!} \sum_{m_i,n_i} 
\Big(\frac{\lambda}{\mu^{4-D}}\Big)^V \!\!
\mu^D A^{(V)}_{m_1n_1;\dots;m_Nn_N}[\Lambda]
\Big(\frac{\phi_{m_1n_1}}{\mu^{\frac{D-2}{2}}}\Big) \cdots 
\Big(\frac{\phi_{m_Nn_N}}{\mu^{\frac{D-2}{2}}}\Big) \;.
\label{LTaylor1}
\end{align}
The functions $A^{(V)}_{m_1n_1;\dots;m_Nn_N}[\Lambda]$ are assumed to
be symmetric in their indices $m_in_i$. 
Inserted into (\ref{polL}) we get 
\begin{align}
&\Lambda \frac{\partial}{\partial \Lambda} 
 A^{(V)}_{m_1n_1;\dots;m_Nn_N}[\Lambda] 
\nonumber
\\*
&= \sum_{m,n,k,l} \frac{1}{2} Q_{nm;lk}(\Lambda) \bigg\{
\sum_{N_1=2}^N \sum_{V_1=1}^{V-1}
 A^{(V_1)}_{m_1n_1;\dots;m_{N_1-1}n_{N_1-1};mn}[\Lambda]
 A^{(V-V_1)}_{m_{N_1}n_{N_1};\dots;m_{N}n_{N};kl}[\Lambda]
\nonumber
\\*[-1ex]
&\hspace*{15em} + \Big(\binom{N}{N_1{-}1} -1\Big) \text{ permutations}
\nonumber
\\*
& \hspace*{8em} -
 A^{(V)}_{m_1n_1;\dots;m_{N}n_{N};mn;kl}[\Lambda]\bigg\}\;,
\label{polL2}
\end{align}
where
\begin{align}
Q_{nm;lk}(\Lambda) := \mu^2 \Lambda \frac{\partial}{\partial \Lambda} 
\Delta^K_{nm;lk}(\Lambda)\;.
\label{Q}
\end{align}
The permutations refer to the possibilities to choose $N_1-1$ of
the pairs of indices $m_1n_1,\dots,m_Nn_N$ which label the external
legs of the first $A$-function.

The cut-off function $K$ in (\ref{GKDK1}) has to be chosen such that
for finite $\Lambda$ there is a finite number of indices $m,n,k,l$
with $Q_{nm;lk}(\Lambda)\neq 0$. By suitable normalisation we can
achieve that the volume of the support of $Q_{nm;lk}(\Lambda)$ with
respect to a chosen index scales as $\Lambda^D$:
\begin{align}
\sum_{m} \text{sign}  \big|K[m,\Lambda]\big| 
\leq  C_D \Big(\frac{\Lambda}{\mu}\Big)^D\;,
\label{volumefactor}
\end{align}
for some constant $C_D$ independent of $\Lambda$. For such a 
normalisation we define two exponents $\delta_0,\delta_1$ by
\begin{align}
\max_{m,n,k,l} |Q_{nm;lk}(\Lambda)| & \leq C_0
\Big(\frac{\mu}{\Lambda}\Big)^{\delta_0} \delta_{m+k,n+l}\;,
\label{est0}
\\*
\max_n \Big(\sum_k \Big(\max_{m,l} |Q_{nm;lk}(\Lambda)| \Big)\Big)
& \leq C_1 \Big(\frac{\mu}{\Lambda}\Big)^{\delta_1} \;.
\label{est1}
\end{align}
In (\ref{est1}) the index $n$ is kept constant for the summation over
$k$. It is convenient to encode the dimension $D$ in a further
exponent $\delta_2$ which describes the product of
(\ref{volumefactor}) with (\ref{est0}):
\begin{align}
\max_{m,n,k,l} |Q_{nm;lk}(\Lambda)|\;
\sum_{m} \text{sign} \big|K[m,\Lambda]  \big|
& \leq C_2 \Big(\frac{\Lambda}{\mu}\Big)^{\delta_2} \;.
\label{est2}
\end{align}
We have obviously $C_2=C_D C_0$ and $\delta_2= D-\delta_0$.

\begin{dfn}
A non-local matrix model defined by the cut-off propagator
$Q_{nm;kl}$ given by (\ref{GKDK1}) and (\ref{Q}) and the normalisation
(\ref{volumefactor}) of the cut-off function is called \underline{regular}
if $\delta_0=\delta_1=2$, otherwise \underline{anomalous}.
\label{def-regular}
\end{dfn}

The three exponents $\delta_0,\delta_1,\delta_2$ play an essential
r\^ole in the power-counting theorem which yields the $\Lambda$-scaling
of a homogeneous part $A^{(V,V^e,B,g,\iota)}_{
m_1n_1;\dots;m_Nn_N}[\Lambda]$ of the interaction coefficients 
\begin{align}
&A^{(V)}_{m_1n_1;\dots;m_Nn_N}[\Lambda] 
\nonumber
\\*
&= \sum_{1 \leq V^e \leq V}\; \sum_{1 \leq B \leq N} \;
\sum_{0 \leq g \leq 1+\frac{V}{2}-\frac{N}{4}-\frac{B}{2}}  \;
\sum_{0 \leq \iota\leq B-1} A^{(V,V^e,B,g,\iota)}_{
m_1n_1;\dots;m_Nn_N}[\Lambda]\Big|_{2 \leq N \leq 2V+2}\;.
\label{hompart}
\end{align}
It is important that the sums over the graphical (topological) data
$V^e,B,g,\iota$ in (\ref{hompart}) are finite. We are going to prove
\begin{thm} 
\label{thm1}
The homogeneous parts $A^{(V,V^e,B,g,\iota)}_{
  m_1n_1;\dots;m_Nn_N}[\Lambda]$ of the coefficients of the effective
action describing a $\phi^4$-matrix model with initial interaction
(\ref{L4start}) and cut-off propagator characterised by the three
exponents $\delta_0,\delta_1,\delta_2$ are for $2 \leq N\leq 2V{+}2$
and $\sum_{i=1}^N (m_i{-}n_i)=0$ bounded by
\begin{align}
\sum_{\mathcal{E}^{s}} \big|A^{(V,V^e,B,g,\iota)}_{
    m_1n_1;\dots;m_Nn_N}[\Lambda] \big| 
  &\leq \Big(\frac{\Lambda}{\mu}\Big)^{\delta_2(
    V-\frac{N}{2}+2-2g-B)}
  \Big(\frac{\mu}{\Lambda}\Big)^{\delta_1(V-V^e-\iota+2g+B-1+s)}
\nonumber
  \\*
& \times  \Big(\frac{\mu}{\Lambda}\Big)^{\delta_0(V^e+\iota-1-s)}
  \,P^{2V-\frac{N}{2}}\Big[\ln \frac{\Lambda}{\Lambda_R}\Big]\;,
\label{ANnorm1}
\end{align}
provided that for all $V'{<}V,\;2\leq N'\leq 2V'{+}2$ and
$V'{=}V,\;N{+}2\leq N'\leq 2V{+}2$ the initial conditions for
relevant\,/\,marginal (irrelevant) $A^{(V',V^{\prime
    e},B',g',\iota')}_{ m_1'n_1';\dots;m'_{N'}n'_{N'}}[\Lambda]$ are
imposed at $\Lambda_R$ ($\Lambda_0$), respectively, according to
Definition/Lemma \ref{def-lem1}. The bound (\ref{ANnorm1}) is
independent of the unsummed indices $m_i,n_i \notin \mathcal{E}^s$. We
have $A^{(V,V^e,B,g,\iota)}_{ m_1n_1;\dots;m_Nn_N}[\Lambda]\equiv 0$
for $N>2V{+}2$ or $\sum_{i=1}^N (m_i{-}n_i)\neq 0$.
\end{thm}
The proof will be given in Section~\ref{appD}. We remark that
$\tilde{L}_0= V-\frac{N}{2}+2-2g-B$ is the number of inner loops of a
graph.

The power-counting estimation (\ref{ANnorm1}) does not make any
reference to the initial scale $\Lambda_0$ \cite{Keller:1992ej} so
that we can safely take the limit $\Lambda_0 \to \infty$. In this way
we have constructed a regular solution of the Polchinski equation
(\ref{polL}) associated with the non-local matrix model. However, this
solution remains useless unless it can be achieved by a \emph{finite
  number} of integrations from $\Lambda_R$ to $\Lambda$ depending on a
finite number of initial conditions at $\Lambda_R$. We refer to the
remarks following Definition/Lemma~\ref{def-lem1}. A first step would
be to achieve regular scaling dimensions:
\begin{cor}
  For regular matrix models according to Definition~\ref{def-regular}
  we have independently of the segmentation index and the numbers of
  external vertices
\begin{align}
\sum_{\mathcal{E}^s} \big|A^{(V,B,g)}_{
    m_1n_1;\dots;m_Nn_N}[\Lambda] \big|
\leq \Big(\frac{\Lambda}{\mu}\Big)^{\omega-D(2g+B-1)}
P^{2V-\frac{N}{2}}\Big[\ln \frac{\Lambda}{\Lambda_R}\Big]\;,
\label{cor1eq}
\end{align}
where $\omega=D+V(D-4)-N\frac{D-2}{2}$ is the classical power-counting
degree of divergence.
\label{cor1}
\end{cor}
We have derived the relation (\ref{cor1eq}) with respect to the
classical power-counting degree of divergence only for $\phi^4$-matrix
models, but it is plausible that it also holds for more general
interactions.

\section{Proof of the power-counting theorem}

\label{appD}

We provide here the proof of Theorem~\ref{thm1}, which is quite long
and technical. The proof amounts to study all possible connections of
two external legs of either different graphs or the same graph. It
will be essential how the legs to connect are situated with respect to
the remaining part of the graph. There are the following arrangements
of the external legs at the distinguished vertex one (or two) of which
we are going to connect:
\begin{align}
  \parbox{35mm}{\begin{picture}(20,20)
      \put(0,0){\epsfig{scale=.9,file=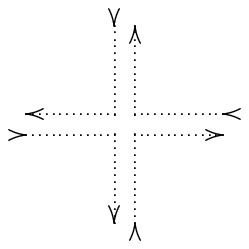,bb=71 625 130 684}}
\end{picture}}
\parbox{40mm}{\begin{picture}(30,20) 
      \put(0,0){\epsfig{scale=.9,file=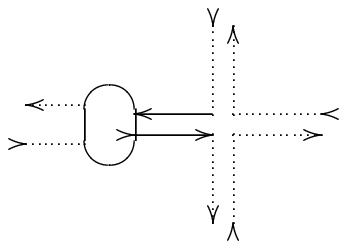,bb=71 625 158 684}}
\end{picture}}
\parbox{35mm}{
\begin{picture}(30,25)
\put(0,0){\epsfig{scale=.9,file=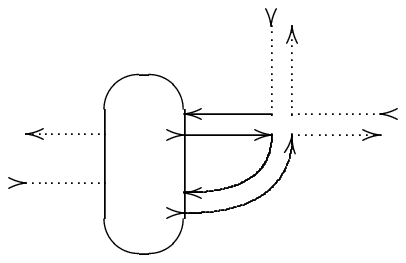,bb=71 616 175 684}}
\end{picture}}
\nonumber
\\
\parbox{60mm}{
\begin{picture}(40,28)
\put(0,0){\epsfig{scale=.9,file=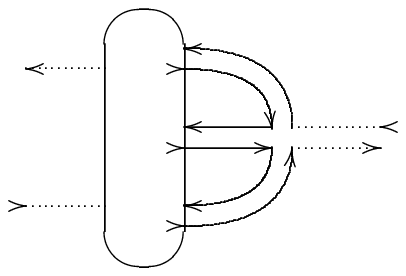,bb=71 608 175 684}}
\end{picture}}
\parbox{50mm}{
\begin{picture}(40,28)
\put(0,0){\epsfig{scale=.9,file=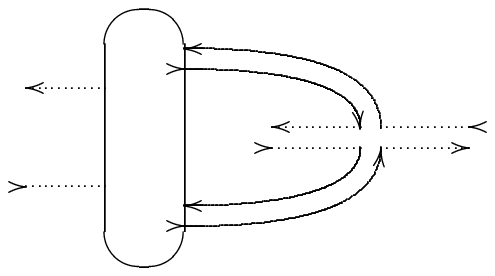,bb=71 608 201 684}}
\end{picture}}
\label{as}
\end{align}
A big oval stands for other parts of the graph the specification of
which is not necessary for the proof.  Dotted lines entering and
leaving the oval stand for the set of all external legs different from
the external legs of the distinguished vertex to contract. If two or
three internal lines are connected to the oval this does not
necessarily mean that these two lines are part of an inner loop.

We are going to integrate the Polchinski equation (\ref{polL2}) by
induction upward in $V$ and for constant $V$ downward in $N$. Due to
the grading $(V,N)$, the differential equation (\ref{polL2}) is
actually constructive. We consider in Section~\ref{tree-contr} the
connection of two smaller graphs of $(V_1,N_1)$ and $(V_2,N_2)$
vertices and external legs and in Sections~\ref{self-contr-1} and
\ref{self-contr-2} the self-contraction of a graph with
$(V_1=V,N_1=N+2)$ vertices and external legs. These graphs are further
characterised by $V^e_i,B_i,g_i,\iota_i$ external vertices, boundary
components, genera and segmentation indices, respectively. Since the
sums in (\ref{hompart}) and the number of arrangements of legs in
(\ref{as}) are finite, it is sufficient to regard the contraction of
subgraphs individually. That is, we consider individual subgraphs
$\gamma_1,\gamma_2$ the contraction of which produces an individual
graph $\gamma$. We also ignore the problem of making the graphs
symmetric in the indices $m_in_i$ of the external legs.  At the very
end we project the sum of graphs $\gamma$ to homogeneous degree
$(V,V^e,B,g,\iota)$. To these homogeneous parts there contributes
according to (\ref{hompart}) a finite number of contractions of
$\gamma_i$. We thus get the bound (\ref{ANnorm1}) if we can prove it
for any individual contraction.

The Theorem is certainly correct for the initial $\phi^4$-interaction
(\ref{L4start}) which due to (\ref{LTaylor1}) gives 
$|A^{(1,1,1,0,0)}_{m_1n_1;\dots;m_4n_4}[\Lambda]| \leq 1$. 

\subsection{Tree-contractions of two subgraphs}
\label{tree-contr}

We start with the first term on the rhs of (\ref{polL2}) which
describes the connection of two smaller subgraphs $\gamma_1,\gamma_2$
of $V_1, V_2$ vertices and $N_1,N_2$ external legs via a propagator.
The total graph $\gamma$ for a tree-contraction has
\begin{align}
  V&=V_1{+}V_2 \text{ vertices}\,, & N&= N_1{+}N_2{-}2 \text{
    external legs}\;, 
\nonumber
\\* 
  I& =I_1{+}I_2{+}1 \text{ propagators}\;, & 
  \tilde{L}& =\tilde{L}_1{+}\tilde{L}_2{-}1 \text{ loops}\;,
\end{align}
because two loops of the subgraphs are merged to a new loop in the
total graph. It follows from (\ref{Euler}) that for tree-contractions
we always have additivity of the genus,
\begin{align}
  g=g_1+g_2 \;.
\label{ggg}
\end{align}

As an example for a contraction between graphs in the first line of
(\ref{as}) let us consider
\begin{align}
  \parbox{50mm}{
\begin{picture}(40,28)
\put(0,0){\epsfig{scale=.9,file=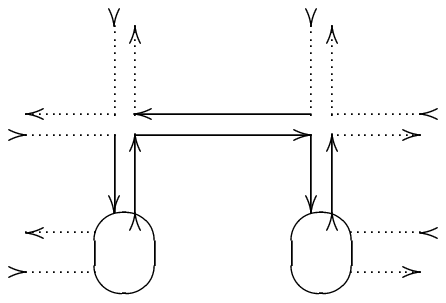,bb=71 605 187 684}}
\put(6,26){\mbox{\scriptsize$n_1$}}
\put(12.5,24){\mbox{\scriptsize$m_1$}}
\put(0,14.5){\mbox{\scriptsize$n_2$}}
\put(2,20){\mbox{\scriptsize$m_2$}}
\put(0,0.5){\mbox{\scriptsize$\sigma n$}}
\put(2,8){\mbox{\scriptsize$\sigma m$}}
\put(26.5,26){\mbox{\scriptsize$l_1$}}
\put(32.5,24){\mbox{\scriptsize$k_1$}}
\put(36,14.5){\mbox{\scriptsize$k_2$}}
\put(38,20){\mbox{\scriptsize$l_2$}}
\put(35,0){\mbox{\scriptsize$\sigma k$}}
\put(38,8){\mbox{\scriptsize$\sigma l$}}
\put(12.5,14){\mbox{\scriptsize$m$}}
\put(26,14){\mbox{\scriptsize$l$}}
\end{picture}}
\label{aa1}
\end{align}
where $\sigma m$ and $\sigma n$ stand for the set of all other
outgoing and incoming indices via external legs at the remaining part
of the left subgraph $\gamma_1$ and similarly for $\sigma k$ and
$\sigma l$ for the right subgraph $\gamma_2$. The two boundary
components to which the contracted vertices belong are joint in the
total graph, i.e.\ $B=B_1+B_2-1$. Moreover, we obviously have
$V^e=V^e_1+V^e_2$ and $\iota=\iota_1+\iota_2$. The graph (\ref{aa1})
determines the $\Lambda$-scaling
\begin{align}
  &\Lambda \frac{\partial}{\partial \Lambda}
  A^{(V,V^e,B,g,\iota)\gamma}_{m_1n_1;m_2n_2;
    \sigma m \,\sigma n;\sigma k \,\sigma l;k_2l_2;k_1l_1}[\Lambda]
  \nonumber
  \\*
  &= \frac{1}{2} \sum_{m,l}
  A^{(V_1,V^e_1,B_1,g_1,\iota_1)\gamma_1}_{
    m_1n_1;m_2n_2; \sigma m \,\sigma n;m m_1}[\Lambda] \,
  Q_{m_1m;ll_1}(\Lambda)\,
  A^{(V_2,V^e_2,B_2,g_2,\iota_2)\gamma_2}_{
    l_1l;\sigma k \,\sigma l; k_2l_2;k_1l_1}[\Lambda] \;.
\end{align}
Due to the conservation of the total amount of indices in $\gamma_1$
and $\gamma_2$ by induction hypothesis (\ref{ANnorm1}), both
\begin{align}
  m=\sigma n -\sigma m + n_2\qquad \text{and} \qquad l=\sigma k-\sigma
  l+k_2
\label{n'k'}
\end{align}
are completely fixed by the other external indices so that from the
sum over $m$ and $l$ there survives a single term only. Then, because
of the relation $m_1+l=m+l_1$ from the propagator
$Q_{m_1m;ll_1}(\Lambda)$, see (\ref{est0}), it follows that the total
amount of indices for
$A^{(V,V^e,B,g,\iota)\gamma}_{ m_1n_1;m_2n_2;
  \sigma m \,\sigma n;\sigma k \,\sigma l;k_2l_2;k_1l_1} $ is
conserved as well.
 
Let $\bar{V}^e_i$ and $\bar{\iota}_i$ be the numbers of external
vertices and segmentation indices on the segments of the subgraphs
$\gamma_i$ on which the contracted vertices are situated. The
induction hypothesis (\ref{ANnorm1}) gives us the bound if these
segments carry $\bar{s}_i \leq
\bar{V}^e_{i}+\bar{\iota}_i-1$ index summations.  The new segment of
the total graph $\gamma$ created by connecting the boundary components
of $\gamma_i$ carries $\bar{V}^e_{1}+\bar{V}^e_{2}$ external vertices
and $\bar{\iota}_1+\bar{\iota}_2$ segmentation indices and therefore
admits up to $\bar{s}_1+\bar{s}_2+1$ index
summations.  In (\ref{aa1}) that additional index summation will be
the $m_1$-summation.

Due to (\ref{numsum}) (for segments) there has to be an external leg
on each segment the outgoing index of which is not allowed to be
summed. If on the $\gamma_2$-part of the contracted segment there is
an unsummed external leg, we can choose $m$ as that particular index
in $\gamma_1$. In this case we take in the propagator the maximum over
$m,l$ and sum the part $\gamma_2$ for given $l$ over those indices
which belong to $\mathcal{E}^{s}$.  The result is bounded
independently of $l$ and all other incoming indices.  Next, we sum
over the indices in $\mathcal{E}^{s}$ which belong to $\gamma_1$,
regarding $m$ as an unsummed index. There is the possibility of an
$m_1$-summation applied to the propagator in the last step, with $l_1$
kept fixed, for which the bound is given by (\ref{est1}). In this case
we therefore get
\begin{align}
  & \sum_{\mathcal{E}^{s},\;\bar{s}_2 \leq
    \bar{V}^e_{2}+\bar{\iota}_2-1,\;m_1\in \mathcal{E}^{s}}
\Big|  \Lambda \frac{\partial}{\partial \Lambda} 
  A^{(V,V^e,B,g,\iota)\gamma}_{m_1n_1;m_2n_2;
    \sigma m \,\sigma n;\sigma k \,\sigma l;k_2l_2;k_1l_1}[\Lambda]
  \Big|\nonumber
  \\*
  &\qquad \leq \frac{1}{2} \Big( \sum_{\mathcal{E}_1^{s_1}}
  \big|A^{(V_1,V^e_1,B_1,g_1,\iota_1)\gamma_1}_{
    m_1n_1;m_2n_2; \sigma m \,\sigma n;m m_1}[\Lambda]\big| \Big)
  \Big(\max_{l_1} \sum_{m_1}\max_{m,l} \big|Q_{m_1m;l
    l_1}(\Lambda)\big|\Big) \nonumber
  \\*
  &\qquad\qquad\qquad \times \Big( \sum_{\mathcal{E}_2^{s_2}}
  \big|A^{(V_2,V^e_2,B_2,g_2,\iota_2)\gamma_2}_{
    l_1l;\sigma k \,\sigma l; k_2l_2;k_1l_1} [\Lambda]\big|\Big)
  \nonumber
  \\
  &\qquad \leq \frac{1}{2} C_1 \Big(\frac{\Lambda}{\mu}\Big)^{
    \delta_2(V-\frac{N+2}{2}+4-2g-(B+1))}
  \Big(\frac{\mu}{\Lambda}\Big)^{\delta_1(1+ V-V^e-\iota +2g+
    (B+1)-2 +(s-1))} \nonumber
  \\*
  &\qquad\qquad \qquad \times
  \Big(\frac{\mu}{\Lambda}\Big)^{\delta_0(
    V^e+\iota-2-(s-1))} P^{2V-\frac{N+2}{2}}\Big[\ln
  \frac{\Lambda}{\Lambda_R}\Big]\;.
\label{E0}
\end{align}
We have used the induction hypothesis (\ref{ANnorm1}) for the
subgraphs as well as (\ref{est1}) for the propagator and have inserted
$N_1+N_2=N+2$, $V_1+V_2=V$, $V^e_1+V^e_2=V^e$,
$\iota_1+\iota_2=\iota$, $B_1+B_2=B+1$,
$g_1+g_2=g$ and $s_1+s_2=s-1$,
because there is an additional summation over $m_1$ which belongs to
$\mathcal{E}^s$ but not to $\mathcal{E}^{s_i}_i$.  If $m_1 \not\in
\mathcal{E}^{s}$ we take instead the unsummed propagator and replace
in (\ref{E0}) one factor (\ref{est1}) by (\ref{est0}) as well as
$(s-1)$ by $s$. The total exponents of $\gamma$ remain unchanged.

Next, let there be no unsummed external leg on 
the contracted segment of $\gamma_2$ viewed from $\gamma$. Now,
we cannot directly use the induction hypothesis. On the other hand,
for a given index configuration of $\gamma_2$ and the propagator, the
index $k_2$ is not an independent summation index:
\begin{align}
  k_2 = l+\sigma l - \sigma k = m -m_1+l_1+\sigma l-\sigma k\;.
\label{k1l3}
\end{align}
See also (\ref{n'k'}). If $m_1 \in \mathcal{E}^{s}$ there must be an
unsummed outgoing index on the contracted segment of $\gamma_1$. We
can thus realise the $k_2$-summation as a summation over $m$ in
$\gamma_1$ for fixed index configuration of $\gamma_2$ and $m_1,l_1$.
This $m$-summation is applied together summation over the
$\gamma_1$-indices of $\mathcal{E}^{s}$ to $\gamma_1$ as the first
step, taking again the maximum of the propagator over $m,l$. In the
second step we sum over the restriction of $\mathcal{E}^{s}$ to
$\gamma_2$ and the propagator.  It is obvious that the estimation
(\ref{E0}) remains unchanged, in particular,
$s_1{+}s_2=(s_1{+}1)+(s_2{-}1)=s{-}1$. If $m_1$ is the only unsummed
index we realise the $k_2$-summation as a summation of the propagator
over $l$. Here, one has to take into account that the subgraph
$\gamma_2$ is bounded independently of the incoming index $l$.  Again
we get the same exponents as in (\ref{E0}).

We can summarise (\ref{E0}) and its discussed modification to
\begin{align}
  & \sum_{\mathcal{E}^{s}} \Big|\Lambda \frac{\partial}{\partial \Lambda}
  A^{(V,V^e,B,g,\iota)\gamma}_{m_1n_1;m_2n_2;
    \sigma m \,\sigma n;\sigma k \,\sigma l;k_2l_2;k_1l_1}[\Lambda]
  \Big|
\nonumber
\\*
&\leq \Big(\frac{\Lambda}{\mu}\Big)^{\delta_2(V-\frac{N}{2}+2-2g-B)} 
  \Big(\frac{\mu}{\Lambda}\Big)^{\delta_1(V-V^e-\iota+2g+B-1+s)}
  \Big(\frac{\mu}{\Lambda}\Big)^{\delta_0(V^e+\iota-1-s)}
\nonumber
\\*[-0.5ex]
&\qquad \times   
P^{2V-\frac{N}{2}-1}\Big[\ln \frac{\Lambda}{\Lambda_R}\Big]\;.
\label{Eall}
\end{align}
For the choice of the boundary conditions according to
Definition/Lemma~\ref{def-lem1}, the $\Lambda$-integration increases
(again according to Definition/Lemma~\ref{def-lem1}) the degree of the
polynomial in $\ln\frac{\Lambda}{\Lambda_R}$ by $1$. Hence, we have
extended (\ref{ANnorm1}) to a bigger degree $V$ for contractions of
type (\ref{aa1}). In particular, the bound is (by induction starting
with (\ref{est1}), which represents the third graph in
(\ref{examplesofgraphs})) independent of the incoming indices
$n_i,l_i$.

\bigskip

The verification of (\ref{ANnorm1}) for any contraction between graphs
of the first line in (\ref{as}) is performed in a similar manner.
Taking the same subgraphs as in (\ref{aa1}), but with a contraction of
other legs, the discussion is in fact a little easier because there
are no trajectories going through both subgraphs:
\begin{align}
  \parbox{70mm}{\begin{picture}(40,33)
      \put(0,0.3){\epsfig{scale=.9,file=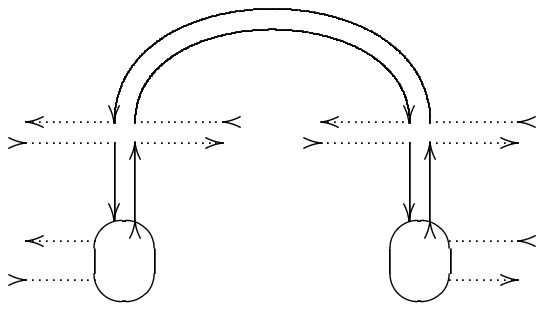,bb=71 591 215 677}}
      \put(15,14.5){\mbox{\scriptsize$m_1$}}
      \put(17,20){\mbox{\scriptsize$n_1$}}
      \put(0,14.5){\mbox{\scriptsize$n_2$}}
      \put(2,20){\mbox{\scriptsize$m_2$}}
      \put(0,0.5){\mbox{\scriptsize$\sigma n$}}
      \put(2,8){\mbox{\scriptsize$\sigma m$}}
      \put(46,14.5){\mbox{\scriptsize$k_2$}}
      \put(48,20){\mbox{\scriptsize$l_2$}}
      \put(31,14.5){\mbox{\scriptsize$l_1$}}
      \put(33,20){\mbox{\scriptsize$k_1$}}
      \put(46,0.5){\mbox{\scriptsize$\sigma k$}}
      \put(48,8){\mbox{\scriptsize$\sigma l$}}
\end{picture}}
\parbox{50mm}{
\begin{picture}(40,33)
\put(0,0.3){\epsfig{scale=.9,file=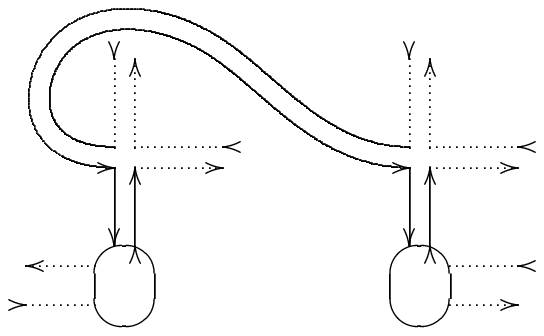,bb=71 591 215 685}}
\put(15,14.5){\mbox{\scriptsize$m_1$}}
\put(17,20){\mbox{\scriptsize$n_1$}}
\put(6,26){\mbox{\scriptsize$n_2$}}
\put(12,24){\mbox{\scriptsize$m_2$}}
\put(5,14.5){\mbox{\scriptsize$n$}}
\put(0,0.5){\mbox{\scriptsize$\sigma n$}}
\put(2,8){\mbox{\scriptsize$\sigma m$}}
\put(46,14.5){\mbox{\scriptsize$k_2$}}
\put(48,20){\mbox{\scriptsize$l_2$}}
\put(36,14.5){\mbox{\scriptsize$l$}}
\put(36,26){\mbox{\scriptsize$l_1$}}
\put(42,24){\mbox{\scriptsize$k_1$}}
\put(46,0.5){\mbox{\scriptsize$\sigma k$}}
\put(48,8){\mbox{\scriptsize$\sigma l$}}
\end{picture}}
\nonumber
\\
\parbox{70mm}{
\begin{picture}(40,30)
\put(0,0.3){\epsfig{scale=.9,file=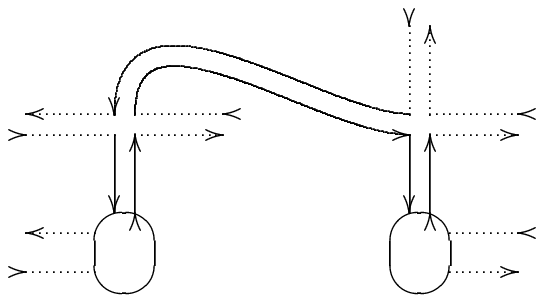,bb=71 591 215 670}}
\put(15,14.5){\mbox{\scriptsize$m_1$}}
\put(17,20){\mbox{\scriptsize$n_1$}}
\put(0,14.5){\mbox{\scriptsize$n_2$}}
\put(2,20){\mbox{\scriptsize$m_2$}}
\put(0,0.5){\mbox{\scriptsize$\sigma n$}}
\put(2,8){\mbox{\scriptsize$\sigma m$}}
\put(46,14.5){\mbox{\scriptsize$k_2$}}
\put(48,20){\mbox{\scriptsize$l_2$}}
\put(36,14.5){\mbox{\scriptsize$l$}}
\put(36,26){\mbox{\scriptsize$l_1$}}
\put(42,24){\mbox{\scriptsize$k_1$}}
\put(46,0.5){\mbox{\scriptsize$\sigma k$}}
\put(48,8){\mbox{\scriptsize$\sigma l$}}
\end{picture}}
\parbox{50mm}{
\begin{picture}(40,30)
\put(0,0.3){\epsfig{scale=.9,file=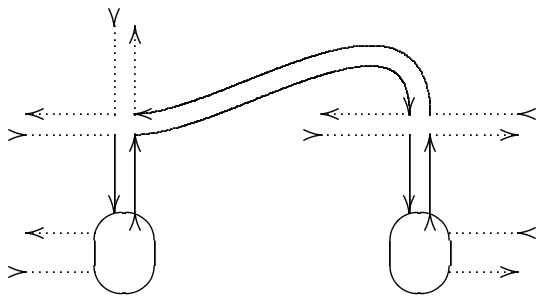,bb=71 591 215 670}}
\put(13,14.5){\mbox{\scriptsize$m$}}
\put(6,26){\mbox{\scriptsize$n_1$}}
\put(12,24){\mbox{\scriptsize$m_1$}}
\put(0,14.5){\mbox{\scriptsize$n_2$}}
\put(2,20){\mbox{\scriptsize$m_2$}}
\put(0,0.5){\mbox{\scriptsize$\sigma n$}}
\put(2,8){\mbox{\scriptsize$\sigma m$}}
\put(31,14.5){\mbox{\scriptsize$l_1$}}
\put(33,20){\mbox{\scriptsize$k_1$}}
\put(46,14.5){\mbox{\scriptsize$k_2$}}
\put(48,20){\mbox{\scriptsize$l_2$}}
\put(46,0.5){\mbox{\scriptsize$\sigma k$}}
\put(48,8){\mbox{\scriptsize$\sigma l$}}
\end{picture}}
\label{aa1a}
\end{align}
The contractions
\begin{align}
  \parbox{77mm}{
\begin{picture}(70,25)
\put(0,0.2){\epsfig{scale=.9,file=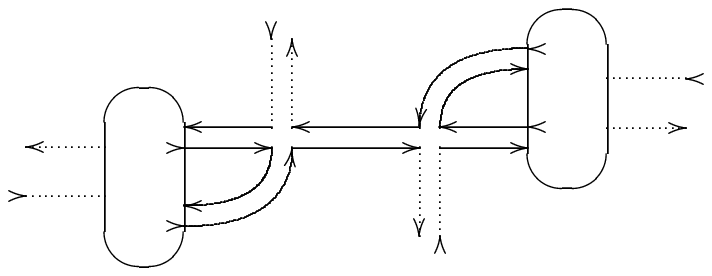,bb=71 608 263 684}}
\put(22,22){\mbox{\scriptsize$n_1$}}
\put(28,20){\mbox{\scriptsize$m_1$}}
\put(0,6){\mbox{\scriptsize$\sigma n$}}
\put(2,13.5){\mbox{\scriptsize$\sigma m$}}
\put(43,4){\mbox{\scriptsize$l_1$}}
\put(36.5,6){\mbox{\scriptsize$k_1$}}
\put(62,12.5){\mbox{\scriptsize$\sigma k$}}
\put(64,20.5){\mbox{\scriptsize$\sigma l$}}
\put(28.5,10){\mbox{\scriptsize$m$}}
\put(37,16){\mbox{\scriptsize$k$}}
\end{picture}}
\nonumber
\\*
\parbox{73mm}{
\begin{picture}(73,25)
\put(0,0.2){\epsfig{scale=.9,file=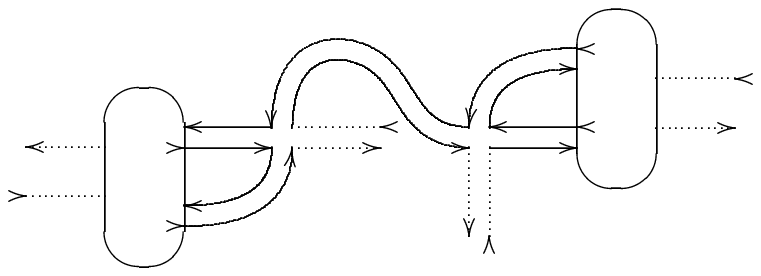,bb=71 608 263 684}}
\put(22,16){\mbox{\scriptsize$n$}}
\put(0,6){\mbox{\scriptsize$\sigma n$}}
\put(2,13.5){\mbox{\scriptsize$\sigma m$}}
\put(48,4){\mbox{\scriptsize$l_1$}}
\put(41.5,6){\mbox{\scriptsize$k_1$}}
\put(67,12.5){\mbox{\scriptsize$\sigma k$}}
\put(69,20.5){\mbox{\scriptsize$\sigma l$}}
\put(31,10){\mbox{\scriptsize$m_1$}}
\put(33,16){\mbox{\scriptsize$n_1$}}
\put(43,16){\mbox{\scriptsize$k$}}
\end{picture}}
\label{aa2}
\end{align}
are treated in the same way. The point is that the summation indices
of the propagator ($m,k$ for the upper graph and $n,k$ for the lower
graph in (\ref{aa2})) are fixed by index conservation for the
subgraphs.  In the same way one also discusses any contraction between
the second and third graph in (\ref{as}).

\bigskip

Let us now contract the left graph in the second line of (\ref{as}) with
any graph of the first line of (\ref{as}), e.g.\ 
\begin{align}
  \parbox{80mm}{
\begin{picture}(70,27)
\put(0,0){\epsfig{scale=.9,file=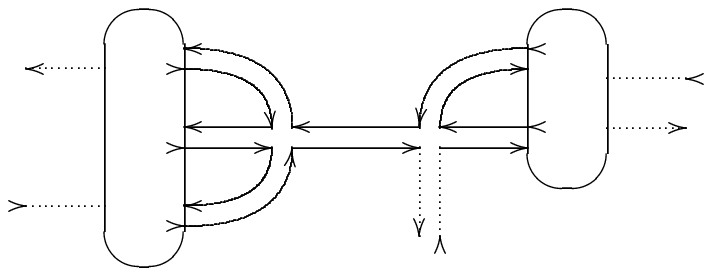,bb=71 608 263 684}}
\put(0,4){\mbox{\scriptsize$\sigma n$}}
\put(2,22){\mbox{\scriptsize$\sigma m$}}
\put(43,4){\mbox{\scriptsize$l_1$}}
\put(36.5,6){\mbox{\scriptsize$k_1$}}
\put(62,12){\mbox{\scriptsize$\sigma k$}}
\put(64,20.5){\mbox{\scriptsize$\sigma l$}}
\put(28.5,10){\mbox{\scriptsize$m$}}
\put(28,16){\mbox{\scriptsize$n$}}
\put(37,16){\mbox{\scriptsize$k$}}
\end{picture}}
\label{aa3}
\end{align}
The number of boundary components is reduced by $1$, giving
$B_1+B_2=B+1$.  We clearly have $\iota=\iota_1+\iota_2$, but there is
now one external vertex less on which we can apply an index summation,
$V^e=V^e_1+V^e_2-1$. At the same time we
need the index summation from the subgraph, because in the
$\Lambda$-scaling
\begin{align}
 \Lambda \frac{\partial}{\partial \Lambda} &
  A^{(V,V^e,B,g,\iota) \gamma}_{ k_1l_1;\sigma
    k\,\sigma l;\sigma m\,\sigma n}[\Lambda] 
\nonumber
\\*
&= \frac{1}{2} \sum_{m,n,k} A^{(V_1,V^e_1,B_1,
    g_1, \iota_1)\gamma_1}_{mn;\sigma m \, \sigma n}[\Lambda] \,
  Q_{nm;k_1k}(\Lambda)\, A^{(V_2,V^e_2,B_2,g_2,
    \iota_2)\gamma_2}_{ kk_1;k_1l_1;\sigma k \, \sigma l}[\Lambda]
\label{aa3L}
\end{align}
there is now one undetermined summation index:
\begin{align}
  k= l_1+\sigma l - \sigma k\;,\qquad m(n)=n + \sigma n - \sigma m\;.
\end{align}

First, let there be an additional unsummed external leg on the segment
of $m,n$ in $\gamma_1$. Then, the induction hypothesis (\ref{ANnorm1})
gives the bound for a summation over $m$. We thus fix $n,k$ and all
indices of $\gamma_2$ in the first step and realise a possible
$k_1$-summation due to $k_1=m+k-n$ as an $m$-summation, which is
applied together with the summation over the $\gamma_1$-indices of
$\mathcal{E}^{s}$, after maximising the propagator over $m,k_1$. The
result is independent of $n$. We thus restrict the $n$-summation to
the propagator, see (\ref{est1}), and apply the remaining
$\mathcal{E}^{s}$-summations to $\gamma_2$, where $k$ remains
unsummed.  We have $s_1+s_2=s$ and get the estimation
\begin{align}
  &\sum_{\mathcal{E}^{s} \ni k_1,\;\bar{s}_1\leq
    \bar{V}^e_1+\bar{\iota}_1-2} \Big|\Lambda
  \frac{\partial}{\partial \Lambda}
  A^{(V,V^e,B,g,\iota) \gamma}_{k_1l_1;\sigma
    k\,\sigma l;\sigma m\,\sigma n}[\Lambda] \Big|\nonumber
  \\*
  &\leq \frac{1}{2} \Big(\sum_{m,\mathcal{E}_1^{s_1}} \big|
  A^{(V_1,V^e_1,B_1,g_1, \iota_1)\gamma_1}_{
    mn;\sigma m \, \sigma n}[\Lambda] \big|\Big) \Big(
  \max_{k}\sum_{n} \max_{m,k_1} \big|Q_{nm;k_1k}(\Lambda) \big|\Big)
  \nonumber
  \\*
  & \qquad\qquad \times \Big( \sum_{\mathcal{E}_2^{s_2}\not\ni
    k_1} \big| A^{(V_2,V^e_2,B_2,g_2,
    \iota_2)\gamma_2}_{kk_1; k_1l_1;\sigma k \, \sigma l} [\Lambda]\big|
  \Big) \nonumber
  \\
  &\leq \frac{1}{2} C_1 \Big(
  \frac{\Lambda}{\mu}\Big)^{\delta_2(V-\frac{N+2}{2}+4 -2g-(B+1))}
  \Big(\frac{\mu}{\Lambda}\Big)^{\delta_1(1+ V-(V^e+1)-\iota 
+2g +(B+1)-2+s) } \nonumber
  \\*
  &\qquad\qquad \times \Big(\frac{\mu}{\Lambda}\Big)^{\delta_0(
    (V^e+1)+\iota-2-s)} P^{2V-\frac{N+2}{2}}\Big[\ln
  \frac{\Lambda}{\Lambda_R}\Big]\;.
\label{aa3LE1}
\end{align}
If $k_1 \not\in \mathcal{E}^{s}$ we do not need the $m$-summation on
$\gamma_1$. Again we have $s=s_1+s_2$ and (\ref{aa3LE1}) remains
unchanged. Here, we may allow for index summations at all other
external legs on the segment of $m,n$ in $\gamma_1$.

If there is no unsummed external leg on the segment of $m,n$ in
$\gamma_1$, we must realise the $k_1$-summation as follows: We proceed
as before up to the step where we sum the propagator for given $k$
over $n$. For each term in this sum we have $k_1=k+\sigma n-\sigma m$.
We thus achieve a different $k_1$ if for given $\sigma m,\sigma n$ we
start from a different $k$. Since the result of the summations over
$\gamma_1$ and the propagator is independent of $k$, see (\ref{est1}),
we realise the $k_1$-summation as a sum over $k$ restricted to
$\gamma_2$.  We now get the same exponents as in (\ref{aa3LE1}) also
for this case.  According to Definition/Lemma \ref{def-lem1}, the
$\Lambda$-integration extends for contractions of type (\ref{aa3}) the
bound (\ref{ANnorm1}) to a bigger order $V$.

\bigskip

The contraction of the other leg of the right vertex
\begin{align}
  \parbox{80mm}{
\begin{picture}(70,27)
\put(0,0){\epsfig{scale=.9,file=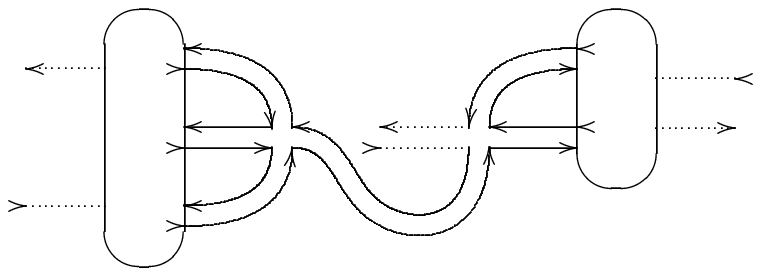,bb=71 608 277 684}}
\put(0,4){\mbox{\scriptsize$\sigma n$}}
\put(2,22){\mbox{\scriptsize$\sigma m$}}
\put(48,10){\mbox{\scriptsize$l$}}
\put(67,12.5){\mbox{\scriptsize$\sigma k$}}
\put(69,20.5){\mbox{\scriptsize$\sigma l$}}
\put(28.5,9){\mbox{\scriptsize$m$}}
\put(28,16){\mbox{\scriptsize$n$}}
\put(38,16){\mbox{\scriptsize$k_1$}}
\put(36,10){\mbox{\scriptsize$l_1$}}
\end{picture}}
\label{aa3a}
\end{align}
is easier to discuss because the $k_1$-summation is directly applied
to $\gamma_2$. Taking the second vertex of the first line of (\ref{as})
instead, we have two contractions which are identical to (\ref{aa3})
and (\ref{aa3a}) and a third one with contractions as in the first and
last graphs of (\ref{aa1a}) where $\gamma_1$ and $\gamma_2$ form different
segments in $\gamma$. This case is much easier because there is no
trajectory involving both subgraphs.

Moreover, contracting the last instead of the first vertex of the second
line of (\ref{as}) gives the same estimates if the two propagators
between the vertex and the oval belong to the same segment:
\begin{align}
  \parbox{82mm}{
\begin{picture}(82,30)
\put(0,0.2){\epsfig{scale=.9,file=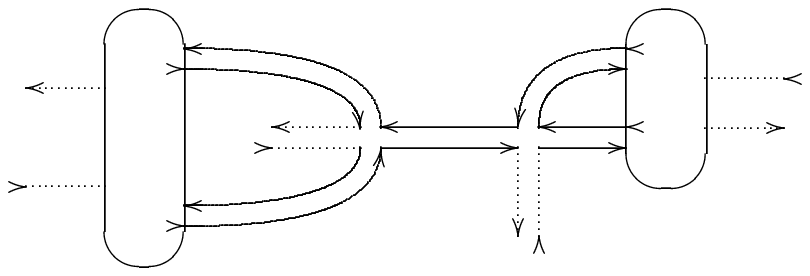,bb=71 608 291 684}}
\put(1,7){\mbox{\scriptsize$\sigma n$}}
\put(3,19){\mbox{\scriptsize$\sigma m$}}
\put(25.5,10){\mbox{\scriptsize$n_1$}}
\put(27.5,16){\mbox{\scriptsize$m_1$}}
\put(53.5,4){\mbox{\scriptsize$l_1$}}
\put(47,6){\mbox{\scriptsize$k_1$}}
\put(72.5,12.5){\mbox{\scriptsize$\sigma k$}}
\put(74.5,20.5){\mbox{\scriptsize$\sigma l$}}
\put(38,10){\mbox{\scriptsize$n$}}
\put(38,16){\mbox{\scriptsize$m$}}
\put(47.5,15.5){\mbox{\scriptsize$l$}}
\end{picture}}
\label{aa7}
\end{align}
The only modification to (\ref{aa3LE1}) and its variants is to replace
$(V^e+1)$ by $V$ and $\iota$ by $(\iota+1)$, because the
total number of external vertices is unchanged whereas the total
segmentation index is reduced by $1$.

\bigskip

If we contract the second vertex of the last line in (\ref{as}) in
such a way that the contracted indices $m,n$ belong to different
segments of $\gamma_1$, e.g.\
\begin{align}
  \parbox{97mm}{
\begin{picture}(60,36)
\put(0,0){\epsfig{scale=.9,file=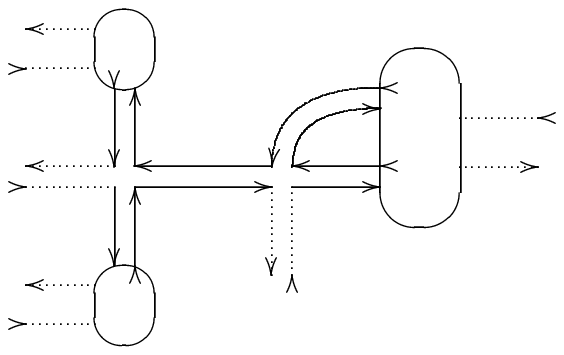,bb=71 585 221 684}}
\put(0,1){\mbox{\scriptsize$\sigma n$}}
\put(2,8){\mbox{\scriptsize$\sigma m$}}
\put(0,26){\mbox{\scriptsize$\sigma n'$}}
\put(2,34.5){\mbox{\scriptsize$\sigma m'$}}
\put(0,14){\mbox{\scriptsize$n_1$}}
\put(2,20){\mbox{\scriptsize$m_1$}}
\put(28,8){\mbox{\scriptsize$l_1$}}
\put(21.5,10){\mbox{\scriptsize$k_1$}}
\put(47,16){\mbox{\scriptsize$\sigma k$}}
\put(49,24.5){\mbox{\scriptsize$\sigma l$}}
\put(12.5,14){\mbox{\scriptsize$m$}}
\put(12.5,20){\mbox{\scriptsize$n$}}
\put(22,19.5){\mbox{\scriptsize$k$}}
\end{picture}}
\nonumber
\\*[-5ex]
\parbox{75mm}{
\begin{picture}(75,32)
\put(0,0){\epsfig{scale=.9,file=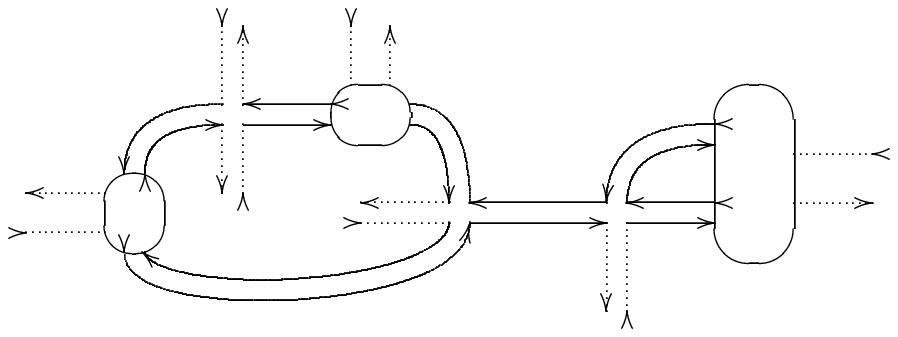,bb=71 600 317 684}}
\put(0,6.5){\mbox{\scriptsize$\sigma n$}}
\put(2,13.5){\mbox{\scriptsize$\sigma m$}}
\put(28.5,28){\mbox{\scriptsize$\sigma n'$}}
\put(38,26){\mbox{\scriptsize$\sigma m'$}}
\put(34.5,7){\mbox{\scriptsize$n_1$}}
\put(36.5,13){\mbox{\scriptsize$m_1$}}
\put(23.5,12.5){\mbox{\scriptsize$n_3$}}
\put(15.5,14.5){\mbox{\scriptsize$m_3$}}
\put(16.5,27){\mbox{\scriptsize$n_2$}}
\put(23.5,25){\mbox{\scriptsize$m_2$}}
\put(62.5,1){\mbox{\scriptsize$l_1$}}
\put(56,3){\mbox{\scriptsize$k_1$}}
\put(81.5,9){\mbox{\scriptsize$\sigma k$}}
\put(83.5,17.5){\mbox{\scriptsize$\sigma l$}}
\put(47,6.5){\mbox{\scriptsize$m$}}
\put(47,12.5){\mbox{\scriptsize$n$}}
\put(56.5,12.5){\mbox{\scriptsize$k$}}
\end{picture}}
\label{aa5}
\end{align}
they are actually determined by index conservation for the segments.
The entire discussion of these examples is therefore similar to the
graph (\ref{aa1}) with bound (\ref{E0}) and its modifications.  Note
that we have $V^e=V^e_1+V^e_2$ and
$\iota=\iota_1+\iota_2$ in (\ref{aa5}).

Accordingly, we can replace in all previous examples a vertex of the
first line of (\ref{as}) by the composed vertex under the condition
that the two contracted trajectories at the composed vertex belong to
different segments.

\bigskip

It remains to study the contraction
\begin{align}
  \parbox{80mm}{
\begin{picture}(70,27)
\put(0,0){\epsfig{scale=.9,file=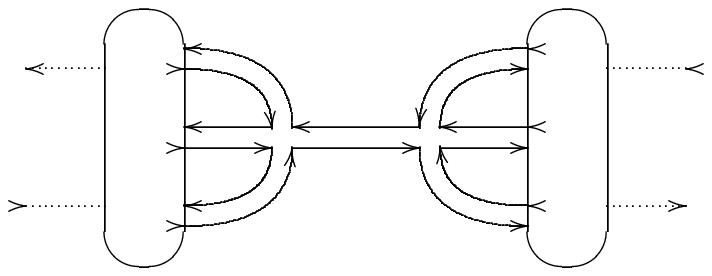,bb=71 608 263 684}}
\put(0,4){\mbox{\scriptsize$\sigma n$}}
\put(2,22){\mbox{\scriptsize$\sigma m$}}
\put(62,4){\mbox{\scriptsize$\sigma k$}}
\put(64,22){\mbox{\scriptsize$\sigma l$}}
\put(28.5,10){\mbox{\scriptsize$m$}}
\put(28,16){\mbox{\scriptsize$n$}}
\put(37,16){\mbox{\scriptsize$k$}}
\put(37,9.5){\mbox{\scriptsize$l$}}
\end{picture}}
\label{aa4}
\end{align}
where two contraction indices ($m$ or $n$ and $k$ or $l$) are
undetermined. We have $V^e=V^e_1+V^e_2-2$ and $\iota=\iota_1+\iota_2$.
We first assume that at least one of the boundary components of
$\gamma_i$ to contract carries more than one external vertex. In this
case we have $B=B_1+B_2-1$.  There has to be at least one unsummed
external vertex on the segment, say on $\gamma_2$. We fix the indices
of $\gamma_2$ as well as $n$ in the first step, take in the propagator
the maximum over $m,l$ and sum over the $\gamma_1$-indices of
$\mathcal{E}^{s}$. Here, $m$ can be regarded as an unsummed index. We
take the maximum of $\gamma_1$ over $n$ so that the $n$-summation
restricts to the propagator only. We take in the summed propagator the
maximum over $k$ so that the remaining $k$-summation is applied
together with the summation over the $\gamma_2$-indices of
$\mathcal{E}^{s}$. We thus need $s_1+s_2=s+1$ summations
and the bound (\ref{est1}) for the propagator:
\begin{align}
  &\sum_{\mathcal{E}^{s} } \Big|\Lambda
  \frac{\partial}{\partial \Lambda}
  A^{(V,V^e,B,g,\iota) \gamma}_{\sigma k\,\sigma
    l;\sigma m\,\sigma n}[\Lambda] \Big|\nonumber
  \\*
  &\leq \frac{1}{2} \Big(\max_n \sum_{\mathcal{E}_1^{s_1}} \big|
  A^{(V_1,V^e_1,B_1,g_1, \iota_1)\gamma_1}_{
    mn;\sigma m \, \sigma n}[\Lambda] \big|\Big) \Big(
  \max_{k}\sum_{n} \max_{m,l} \big|Q_{nm;lk}(\Lambda) \big|\Big)
  \nonumber
  \\*
  & \qquad\qquad \times \Big( \sum_{k,\mathcal{E}_2^{s_2} } \big|
  A^{(V_2,V^e_2,B_2,g_2,
    \iota_2)\gamma_2}_{kl;\sigma k \, \sigma l} [\Lambda]\big| \Big)
  \nonumber
  \\
  &\leq \frac{1}{2} C_1 \Big(
  \frac{\Lambda}{\mu}\Big)^{\delta_2(V-\frac{N+2}{2}+4 -2g-(B+1))}
  \Big(\frac{\mu}{\Lambda}\Big)^{\delta_1(1+
    V-(V^e+2)-\iota+2g+(B+1)-2+(s+1)) } \nonumber
  \\*
  &\qquad\qquad \times \Big(\frac{\mu}{\Lambda}\Big)^{\delta_0(
    (V^e+2)+\iota-2-(s+1))}
  P^{2V-\frac{N+2}{2}}\Big[\ln \frac{\Lambda}{\Lambda_R}\Big]\;.
\label{aa4E}
\end{align}

Finally, we have to consider the case where the only external vertices
of both boundary components of $\gamma_i$ to contract are just the
contracted vertices. In this case the contraction removes these two
boundary components at expense of a completely inner loop, giving
$B=B_1+B_2-2$.  The differences $n-m$ and $k-l$ are fixed by the
remaining indices of $\gamma_i$. For given $m$ we may thus take the
maximum of $\gamma_2$ over $l$ and realise the $l$-summation as a
summation (\ref{est1}) over the propagator. We thus exhaust all
differences $m-l$. In order to exhaust all values of $m$ we take the
maximum of $\gamma_1$ over $m,n$ and multiply the result by a volume
factor (\ref{volumefactor}).  We thus replace in (\ref{aa4E})
$(s+1)\mapsto s$ and $(B+1)\mapsto (B+2)$, and combine one factor
(\ref{est0}) and a volume factor (\ref{volumefactor}) to (\ref{est2}).
We thus get the same total exponents as in (\ref{ANnorm1}) so that the
$\Lambda$-integration extends (\ref{ANnorm1}) to a bigger order $V$
for all contractions represented by (\ref{aa4}).

\bigskip

The contractions
\begin{align}
  \parbox{92mm}{
\begin{picture}(82,30)
\put(0,0.1){\epsfig{scale=.9,file=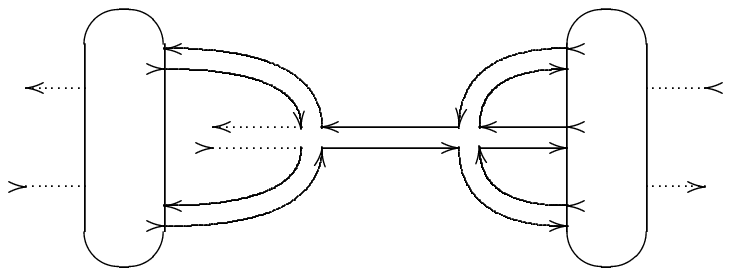,bb=71 608 269 684}}
\put(1,7){\mbox{\scriptsize$\sigma n$}}
\put(1.5,19.5){\mbox{\scriptsize$\sigma m$}}
\put(19.5,10){\mbox{\scriptsize$n_1$}}
\put(21.5,16){\mbox{\scriptsize$m_1$}}
\put(65.5,6){\mbox{\scriptsize$\sigma k$}}
\put(67,19.5){\mbox{\scriptsize$\sigma l$}}
\put(32,10.5){\mbox{\scriptsize$m$}}
\put(32,16){\mbox{\scriptsize$n$}}
\put(41.5,15.5){\mbox{\scriptsize$k$}}
\put(41.5,10){\mbox{\scriptsize$l$}}
\end{picture}}
\nonumber
\\*[-3ex]
\parbox{73mm}{
\begin{picture}(73,30)
\put(0,0.1){\epsfig{scale=.9,file=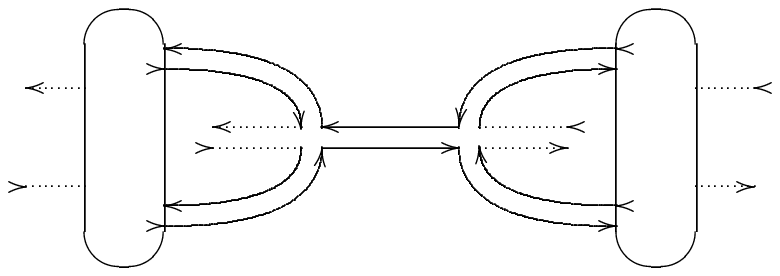,bb=71 608 283 684}}
\put(1,7){\mbox{\scriptsize$\sigma n$}}
\put(1.5,19.5){\mbox{\scriptsize$\sigma m$}}
\put(19.5,10.5){\mbox{\scriptsize$n_1$}}
\put(21.5,16){\mbox{\scriptsize$m_1$}}
\put(71,6){\mbox{\scriptsize$\sigma k$}}
\put(72,19.5){\mbox{\scriptsize$\sigma l$}}
\put(50.5,10){\mbox{\scriptsize$k_1$}}
\put(52.5,16){\mbox{\scriptsize$l_1$}}
\put(32,10){\mbox{\scriptsize$m$}}
\put(32,16){\mbox{\scriptsize$n$}}
\put(41.5,15.5){\mbox{\scriptsize$k$}}
\put(41.5,10){\mbox{\scriptsize$l$}}
\end{picture}}
\label{aa8}
\end{align}
are treated in the same way as (\ref{aa4}), now with the two unknown
summation indices taken into account by a reduction of $V^e+\iota
=(V^e_1+\iota_1)+(V^e_2+\iota_2)-2$. In particular, there is also the
situation where $m,n$ and $k,l$ are the only external legs of their
boundary components before the contraction. In this case the number of
boundary components drops by $2$, which requires a volume factor in
order to realise the sum over the starting point of the inner loop.

\bigskip

Thus, (\ref{ANnorm1}) is proven for any contractions produced by the
first (bilinear) term on the rhs of (\ref{polL2}).

\subsection{Loop-contractions at the same vertex}
\label{self-contr-1}

It remains to verify the scaling formula (\ref{ANnorm1}) for the
second term (the last line) on the rhs of the Polchinski equation
(\ref{polL2}), which describes self-contractions of graphs. The
graphical data for the subgraph will obtain a subscript 1, such as the
number of external vertices $V^e_1$, the segmentation index $\iota_1$
and the set $\mathcal{E}_1^{s_1}$ of summation indices.  We always
have $V_1=V$ and $N_1=N+2$.  We first consider contractions of
external lines at the same vertex, for which we have the possibilities
shown in (\ref{as}).

The very first vertex leads to two different self-contractions:
\begin{align}
  \parbox{20mm}{\begin{picture}(20,30)
      \put(0,0){\epsfig{scale=.9,file=a12,bb=71 609 134 687}}
      \put(-1,6){\mbox{\scriptsize$m_1$}}
      \put(3,0){\mbox{\scriptsize$n_1$}}
      \put(14,2){\mbox{\scriptsize$m_2$}}
      \put(21,4){\mbox{\scriptsize$n_2$}}
      \put(10,16){\mbox{\scriptsize$l$}}
    \end{picture}}
+~~ \parbox{20mm}{\begin{picture}(20,30)
    \put(0,0){\epsfig{scale=.9,file=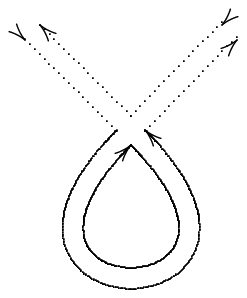,bb=71 606 134 684}}
    \put(-1,22){\mbox{\scriptsize$n_1$}}
    \put(4,26){\mbox{\scriptsize$m_1$}}
    \put(16,28){\mbox{\scriptsize$n_2$}}
    \put(21,22){\mbox{\scriptsize$m_2$}}
    \put(10,11){\mbox{\scriptsize$l$}}
\end{picture}}
&& \Big|\Lambda \frac{\partial}{\partial \Lambda} 
A^{(1,1,1,0,0)}_{m_1n_1;m_2n_2}[\Lambda] \Big| 
&=  \sum_l |Q_{m_1 l;ln_2}(\Lambda)|\, \delta_{n_1m_2} 
\nonumber \\[-8mm]
&&&+ \sum_l |Q_{ln_1;m_2l}(\Lambda)|\, \delta_{n_2m_1} \;,
\label{A2planar}
\\
    \parbox{40mm}{\begin{picture}(30,25)
    \put(0,0){\epsfig{scale=.9,file=a12b,bb=71 626 158 684}}
    \put(0,6.5){\mbox{\scriptsize$n_1$}}
    \put(3,12.5){\mbox{\scriptsize$m_1$}}
    \put(15,6.5){\mbox{\scriptsize$m_2$}}
    \put(17,12.5){\mbox{\scriptsize$n_2$}}
  \end{picture}}
&& \Big|\Lambda \frac{\partial}{\partial \Lambda} 
A^{(1,1,2,0,1)}_{m_1n_1;m_2n_2}[\Lambda]\Big| 
&= |Q_{m_1 n_2;m_2 n_1}(\Lambda)|\;.
\label{A2nonplanar}
\end{align}
For the planar contraction (\ref{A2planar}) we estimate the
$l$-summation by a volume factor so that we obtain (\ref{ANnorm1})
from (\ref{est2}). For the non-planar graph (\ref{A2nonplanar}) we
obtain (\ref{ANnorm1}) for $s=0$ directly from (\ref{est0}). According
to (\ref{numsum}) we can apply one index summation which yields
(\ref{ANnorm1}) via (\ref{est1}).

\bigskip

For the second graph in the first line of (\ref{as}) we first
investigate the contraction
\begin{align}
  {} \parbox{50mm}{
\begin{picture}(50,20)
\put(0,0){\epsfig{scale=.9,file=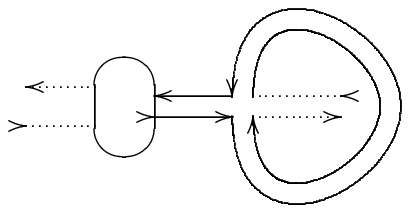,bb=71 611 181 670}}
\put(0,6.5){\mbox{\scriptsize$\sigma n$}}
\put(2,13.5){\mbox{\scriptsize$\sigma m$}}
\put(30,12.5){\mbox{\scriptsize$n_1$}}
\put(28,7){\mbox{\scriptsize$m_1$}}
\put(19,13){\mbox{\scriptsize$l$}}
\put(18,6){\mbox{\scriptsize$k$}}
\end{picture}}
\label{al1}
\end{align}
The number of loops of the amputated graph is increased by $1$,
$\tilde{L}=\tilde{L}_1+1$, so that due to (\ref{Euler}) and $I=I_1+1$
we get $g=g_1$.  The graph (\ref{al1}) determines the
$\Lambda$-variation
\begin{align}
  \Lambda \frac{\partial}{\partial \Lambda}
  A^{(V,V^e,B,g,\iota)\gamma}_{m_1n_1;\sigma m\, \sigma
    n}[\Lambda] &= -\frac{1}{2} \sum_{k,l} Q_{m_1k;ln_1}(\Lambda)\,
  A^{(V_1,V^e_1,B_1,g_1,\iota_1)
    \gamma_1}_{m_1n_1;n_1l;\sigma m\, \sigma n;km_1} [\Lambda]\;,
\end{align}
with one of the indices $k,l$ being undetermined. First, let there be
at least one further external leg on the same boundary component as
$l,k$. In this case the number of boundary components is increased by
$1$, $B=B_1+1$. If there is an unsummed index on the segment
of $k,l$ we can realise the
$k$-summation in $\gamma$ as a summation in $\gamma_1$ after taking in
the propagator the maximum over $k,l$. We thus have $s_1=s+1$ and
consequently
\begin{align}
  \sum_{\mathcal E^{s},\;m_1\notin \mathcal{E}^{s}} &\Big|\Lambda
  \frac{\partial}{\partial \Lambda}
  A^{(V,V^e,B,g,\iota)\gamma}_{m_1n_1;\sigma m\, \sigma n}
[\Lambda] \Big| \nonumber
  \\*
  &\leq \frac{1}{2} \Big(\max_{k,l}
  \big|Q_{m_1k;ln_1}(\Lambda)\big|\Big) \Big(
  \sum_{k,\mathcal{E}_1^{s}}
  \big|A^{(V,V^e_1,B_1,g_1,\iota_1)\gamma_1}_{m_1n_1; n_1l;\sigma m\,
    \sigma n;km_1} [\Lambda]\Big) \nonumber
  \\
  &\leq \frac{1}{2} C_0 \Big(\frac{\Lambda}{\mu_0}\Big)^{\delta_2(
    V-\frac{N+2}{2}+2-2g-(B-1))}
  \Big(\frac{\mu}{\Lambda}\Big)^{\delta_1(V-V^e-\iota +2
  g+(B-1)-1+(s+1))} 
  \nonumber
  \\*
  &\qquad\qquad \times
  \Big(\frac{\mu}{\Lambda}\Big)^{\delta_0(1+V^e+\iota
-1-(s+1))}
  P^{2V-\frac{N+2}{2}}\Big[\ln \frac{\Lambda}{\Lambda_R}\Big]\;.
\label{al1E}
\end{align}
We can sum the contracting propagator over $m_1$ for fixed $n_1$,
which amounts to replace one factor (\ref{est0}) by (\ref{est1})
compensated by $s=s_1$ replacing $s=s_1-1$.

If $k$ cannot be a summation index in $\gamma_1$ then $m_1$ must be
unsummed in $\gamma$. We first apply the summation over
$\mathfrak{o}[l]$ for given $l$ in $\gamma_1$. The result is
independent of $l$ so that, for given $k$, the $l$-summation can be
restricted to the contracting propagator maximised over $m_1,n_1$.
Finally, the remaining $\mathcal{E}^s$-summations are applied. We have
to replace in (\ref{al1E}) $(s+1)$ by $s$ and one factor (\ref{est0})
by (\ref{est1}).

Finally, let there be no further external leg on the same boundary
component as $l,k$. Now the number of boundary components remains
constant, $B=B_1$.  Since $k-l=n_1-m_1$ is a constant, the required
summation over e.g.\ $k$ has to be estimated by a volume factor
(\ref{volumefactor}). We thus replace in (\ref{al1E}) $(B-1)\mapsto B$
and $(s+1) \mapsto s$ and combine one factor (\ref{est0}) and the
volume factor to (\ref{est2}).

In summary, we extend after $\Lambda$-integration the scaling law
(\ref{ANnorm1}) for the same degree $V$ to a reduced number $N$ of
external lines.

\bigskip

Next, we study the following contraction of the second graph in
(\ref{as}) which gives rise to an inner loop:
\begin{align}
  \parbox{50mm}{
\begin{picture}(50,20)
\put(0,0){\epsfig{scale=.9,file=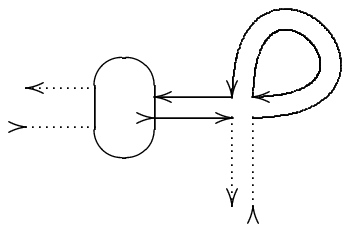,bb=71 625 164 684}}
\put(0,6.5){\mbox{\scriptsize$\sigma n$}}
\put(2,13.5){\mbox{\scriptsize$\sigma m$}}
\put(17,12.5){\mbox{\scriptsize$n_1$}}
\put(24,13){\mbox{\scriptsize$l$}}
\put(17,3){\mbox{\scriptsize$m_1$}}
\put(24,1){\mbox{\scriptsize$n_1$}}
\end{picture}}
\label{al2}
\end{align}
It describes the $\Lambda$-variation
\begin{align}
  \Lambda \frac{\partial}{\partial \Lambda}
  A^{(V,V^e,B,g,\iota)\gamma}_{ m_1n_1;\sigma m\,
    \sigma n}[\Lambda] &= -\frac{1}{2} \Big(\sum_{l}
  Q_{n_1l;ln_1}(\Lambda)\Big)\,
  A^{(V_1,V^e_1,B_1,g_1,\iota_1)\gamma_1}_{
    m_1n_1;n_1l;ln_1;\sigma m\,\sigma n} [\Lambda]\;.
\end{align}
The number of loops of the amputated graph is increased by $1$ and
the number of boundary components remains constant, giving
$g=g_1$ and $B=B_1$. Note that
$A^{(V,V^e_1,B_1,g_1,\iota_1)\gamma_1}_{m_1n_1;n_1l;
  ln_1;\sigma m\,\sigma n}$ is independent of $l$ so that the
$l$-summation acts on the propagator only. We estimate the $l$-summed
propagator by (\ref{est2}) for the
product of (\ref{est0}) with a volume factor (\ref{volumefactor}).
The factor (\ref{est2}) compensates the decrease
$N=N_1-2$, all other exponents remain unchanged when passing from
$\gamma_1$ to $\gamma$.  Now the $\Lambda$-integration extends the
scaling law (\ref{ANnorm1}) to a reduced $N$.

\bigskip

The third graph in the first line of (\ref{as}) leads to the
contracted graph
\begin{align}
  \parbox{50mm}{
\begin{picture}(50,23)
\put(0,0){\epsfig{scale=.9,file=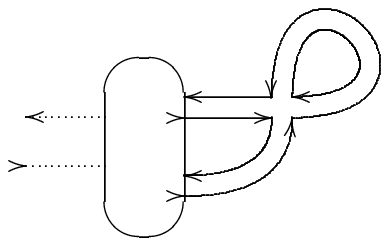,bb=71 616 175 684}}
\put(0,6){\mbox{\scriptsize$\sigma n$}}
\put(2,13.5){\mbox{\scriptsize$\sigma m$}}
\put(22,15.5){\mbox{\scriptsize$n$}}
\put(28,10){\mbox{\scriptsize$n$}}
\put(28,15.5){\mbox{\scriptsize$l$}}
\end{picture}}
\label{al3}
\end{align}
There is one additional loop of the amputated graph, giving
$g=g_1$. We have $B=B_1$ 
if there are further external legs on the boundary component
of $n$ and $B=B_1-1$ if no
further external leg exists on the contracted boundary component. Very
similar to (\ref{al2}), the $l$-summation is restricted to the
propagator maximised over $n$, giving a factor (\ref{est2}) which
compensates $N=N_1-2$ in the first exponent of (\ref{ANnorm1}). For
$B=B_1$ the $n$-summation in (\ref{al3}) is provided by the subgraph
$\gamma_1$, where the additional summation $s_1=s+1$ compared with $\gamma$
compensates the change $V^e_1=V^e+1$ of external
vertices in the second and third exponent of (\ref{ANnorm1}).

On the other hand, if $B_1=B+1$ we have $s=s_1$ and the summation over
$n$ has to come from a volume factor (\ref{volumefactor}) combined
with one factor (\ref{est0}) to (\ref{est2}). This verifies
(\ref{ANnorm1}) for the contraction (\ref{al3}).

\bigskip

The last case for which contractions of two external lines at the same
vertex are to investigate is the last vertex in the second line of
(\ref{as}). As before in the proof for tree-contractions, we have to
distinguish whether the composed vertex under consideration appears
inside a tree, in a loop but together with further composed vertices,
or in a loop but as the single composed vertex. In the first case we
have to analyse the graph
\begin{align}
  \parbox{50mm}{
\begin{picture}(50,28)
\put(0,0){\epsfig{scale=.9,file=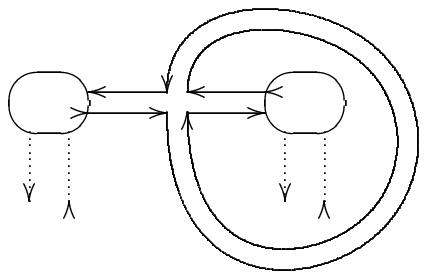,bb=71 605 191 683}}
\put(-3,10){\mbox{\scriptsize$\sigma m$}}
\put(7,8){\mbox{\scriptsize$\sigma n$}}
\put(22.5,10){\mbox{\scriptsize$\sigma m'$}}
\put(33,9){\mbox{\scriptsize$\sigma n'$}}
\put(20,19.5){\mbox{\scriptsize$k$}}
\put(20,14){\mbox{\scriptsize$n$}}
\put(13.5,19.5){\mbox{\scriptsize$l$}}
\put(13.5,14){\mbox{\scriptsize$m$}}
\end{picture}}
\label{al4}
\end{align}
Before the contraction, the indices $m,n,k,l$ were all located on the
same loop of the amputated graph and the same boundary component.
After the contraction they are split into two loops,
$g=g_1$. The number of boundary components is
increased by $1$ if both resulting boundary components of $l,m$ and
$k,n$ carry further external legs, $B=B_1+1$. We have $B=B_1$ if only
one of the resulting boundary components of $l,m$ or $k,n$ carries
further external legs and $B=B_1-1$ if there are no further external
legs on these boundary components. We clearly have $\iota=\iota_1$
and $V^e=V^e_1-1$. Due to index conservation for segments, either $k$
or $n$ is an unknown summation index, and either $l$ or $m$.

We first consider the case $B=B_1+1$.
In both segments of $\gamma_1$ to contract there must be
at least one unsummed outgoing index, which we can choose to be
different from the vertex to contract. We thus take in the propagator
the maximum (\ref{est0}) over all indices and restrict the required
index summations over $k,m$ to the segments of the subgraphs. This
means that we have $s_1=s+2$ summations, which compensates
the change of the numbers of boundary components $B_1=B-1$, external
legs $N_1=N+2$ and external vertices $V^e_1=V^e+1$:
\begin{align}
  & \sum_{\mathcal{E}^{s},B=B_1+1} \Big|\Lambda
  \frac{\partial}{\partial \Lambda}
  A^{(V,V^e,B,g,\iota)\gamma}_{ \sigma m\,
    \sigma n;\sigma m' \,\sigma n'}[\Lambda] \Big| \nonumber
  \\*
  & \leq \frac{1}{2} \Big( \max_{m,n,k,l}
  \big|Q_{nm;lk}(\Lambda)\big|\Big) \Big(\max_{l,n}
  \sum_{k,m,\mathcal{E}^{s} }
  \big|A^{(V_1,V^e_1,B_1,g_1,\iota_1)\gamma_1}_{
    \sigma m\,\sigma n;mn;\sigma m'\, \sigma' n;kl }
  [\Lambda]\big| \Big) \nonumber
  \\
  & \leq \frac{1}{2} C_0 \Big(\frac{\Lambda}{\mu}\Big)^{\delta_2(
    V-\frac{N+2}{2}+2-2g-(B-1))}
  \Big(\frac{\mu}{\Lambda}\Big)^{\delta_1(
    V-(V^e+1)-\iota+2g+(B-1)-1+(s+2))} \nonumber
  \\*
  &\qquad \times \Big(\frac{\mu}{\Lambda}\Big)^{\delta_0(
    1+(V^e+1)+\iota-1-(s+2))}
  P^{2V-\frac{N+2}{2}}\Big[\ln
  \frac{\Lambda}{\Lambda_R}\Big]\;.
\label{al4E}
\end{align}
We immediately confirm (\ref{ANnorm1}). Alternatively, instead of
consuming a $\gamma_1$-summation to get the $k$-summation
we can also sum the propagator for maximised $l,m$ and given $k$ over
$n$. Compared with (\ref{al4E}) we have to replace $(s+2)$ by $(s+1)$
and one factor (\ref{est0}) by (\ref{est1}), ending up in the same
exponents.  

Next, we investigate the case $B=B_1$ where, for example, the
restriction of the boundary component to the left segment does not
carry another external leg than $m,l$. The summation over $m$ in
$\gamma_1$ is now provided by a volume factor, which means that in
(\ref{al4E}) we have to replace $(s+2)$ by $(s+1)$, $(B-1)$ by $B$ and
one factor (\ref{est0}) by (\ref{est2}). All
exponents match again (\ref{ANnorm1}). 

Finally, let us look at the possibility $B=B_1-1$ where the indices
$m,n,k,l$ to contract were the only external indices of the boundary
component. We thus combine two volume factors (\ref{volumefactor}) and
two factors (\ref{est0}) to two factors (\ref{est2}), compensating
$(B-1)\mapsto (B+1)$ and $(s+2) \mapsto s$.  After
$\Lambda$-integration we extend (\ref{ANnorm1}) to a reduced $N$.

\bigskip

The case that the two sides of the composed vertex to contract are
connected but belong to different segments, e.g.\ 
\begin{align}
  \parbox{60mm}{
\begin{picture}(50,28)
\put(0,0){\epsfig{scale=.9,file=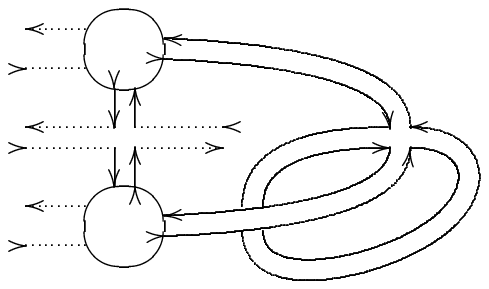,bb=71 604 204 684}}
\put(1,20){\mbox{\scriptsize$\sigma n$}}
\put(2,27){\mbox{\scriptsize$\sigma m$}}
\put(2,9){\mbox{\scriptsize$\sigma m'$}}
\put(1,1.5){\mbox{\scriptsize$\sigma n'$}}
\put(1,12){\mbox{\scriptsize$n_1$}}
\put(3,17){\mbox{\scriptsize$m_1$}}
\put(15,12){\mbox{\scriptsize$m_2$}}
\put(17,17){\mbox{\scriptsize$n_2$}}
\put(33,11.5){\mbox{\scriptsize$n$}}
\put(33,16.5){\mbox{\scriptsize$m$}}
\put(40,17){\mbox{\scriptsize$l$}}
\put(40,10.5){\mbox{\scriptsize$k$}}
\end{picture}}
\label{al4a}
\end{align}
is similar to treat concerning index summations, but for the
interpretation of the genus there is a different situation possible.
In the amputated subgraph $\gamma_1$ the indices $m,n$ and $k,l$ may be
situated on \emph{different} loops and thus different boundary
components. The contraction joins in this case the two loops,
$\tilde{L}=\tilde{L}_1-1$, which results due to (\ref{Euler}) in
$g=g_1+1$ and $B=B_1-1$.
There is at least one additional external leg on each of the
boundary components of $m,n$ and $k,l$ before the contraction, because
in order to close the loop we have to pass through the vertex
$m_1,n_1,m_2,n_2$. Now we have to replace in (\ref{al4E}) $(B-1)$ by $(B+1)$
and $g$ by $(g-1)$, confirming (\ref{ANnorm1}) also in
this case. If all indices $m,n,k,l$ are on the same loop
in $\gamma_1$, the contraction splits it into two and the entire
discussion of (\ref{al4}) can be used without modification to the
present example.

\bigskip

It remains the case  
\begin{align}
  \parbox{60mm}{
\begin{picture}(50,25)
\put(0,0){\epsfig{scale=.9,file=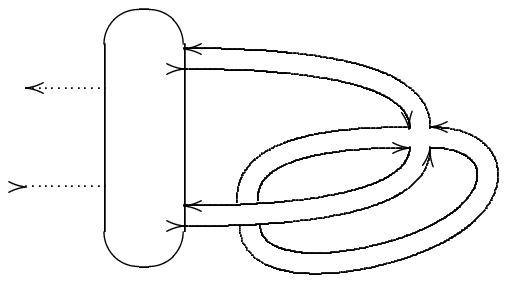,bb=71 606 209 684}}
\put(1,7){\mbox{\scriptsize$\sigma n$}}
\put(3,19.5){\mbox{\scriptsize$\sigma m$}}
\put(35,11){\mbox{\scriptsize$n$}}
\put(35,16){\mbox{\scriptsize$m$}}
\put(42,16.5){\mbox{\scriptsize$l$}}
\put(42,10){\mbox{\scriptsize$k$}}
\end{picture}}
\label{al5}
\end{align}
where the two halves of the composed vertex to contract belong to the
same segment. Three of the indices $m,n,k,l$ are now summation
indices. We have $\iota=\iota_1-1$ and $V^e=V^e_1-1$. Let first the
indices $m,n$ on one hand and $k,l$ on the other hand be situated on
different loops of the amputated graph $\gamma_1$. These are joint by
the contraction, yielding $g=g_1+1$. If there remain further external
legs on the contracted loop we have $B=B_1-1$, otherwise $B=B_1-2$.
We start with $B=B_1-1$.  Due to the segmentation index present in
$\gamma_1$, the induction hypothesis for $\gamma_1$ gives us the bound
for two additional summations over $m,k$ not present in $\gamma$. The
third summation is provided by the propagator via (\ref{est1}).
Assuming $\mathfrak{i}[k] \neq l,n$ in $\gamma_1$ we first take in the
contracting propagator the maximum over $m,l$, then sum the
$m,n$-boundary component over $m$ and those indices of
$\mathcal{E}^{s}$ which belong to the $m,n$-boundary component,
followed by the summation of the propagator over $n$ for given $k$.
Finally, we sum $\gamma_1$ over the remaining indices of
$\mathcal{E}^{s}$ and over $k$:
\begin{align}
  \sum_{\mathcal{E}^{s}}& \Big| \Lambda \frac{\partial}{\partial \Lambda}
  A^{(V,V^e,B,g,\iota)\gamma}_{ \sigma m\,
    \sigma n}[\Lambda] \Big| \nonumber
  \\*
  & \leq \frac{1}{2} \Big( \max_k \sum_{n} \max_{l,m}
  \big|Q_{mn;kl}(\Lambda)\big|\Big) \sum_{k,m,\mathcal{E}^{s} }
  \big|A^{(V_1,V^e_1,B_1,g_1,\iota_1)\gamma_1}_{
    \sigma m\,\sigma n;mn;kl} [\Lambda]\big|\Big) \nonumber
  \\
  &\leq \frac{1}{2} C_1 \Big(\frac{\Lambda}{\mu}\Big)^{\delta_2(
    V-\frac{N+2}{2}+2- 2(g-1)-(B+1))}
 \nonumber
  \\*
  &\qquad\qquad \times
  \Big(\frac{\mu}{\Lambda}\Big)^{\delta_1(1+
    V-(V^e+1)-(\iota+1)+2(g-1)+(B+1)-1+(s+2))}
  \nonumber
  \\*
  &\qquad\qquad \times \Big(\frac{\mu}{\Lambda}\Big)^{\delta_0(
    (V^e+1)+(\iota+1)-1-(s+2))}
  P^{2V-\frac{N+2}{2}}\Big[\ln
  \frac{\Lambda}{\Lambda_R}\Big]\;.
\label{al5E}
\end{align}
It is essential that the summation over $k-\mathfrak{i}[k]$ is independent.

If there are no external legs on the contracted loop, $B=B_1-2$, then
we have in $\gamma_1$ either $\mathfrak{i}[m]=n,\;\mathfrak{i}[k]=l$
or $\mathfrak{i}[m]=l,\;\mathfrak{i}[k]=n$.  In the first case we
would first fix $n,k$ and maximise the propagator over $m,l$. Now the
$m$-summation restricts to $\gamma_1$ with bound independent of $n$.
Thus, the $n$-summation for given $k$ restricts to the propagator and
delivers a factor (\ref{est1}), independent of $k$.  However, since
$k-\mathfrak{i}[k]\equiv k-l= n-m$ is already exhausted in $\gamma_1$,
the remaining $k$-summation has to come from a volume factor.  We thus
make in (\ref{al5E}) the replacements $(B{+}1)\mapsto (B{+}2)$,
$(s{+}2)\mapsto (s{+}1)$ and combine one factor (\ref{est0}) with a
volume factor (\ref{volumefactor}) to (\ref{est2}). The exponents
match again (\ref{ANnorm1}).

Next, we investigate the situation where all indices $m,n,k,l$ are
located on the same loop of the amputated subgraph $\gamma_1$. In this
case the contraction to $\gamma$ splits that loop into two so that we
have $g=g_1$.  As before we have $B=B_1+1$ 
if both split loops contain further external legs, $B=B_1$
if only one of the split loops contains further external
legs, and $B=B_1-1$ if the split loops do not contain
further external legs. The discussion is similar as for (\ref{al4}),
the difference is that three of $m,n,k,l$ are now summations
indices, which is taken into account by the replacement of $\iota$ in
(\ref{al4E}) by $(\iota+1)$. We thus finish the verification of
(\ref{ANnorm1}) for self-contractions of a vertex.

\subsection{Loop-contractions at different vertices}
\label{self-contr-2}

It remains to check (\ref{ANnorm1}) for contractions of different
vertices of the same graph. The external lines of the two vertices are
arranged according to (\ref{as}). We start with two vertices of the
type shown as the second graph in (\ref{as}). One possible contraction
of their external lines is
\begin{align}
  \parbox{50mm}{
\begin{picture}(40,30)
\put(0,0){\epsfig{scale=.9,file=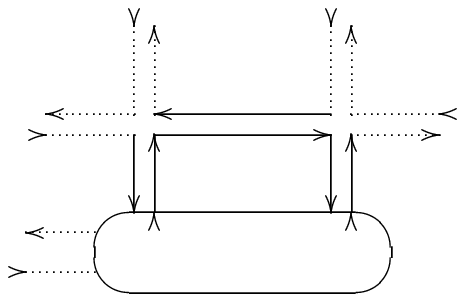,bb=71 605 192 684}}
\put(8,26){\mbox{\scriptsize$n_1$}}
\put(14.5,24){\mbox{\scriptsize$m_1$}}
\put(2,14.5){\mbox{\scriptsize$n_2$}}
\put(4,20){\mbox{\scriptsize$m_2$}}
\put(0,0.5){\mbox{\scriptsize$\sigma n$}}
\put(2,8){\mbox{\scriptsize$\sigma m$}}
\put(28.5,26){\mbox{\scriptsize$l_1$}}
\put(34.5,24){\mbox{\scriptsize$k_1$}}
\put(38,14){\mbox{\scriptsize$k_2$}}
\put(40,20){\mbox{\scriptsize$l_2$}}
\put(14.5,14){\mbox{\scriptsize$m$}}
\put(28,14){\mbox{\scriptsize$l$}}
\end{picture}}
\label{ab1}
\end{align}
assuming that the vertices to contract are located on the same segment
in $\gamma_1$.  One of the indices $m,l$ is a summation index. We
first consider the case that the two vertices to contract are located
on the same loop of the amputated graph $\gamma_1$.  The contraction
to $\gamma$ splits that loop into two, giving $g=g_1$. We have
$B=B_1+1$ if the trajectory starting at $l$ does not leave $\gamma_1$
(and $\gamma$) in $m$, whereas $B=B_1$ if $m,l$ are on the same
trajectory in $\gamma_1$. In case of $B=B_1+1$ we keep
$\mathfrak{i}[m]$ in $\gamma_1$ fixed, take in the propagator the
maximum over $m,l$ and restrict the $m$-summation to $\gamma_1$.  Due
to $V^e_1=V^e$, $\iota_1=\iota$ and $B_1=B-1$ we have in the case that
$m_1$ remains unsummed
\begin{align}
  &\sum_{\mathcal{E}^{s} \not\ni m_1,B=B_1+1} \Big|\Lambda
  \frac{\partial}{\partial \Lambda}
  A^{(V,V^e,B,g,\iota)\gamma}_{
    k_2l_2;k_1l_1;m_1n_1;m_2n_2;\sigma m\, \sigma n} [\Lambda]\Big|
  \nonumber
  \\*
  &\leq \frac{1}{2} \Big( \max_{m,l,m_1,l_1} \big|Q_{m_1m;ll_1}
  (\Lambda)\big|\Big)\,
\Big( \sum_{m,\mathcal{E}^{s}} \big|
  A^{(V_1,V^e_1,B_1,g_1,\iota_1)\gamma_1}_{
    k_2l_2;k_1l_1;l_1 l;m m_1;m_1n_1;m_2n_2; \sigma m\,
    \sigma n}[\Lambda] \big|\Big) \nonumber
  \\*
  & \leq \frac{1}{2} C_0 \Big(\frac{\Lambda}{\mu}\Big)^{\delta_2(
    V-\frac{N+2}{2}+2-2g-(B-1))}
  \Big(\frac{\mu}{\Lambda}\Big)^{\delta_1(
    V-V^e-\iota+2g+(B-1)+1+(s+1)) } \nonumber
  \\*
  &\qquad\qquad \times \Big(\frac{\mu}{\Lambda}\Big)^{\delta_0(
    1+V^e+\iota-1-(s+1))} P^{2V-\frac{N+2}{2}}\Big[\ln
  \frac{\Lambda}{\Lambda_R}\Big]\;.
\label{ab1E}
\end{align}
Summing additionally over $m_1$ we replace in (\ref{ab1E}) one factor
(\ref{est0}) by (\ref{est1}). It is clear that this reproduces the
exponents of (\ref{ANnorm1}) correctly.

If $l=\mathfrak{i}[m]$ in $\gamma_1$, we have to realise the
$m$-summation by a volume factor. We thus replace in (\ref{ab1E})
$(B-1)\mapsto B$, $(s+1)\mapsto s$ and combine (\ref{volumefactor})
with one factor (\ref{est0}) to (\ref{est2}). 

Finally, the two vertices to contract in (\ref{ab1}) may be located on
different loops of the amputated graph $\gamma_1$. They are joint by
the contraction to $\gamma$, giving $g=g_1+1$, and
because the newly created loop obviously has external legs, we have
$B=B_1-1$. As separated loops in $\gamma_1$, $l$ cannot be the
incoming index of the trajectory through $m$. Therefore, the
$m$-summation gives the same bound as the rhs of (\ref{ab1E}), now
with $(B-1)$ replaced by $(B+1)$ and $g$ by $(g-1)$.
We have thus extended (\ref{ANnorm1}) to a reduced $N$ for all types
of contractions (\ref{ab1}).

\bigskip

If the vertices to contract are located on different segments in
$\gamma_1$, e.g.\ 
\begin{align}
  \parbox{50mm}{
\begin{picture}(40,31)
\put(0,0){\epsfig{scale=.9,file=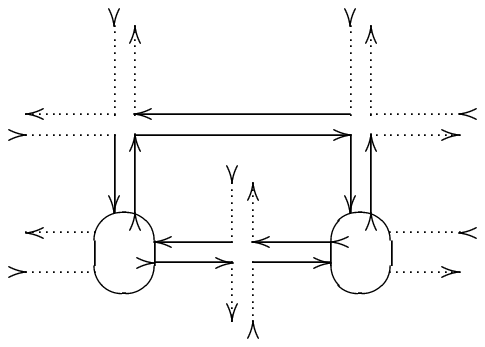,bb=71 597 198 684}}
\put(6,28.5){\mbox{\scriptsize$n_1$}}
\put(12.5,26.5){\mbox{\scriptsize$m_1$}}
\put(0,17){\mbox{\scriptsize$n_2$}}
\put(2,22.5){\mbox{\scriptsize$m_2$}}
\put(0,3){\mbox{\scriptsize$\sigma n$}}
\put(2,10.5){\mbox{\scriptsize$\sigma m$}}
\put(39.5,3){\mbox{\scriptsize$\sigma k$}}
\put(41.5,10.5){\mbox{\scriptsize$\sigma l$}}
\put(30.5,28.5){\mbox{\scriptsize$l_1$}}
\put(36.5,26.5){\mbox{\scriptsize$k_1$}}
\put(40,17){\mbox{\scriptsize$k_2$}}
\put(42,22.5){\mbox{\scriptsize$l_2$}}
\put(12.5,16.5){\mbox{\scriptsize$m$}}
\put(30,16.5){\mbox{\scriptsize$l$}}
\put(17,3){\mbox{\scriptsize$m_3$}}
\put(24.5,0.5){\mbox{\scriptsize$n_3$}}
\put(17.5,13){\mbox{\scriptsize$n_4$}}
\put(24.5,11){\mbox{\scriptsize$m_4$}}
\end{picture}}
\label{ab2}
\end{align}
both indices $m,l$ are determined by index conservation for segments.
We can thus save an index summation compared with (\ref{ab1E}) and
replace there and in its discussed modifications $(s+1)$ by $s$ and
$\iota_1=\iota$ by $\iota_1=(\iota-1)$. Since the $m$-summation is not
required, there is effectively an additional summation possible in
agreement with (\ref{numsum}). It is not possible that $m$ and $l$ are
located on the same trajectory in $\gamma_1$ so that either
$g=g_1,B=B_1+1$ or $g=g_1+1,B=B_1-1$.

Let us make a few more comments on the segmentation index. It is
essential that the contraction joins separated segments.  For
instance, the contraction
\begin{align}
  \parbox{51mm}{
\begin{picture}(51,30)
\put(0,0){\epsfig{scale=.9,file=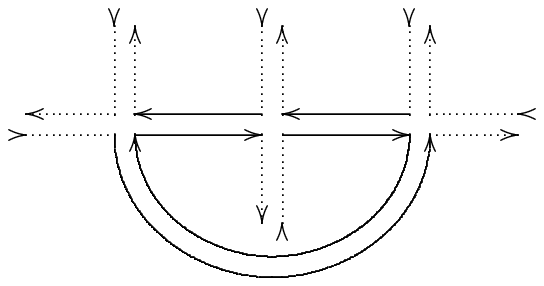,bb=71 610 215 684}}
\put(12.5,12.5){\mbox{\scriptsize$m$}}
\put(35,12){\mbox{\scriptsize$l$}}
\end{picture}}
\mapsto \quad \parbox{50mm}{
\begin{picture}(50,32)
\put(0,0){\epsfig{scale=.9,file=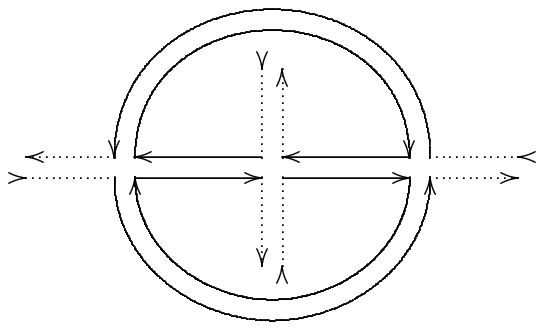,bb=71 595 215 687}}
\put(12.5,13){\mbox{\scriptsize$m$}}
\put(35,12.5){\mbox{\scriptsize$l$}}
\put(12.5,18.5){\mbox{\scriptsize$n$}}
\put(35,18.5){\mbox{\scriptsize$k$}}
\end{picture}}
\end{align}
does not increase the segmentation index, because in agreement with
Definition~\ref{defiota} the number of segments remains constant. The
graph on the left has $\iota=1$, and the internal indices $m,l$ are
determined by the external ones. The graph on the right has $\iota=1$
as well, and now one of the indices $n,k$ becomes a summation index.
Having several composed vertices in the middle link does not change
the segmentation index:
\begin{align}
  \parbox{60mm}{
\begin{picture}(60,30)
\put(0,0){\epsfig{scale=.9,file=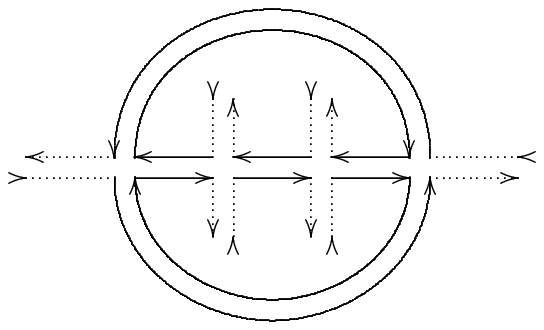,bb=71 595 215 687}}
\put(12.5,13){\mbox{\scriptsize$m$}}
\put(35,12.5){\mbox{\scriptsize$l$}}
\put(12.5,18.5){\mbox{\scriptsize$n$}}
\put(35,18.5){\mbox{\scriptsize$k$}}
\end{picture}}
\end{align}
It makes, however, a difference if the two composed vertices are
situated on different links:
\begin{align}
  \parbox{52mm}{
\begin{picture}(52,31)
\put(0,0){\epsfig{scale=.9,file=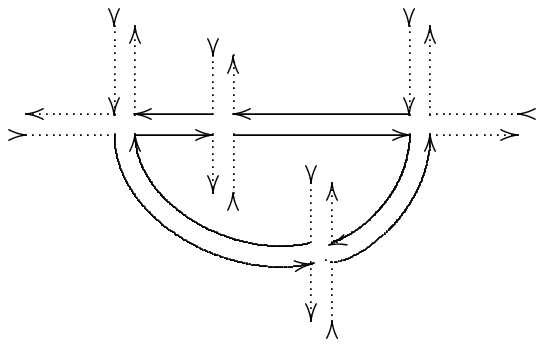,bb=71 597 215 684}}
\put(12.5,17){\mbox{\scriptsize$m$}}
\put(35,16.5){\mbox{\scriptsize$l$}}
\end{picture}}
\mapsto \quad \parbox{50mm}{
\begin{picture}(50,35)
\put(0,0){\epsfig{scale=.9,file=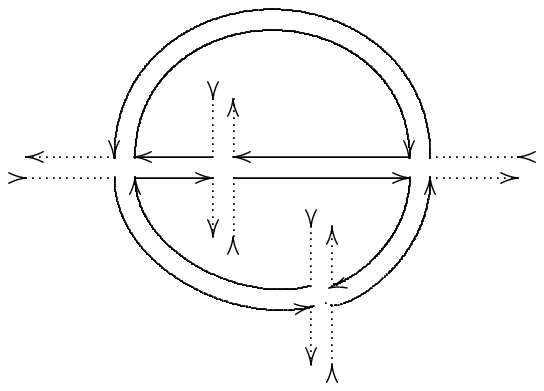,bb=71 583 215 687}}
\put(12.5,17){\mbox{\scriptsize$m$}}
\put(35,17){\mbox{\scriptsize$l$}}
\put(12.5,22.5){\mbox{\scriptsize$n$}}
\put(35,22.5){\mbox{\scriptsize$k$}}
\end{picture}}
\label{iota2}
\end{align}
Here, the segmentation index increases from $\iota=1$ on the left to
$\iota=2$ on the right, in agreement with Definition~\ref{defiota}.

\bigskip

The case
\begin{align}
  \parbox{60mm}{\begin{picture}(60,33)
      \put(0,0){\epsfig{scale=.9,file=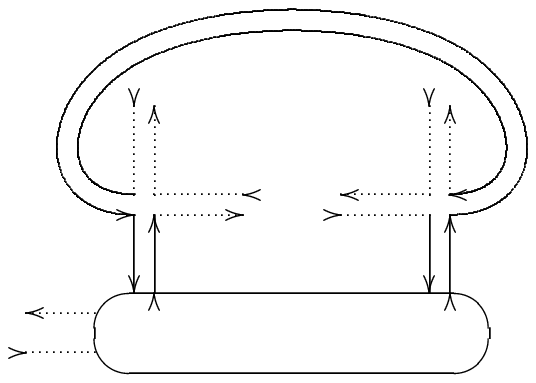,bb=71 577 217 684}}
      \put(8,14){\mbox{\scriptsize$n$}}
      \put(44,14){\mbox{\scriptsize$k$}}
      \put(8,26){\mbox{\scriptsize$n_1$}}
      \put(14.5,24){\mbox{\scriptsize$m_1$}}
      \put(18,14.5){\mbox{\scriptsize$m_2$}}
      \put(20,20){\mbox{\scriptsize$n_2$}}
      \put(0,0.5){\mbox{\scriptsize$\sigma n$}}
      \put(2,8){\mbox{\scriptsize$\sigma m$}}
      \put(38.5,26){\mbox{\scriptsize$l_1$}}
      \put(44.5,24){\mbox{\scriptsize$k_1$}}
      \put(33,14.5){\mbox{\scriptsize$l_2$}}
      \put(35,20){\mbox{\scriptsize$k_2$}}
\end{picture}}
\end{align}
is completely identical to (\ref{ab1}). In the contraction
\begin{align}
  \parbox{60mm}{\begin{picture}(60,32)
      \put(0,0){\epsfig{scale=.9,file=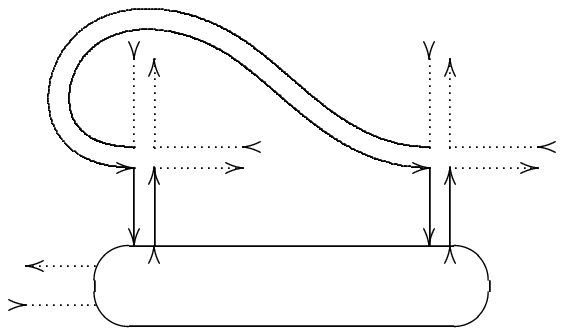,bb=71 591 221 685}}
      \put(8,14){\mbox{\scriptsize$n$}}
      \put(38,14){\mbox{\scriptsize$l$}}
      \put(8,26){\mbox{\scriptsize$n_1$}}
      \put(14.5,24){\mbox{\scriptsize$m_1$}}
      \put(18,14.5){\mbox{\scriptsize$m_2$}}
      \put(20,20){\mbox{\scriptsize$n_2$}}
      \put(0,0.5){\mbox{\scriptsize$\sigma n$}}
      \put(2,8){\mbox{\scriptsize$\sigma m$}}
      \put(38.5,26){\mbox{\scriptsize$l_1$}}
      \put(44.5,24){\mbox{\scriptsize$k_1$}}
      \put(48,14){\mbox{\scriptsize$k_2$}}
      \put(50,20){\mbox{\scriptsize$l_2$}}
\end{picture}}
\end{align}
the summation index $n$ or $l$ is provided by the propagator,
replacing in (\ref{ab1E}) and its modifications $(s+1)$ by $s$ and one
factor (\ref{est0}) by (\ref{est1}).  It is not possible that $n$ and
$l$ are located on the same trajectory in $\gamma_1$ so that either
$g=g_1,B=B_1+1$ or $g=g_1+1,B=B_1-1$. 

In order to treat the contraction
\begin{align}
  \parbox{60mm}{\begin{picture}(60,30)
      \put(0,0){\epsfig{scale=.9,file=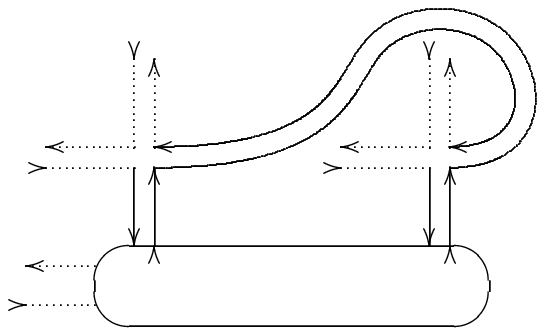,bb=71 591 220 685}}
      \put(8,26){\mbox{\scriptsize$n_1$}}
      \put(14.5,24){\mbox{\scriptsize$m_1$}}
      \put(2,14.5){\mbox{\scriptsize$n_2$}}
      \put(4,20){\mbox{\scriptsize$m_2$}}
      \put(0,0.5){\mbox{\scriptsize$\sigma n$}}
      \put(2,8){\mbox{\scriptsize$\sigma m$}}
      \put(38.5,26){\mbox{\scriptsize$l_1$}}
      \put(44.5,24){\mbox{\scriptsize$k_1$}}
      \put(33,14.5){\mbox{\scriptsize$l_2$}}
      \put(35,20){\mbox{\scriptsize$k_2$}}
      \put(14,14){\mbox{\scriptsize$m$}}
      \put(44,14){\mbox{\scriptsize$k$}}
\end{picture}}
\end{align}
one has to use that the summation over $m_1$ can due to $m_1=k+m-k_1$
be transferred as a $k$-summation of $\gamma_1$. The summation over the
undetermined index $m$ is applied in the last step.

\bigskip

Finally,
\begin{align}
  \parbox{65mm}{\begin{picture}(55,38)
      \put(0,0){\epsfig{scale=.9,file=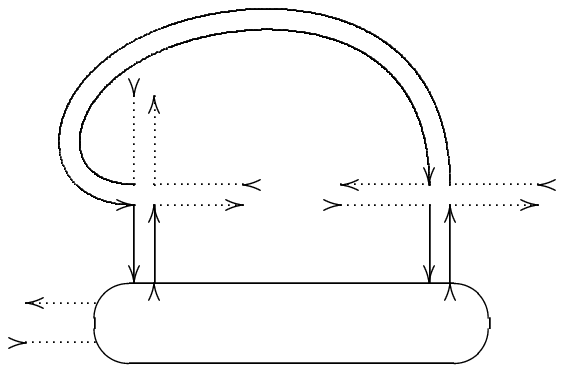,bb=71 577 221 681}}
      \put(7.5,14){\mbox{\scriptsize$n$}}
      \put(8,26){\mbox{\scriptsize$n_1$}}
      \put(14.5,24){\mbox{\scriptsize$m_1$}}
      \put(18,14.5){\mbox{\scriptsize$m_2$}}
      \put(20,20){\mbox{\scriptsize$n_2$}}
      \put(0,0.5){\mbox{\scriptsize$\sigma n$}}
      \put(2,8){\mbox{\scriptsize$\sigma m$}}
      \put(50,20){\mbox{\scriptsize$l_1$}}
      \put(48,14){\mbox{\scriptsize$k_1$}}
      \put(35,20){\mbox{\scriptsize$k_2$}}
      \put(33,14){\mbox{\scriptsize$l_2$}}
\end{picture}}
\parbox{50mm}{\begin{picture}(50,38) 
    \put(0,0){\epsfig{scale=.9,file=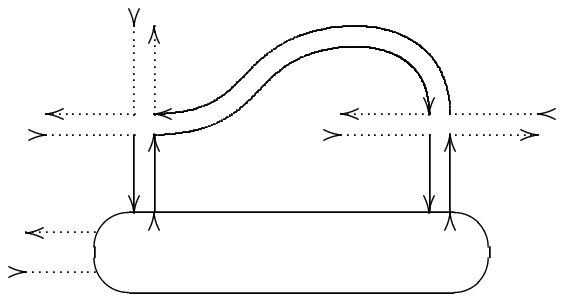,bb=71 605 221 684}}
    \put(14,14){\mbox{\scriptsize$m$}}
      \put(8,26){\mbox{\scriptsize$n_1$}}
      \put(14.5,24){\mbox{\scriptsize$m_1$}}
      \put(2,14.5){\mbox{\scriptsize$n_2$}}
      \put(4,20){\mbox{\scriptsize$m_2$}}
      \put(0,0.5){\mbox{\scriptsize$\sigma n$}}
      \put(2,8){\mbox{\scriptsize$\sigma m$}}
    \put(50,20){\mbox{\scriptsize$l_1$}}
    \put(48,14){\mbox{\scriptsize$k_1$}}
    \put(35,20){\mbox{\scriptsize$k_2$}}
    \put(33,14){\mbox{\scriptsize$l_2$}}
\end{picture}}
\end{align}
are similar to the $\iota$-increased variant (\ref{ab2}). The
contraction
\begin{align}
  \parbox{50mm}{\begin{picture}(50,30)
      \put(0,0){\epsfig{scale=.9,file=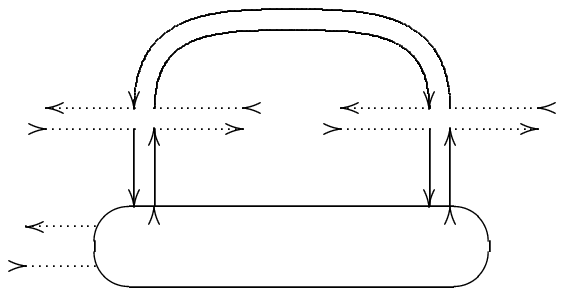,bb=71 605 221 687}}
      \put(2,14.5){\mbox{\scriptsize$n_1$}}
      \put(4,20){\mbox{\scriptsize$m_1$}}
      \put(18,14.5){\mbox{\scriptsize$m_2$}}
      \put(20,20){\mbox{\scriptsize$n_2$}}
      \put(0,0.5){\mbox{\scriptsize$\sigma n$}}
      \put(2,8){\mbox{\scriptsize$\sigma m$}}
    \put(50,20){\mbox{\scriptsize$l_1$}}
    \put(48,14){\mbox{\scriptsize$k_1$}}
    \put(35,20){\mbox{\scriptsize$k_2$}}
    \put(33,14){\mbox{\scriptsize$l_2$}}
\end{picture}}
\end{align}
is an example for a realisation of (\ref{iota}) where $V_c$ is
increased by $2$ and $S$ by $1$, giving again a segmentation index
increased by $1$.

\bigskip

It is obvious that the discussion of contractions involving the second
and third or two of the third vertices of the first line in (\ref{as})
is analogous.

\bigskip

Let us now study loop contractions which involve the first graph in
the second line of (\ref{as}), assuming first that the vertices are
situated on the same segment:
\begin{align}
  \parbox{85mm}{\begin{picture}(55,28)
      \put(0,0){\epsfig{scale=.9,file=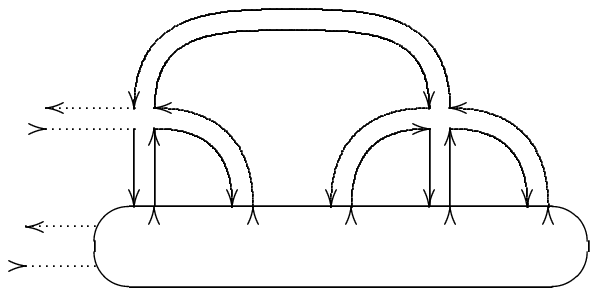,bb=71 605 235 687}}
      \put(2.5,14){\mbox{\scriptsize$n_1$}}
      \put(4.5,20){\mbox{\scriptsize$m_1$}}
      \put(2.5,8){\mbox{\scriptsize$\sigma m$}}
      \put(1,1){\mbox{\scriptsize$\sigma n$}}
      \put(15,20){\mbox{\scriptsize$m$}}
      \put(45,20){\mbox{\scriptsize$k$}}
      \put(38,20){\mbox{\scriptsize$l$}}
\end{picture}}
\label{ab5}
\end{align}
We thus have $\iota=\iota_1$ and $V^e=V^e_1-1$.  Two of the summation
indices $m,k,l$ are undetermined. Let first the two vertices to
contract be located on the same loop of the amputated subgraph
$\gamma_1$. The contraction splits that loop into two, giving $g=g_1$.
Next question concerns the number of boundary components. We have
$B=B_1+1$ if there are further external legs on the loop through $l,m$
and $B=B_1$ if $l=\mathfrak{i}[m]$ in $\gamma_1$. We start with
$B=B_1+1$.  In general, the induction hypothesis provides us
with bounds for summations over $m$ and $k$, because $l \neq
\mathfrak{i}[m]$. If $m_1$ is an unsummed index we thus have
\begin{align}
  &\sum_{\mathcal{E}^{s} \not\ni m_1,B=B_1+1} \Big|\Lambda
  \frac{\partial}{\partial \Lambda}
  A^{(V,V^e,B,g,\iota)\gamma}_{ m_1n_1;\sigma m\,
    \sigma n} [\Lambda] \Big|\nonumber
  \\*
  &\leq \frac{1}{2} \Big( \max_{m,l,k,m_1} \big|Q_{m_1m;lk}
  (\Lambda)\big|\Big) \Big( \sum_{m,k,\mathcal{E}^{s}} \big|
  A^{(V_1,V^e_1,B_1,g_1,\iota_1)\gamma_1}_{
    kl;nm_1;m_1n_1; \sigma m\, \sigma n}[\Lambda] \big|\Big) \nonumber
  \\*
  & \leq \frac{1}{2} C_0 \Big(\frac{\Lambda}{\mu}\Big)^{\delta_2(
    V-\frac{N+2}{2}+2-2g-(B-1))}
  \Big(\frac{\mu}{\Lambda}\Big)^{\delta_1(
    V-(V^e+1)-\iota +2 g+(B-1)+1+(s+2)) } \nonumber
  \\*
  &\qquad\qquad \times \Big(\frac{\mu}{\Lambda}\Big)^{\delta_0(
    1+(V^e+1)+\iota-1-(s+2))}
  P^{2V-\frac{N+2}{2}}\Big[\ln
  \frac{\Lambda}{\Lambda_R}\Big]\;.
\label{ab5E}
\end{align}
Now an additional summation over $m_1$ can immediately be taken into
account by replacing the maximised propagator (\ref{est0}) by the
summed propagator (\ref{est1}), in agreement with $(s+2)$ replaced by
$(s+1)$. The $m_1$-summation is applied before the $k$-summation is
carried out. 

These considerations require an unsummed outgoing index on the
contracted segment of $\gamma_1$. If this is not the case then $m_1$
has to be the unsummed outgoing index. Now the $l$-summation for given
$m$ has to be restricted to the propagator and delivers a factor 
(\ref{est1}). The exponents match again (\ref{ANnorm1}).

Next, for $l =\mathfrak{i}[m]$ in $\gamma_1$ we cannot use a summation
over $m$ in $\gamma_1$ in order to account for the undetermined
contraction index, because the incoming index $l$ would change
simultaneously. Instead we have to use a volume factor
(\ref{volumefactor}) combined with one factor (\ref{est0}) to
(\ref{est2}). Additionally we have to replace in (\ref{ab5E}) $(s+1)$
by $s$ and $(B-1)$ by $B$.

Second, the two vertices to contract may be located on different
loops of the amputated graph $\gamma_1$. They are joint by the
contraction, giving $g=g+1$. Because the loop carries
at least the external leg $m_1n_1$, we necessarily have $B=B_1-1$. Now,
$l \neq \mathfrak{i}[m]$ in $\gamma_1$ so that we use 
summations over $m,k$ in $\gamma_1$, giving the same balance (\ref{ab5E})
for the exponents, with $(B-1)\mapsto (B+1)$ and $g\mapsto
(g-1)$.

\bigskip

The discussion is identical for the contraction
\begin{align}
  \parbox{85mm}{\begin{picture}(55,30)
      \put(0,0){\epsfig{scale=.9,file=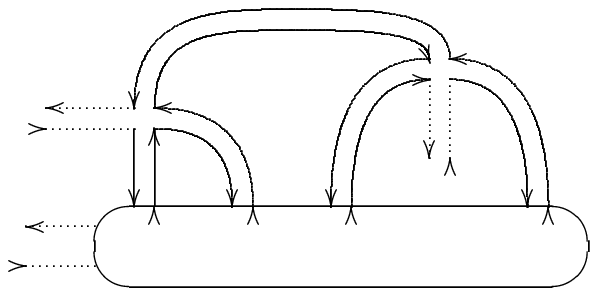,bb=71 605 235 687}}
      \put(2.5,14){\mbox{\scriptsize$n_1$}}
      \put(4.5,20){\mbox{\scriptsize$m_1$}}
      \put(2.5,8){\mbox{\scriptsize$\sigma m$}}
      \put(1,1){\mbox{\scriptsize$\sigma n$}}
      \put(15,20){\mbox{\scriptsize$m$}}
      \put(45,25){\mbox{\scriptsize$k$}}
      \put(36,23.5){\mbox{\scriptsize$l$}}
      \put(44,13){\mbox{\scriptsize$n_2$}}
      \put(36,15){\mbox{\scriptsize$m_2$}}
\end{picture}}
\label{ab7}
\end{align}
which in the case that $k,l$ belong to the same segment in $\gamma$
has two undetermined summation indices as well. We thus proceed as in
(\ref{ab5E}) and its discussed modification and only have to replace
$(V^e+1)$ by $V^e$ and $\iota$ by $(\iota+1)$. If $k,l$ are situated
on different segments in $\gamma$, e.g.\ 
\begin{align}
  \parbox{70mm}{\begin{picture}(55,29)
      \put(0,0){\epsfig{scale=.9,file=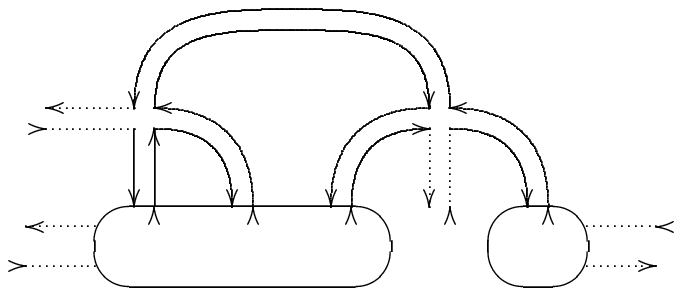,bb=71 605 255 687}}
      \put(2.5,14){\mbox{\scriptsize$n_1$}}
      \put(4.5,20){\mbox{\scriptsize$m_1$}}
      \put(2.5,7.5){\mbox{\scriptsize$\sigma m$}}
      \put(1,0.5){\mbox{\scriptsize$\sigma n$}}
      \put(15,20){\mbox{\scriptsize$m$}}
      \put(45,20){\mbox{\scriptsize$k$}}
      \put(38,20){\mbox{\scriptsize$l$}}
      \put(44,8){\mbox{\scriptsize$n_2$}}
      \put(36,10){\mbox{\scriptsize$m_2$}}
      \put(61,8){\mbox{\scriptsize$\sigma l$}}
      \put(59,0.5){\mbox{\scriptsize$\sigma k$}}
\end{picture}}
\parbox{55mm}{\begin{picture}(55,29) 
    \put(0,0){\epsfig{scale=.9,file=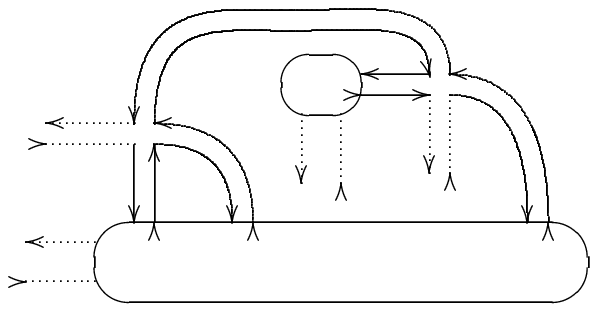,bb=71 597 235 683}}
    \put(2.5,14){\mbox{\scriptsize$n_1$}}
    \put(4.5,20){\mbox{\scriptsize$m_1$}}
    \put(2.5,7.5){\mbox{\scriptsize$\sigma m$}}
    \put(1,0.5){\mbox{\scriptsize$\sigma n$}}
    \put(15,20){\mbox{\scriptsize$m$}}
    \put(45,25){\mbox{\scriptsize$k$}}
    \put(38,24.5){\mbox{\scriptsize$l$}}
    \put(44,14){\mbox{\scriptsize$n_2$}}
    \put(36,16){\mbox{\scriptsize$m_2$}}
    \put(33,12){\mbox{\scriptsize$\sigma l$}}
    \put(24,14){\mbox{\scriptsize$\sigma k$}}
\end{picture}}
\label{ab6b}
\end{align}
there is only one undetermined summation index, which is reflected in
the analogue of (\ref{ab5E}) by the fact that the segmentation index
remains unchanged, $\iota=\iota_1$. Note that in the right graph of
(\ref{ab6b}) we either have $B=B_1-1,g=g+1$ or
$B=B_1+1,g=g$. Of course we get the same estimations
if the segment of $\gamma_1$ with external lines $\sigma k,\sigma l$ are
connected by several composed vertices to the part of $\gamma_1$ with
external lines $\sigma m,\sigma n$.

\bigskip

The contractions
\begin{align}
  \parbox{65mm}{\begin{picture}(65,28)
      \put(0,0){\epsfig{scale=.9,file=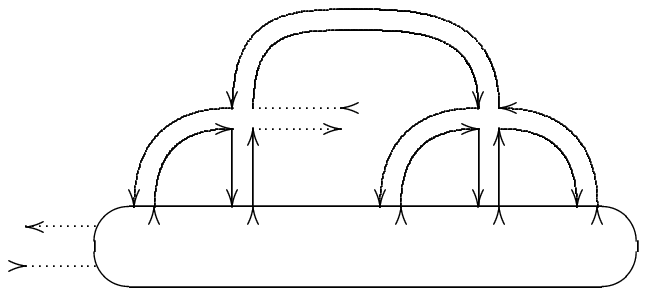,bb=71 605 249 687}}
      \put(27.5,14){\mbox{\scriptsize$m_1$}}
      \put(29.5,20){\mbox{\scriptsize$n_1$}}
      \put(2.5,7.5){\mbox{\scriptsize$\sigma m$}}
      \put(1,0.5){\mbox{\scriptsize$\sigma n$}}
      \put(18.5,20){\mbox{\scriptsize$n$}}
      \put(43,20){\mbox{\scriptsize$l$}}
      \put(50,20){\mbox{\scriptsize$k$}}
\end{picture}}
\parbox{62mm}{\begin{picture}(60,28) 
    \put(0,0){\epsfig{scale=.9,file=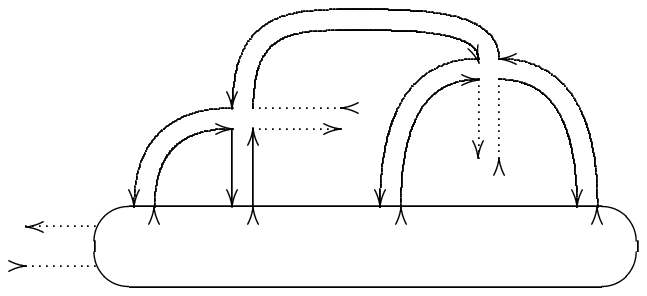,bb=71 605 249 687}}
    \put(27.5,14){\mbox{\scriptsize$m_1$}}
    \put(29.5,20){\mbox{\scriptsize$n_1$}}
    \put(2.5,7.5){\mbox{\scriptsize$\sigma m$}}
    \put(1,0.5){\mbox{\scriptsize$\sigma n$}}
    \put(18.5,20){\mbox{\scriptsize$n$}}
    \put(43,16.5){\mbox{\scriptsize$k_1$}}
    \put(49.5,14.5){\mbox{\scriptsize$l_1$}}
    \put(42,23.8){\mbox{\scriptsize$l$}}
    \put(49.5,24.5){\mbox{\scriptsize$k$}}
\end{picture}}
\label{ab5b}
\end{align}
are a little easier because the contracting propagator does not have
outgoing indices which for certain summations had to be transferred to
the subgraph $\gamma_1$. If $n=\mathfrak{i}[k]$ in $\gamma_1$, the
$k$-summation for given $n,n_1$ can be restricted to $\gamma_1$ after
maximising the propagator over all indices. Since the result for
$\gamma_1$ is independent of the starting point $n$, the $k$-summation
can be regarded as a summation over all differences $k-n$. The final
summation over all pairs $k,n$ with fixed difference $k-n$ is provided
by a volume factor (\ref{volumefactor}) combined with (\ref{est0}) to
(\ref{est2}). The balance of exponents is identical to (\ref{ab5E})
and its discussed variants.

It is clear that the analogue of (\ref{ab6b}) with the left vertex
connected as in (\ref{ab5b}) is similar to treat.

\bigskip

Next, we discuss the variant of (\ref{ab5}) where the two vertices to
contract belong to different segments in the subgraph $\gamma_1$:
\begin{align}
  \parbox{85mm}{\begin{picture}(75,34)
      \put(0,0){\epsfig{scale=.9,file=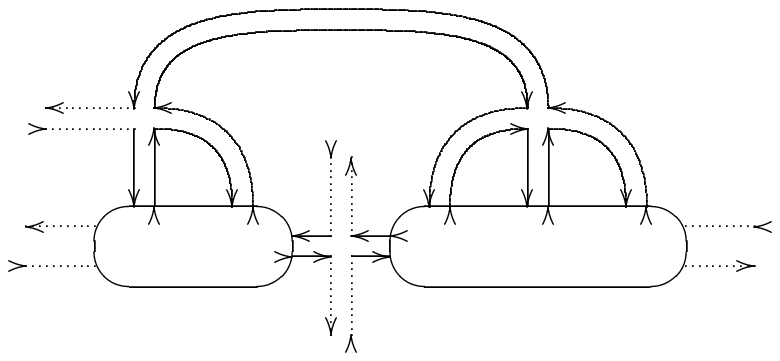,bb=71 591 283 687}}
      \put(2.5,19){\mbox{\scriptsize$n_1$}}
      \put(4.5,25){\mbox{\scriptsize$m_1$}}
      \put(2.5,12.5){\mbox{\scriptsize$\sigma m$}}
      \put(1,5.5){\mbox{\scriptsize$\sigma n$}}
      \put(72,12.5){\mbox{\scriptsize$\sigma l$}}
      \put(70,5.5){\mbox{\scriptsize$\sigma k$}}
      \put(34,1){\mbox{\scriptsize$n_2$}}
      \put(26.5,3){\mbox{\scriptsize$m_2$}}
      \put(34,16){\mbox{\scriptsize$m_3$}}
      \put(27.5,17.5){\mbox{\scriptsize$n_3$}}
      \put(15,25){\mbox{\scriptsize$m$}}
      \put(55,25){\mbox{\scriptsize$k$}}
      \put(48,25){\mbox{\scriptsize$l$}}
\end{picture}}
\label{ab5a}
\end{align}
Now only one of the indices $m,k,l$ is an undetermined summation
index, with $k$ being the most natural choice. We therefore get a
bound for the $\Lambda$-scaling analogous to (\ref{ab5E}) but with
$\iota$ replaced by $\iota-1$, reflecting the increase of the
segmentation index $\iota=\iota_1+1$.  There is now an additional
index summation possible, here via (\ref{est1}) over the index
$m_1$. Note that we have either $B=B_1-1,g=g+1$ or
$B=B_1+1,g=g$.

The discussion of the variants of (\ref{ab5a}) with the right vertex
taken as the second one in the last line of (\ref{as}) and/or the left
vertex arranged as in (\ref{ab5b}) is very similar.

\bigskip

It remains to investigate contractions between two of the vertices in
the second line of (\ref{as}). We discuss in detail the contraction
\begin{align}
  \parbox{75mm}{\begin{picture}(70,28)
      \put(0,0){\epsfig{scale=.9,file=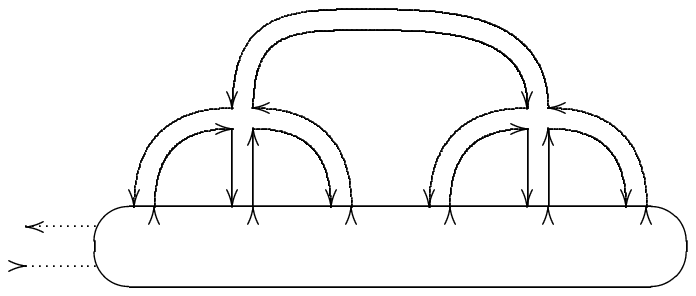,bb=71 605 263 687}}
      \put(2.5,7.5){\mbox{\scriptsize$\sigma m$}}
      \put(1,0.5){\mbox{\scriptsize$\sigma n$}}
      \put(18.5,20){\mbox{\scriptsize$n$}}
      \put(24.5,20){\mbox{\scriptsize$m$}}
      \put(48,20){\mbox{\scriptsize$l$}}
      \put(55,20){\mbox{\scriptsize$k$}}
\end{picture}}
\label{ab8}
\end{align}
All variants are similar as described between (\ref{ab5}) and
(\ref{ab5a}).

Three of the four summation indices $m,n,k,l$ in (\ref{ab8}) are
undetermined.  We clearly have $V^e_1=V^e+2$ and $\iota=\iota_1$.  We
first consider the case where the four indices $m,n,k,l$ are located
on the same loop of the amputated subgraph $\gamma_1$. The
contraction will split that loop into two, giving
$g=g_1$. There are three possibilities for the
change of the number of boundary components after the contraction.
First, if on both paths of trajectories in $\gamma_1$ from $n$ to $k$
and from $l$ to $m$ there are further external legs, we have
$B=B_1+1$. Second, if on one of these paths there is no further
external leg, we have $B=B_1$. Third, if both paths
contain no further external legs, i.e.\ $m$ and $k$ are the outgoing
indices of the trajectories starting at $l$ and $n$, respectively, we
have $B=B_1-1$.

We start with $B=B_1+1$. Then, $\mathfrak{i}[k]$ and $\mathfrak{i}[m]$
are fixed as external indices so that the induction hypothesis for
$\gamma_1$ provides the bounds for two summations over $k,m$. We first
apply a possible summation to the outgoing index of the trajectory
starting at $l$. The result is maximised independently from $l$ so
that we can restrict the $l$-summation to the propagator, maximised
over $k,n$ with $m$ being fixed. Finally, we apply the summations over
$k,m$ and all remaining $\mathcal{E}^{s}$-summations to $\gamma_1$. We
thus obtain
\begin{align}
  &\sum_{\mathcal{E}^{s} ,B=B_1+1} \Big|\Lambda \frac{\partial}{\partial
    \Lambda} A^{(V,V^e,B,g,\iota)\gamma}_{\sigma
    m\, \sigma n} [\Lambda] \Big|\nonumber
  \\*
  &\leq \frac{1}{2} \Big( \max_m \sum_l \max_{n,k} \big|Q_{nm;lk}
  (\Lambda)\big|\Big) \Big( \sum_{m,k,\mathcal{E}^{s}} \big|
  A^{(V_1,V^e_1,B_1,g_1,\iota_1)\gamma_1}_{mn;
    \sigma m\, \sigma n;kl}[\Lambda] \big|\Big) \nonumber
  \\*
  & \leq \frac{1}{2} C_1 \Big(\frac{\Lambda}{\mu}\Big)^{\delta_2(
    V-\frac{N+2}{2}+2- 2g-(B-1))}
  \Big(\frac{\mu}{\Lambda}\Big)^{\delta_1(1+
    V-(V^e+2)+\iota+1+2g+(B-1)-1+(s+2)) } \nonumber
  \\*
  &\qquad\qquad \times \Big(\frac{\mu}{\Lambda}\Big)^{\delta_0(
    (V^e+2)+\iota-1-(s+2))}
  P^{2V-\frac{N+2}{2}}\Big[\ln\frac{\Lambda}{\Lambda_R}\Big]\;.
\label{ab8E}
\end{align}
The $\Lambda$-integration verifies (\ref{ANnorm1}) in the topological
situation under consideration. 

Next, we discuss the case $B=B_1$, assuming e.g.\ $l=\mathfrak{i}[m]$
in $\gamma_1$. We maximise the propagator over $k,n$ for given $l$ so
that the $m$-summation can be restricted to $\gamma_1$. Next, we apply
the $\mathcal{E}^{s}$-summations and the $k$-summation to $\gamma_1$,
still for given $l$. The final $l$-summation counts the number of
graphs with different $l$, giving the bound (\ref{est0}) of the
propagator times a volume factor. In any case the required
modifications of (\ref{ab8E}), in particular $(B-1)\mapsto B$, lead to
the correct exponents of (\ref{ANnorm1}).

If $B=B_1-1$, i.e.\ $l=\mathfrak{i}[m]$ and $n=\mathfrak{i}[k]$, we
take in the propagator the maximum over $n,k$ so that for given $l$
the $m$-summation can be restricted to $\gamma_1$. The result of that
summation is bounded independently of $l$. Thus, each summand only
fixes $m-l=n-k$, and the remaining freedom for the summation indices
is exhausted by two volume factors and the bound (\ref{est0}) for the
propagator. We thus replace in (\ref{ab8E}) $(s+2)\mapsto(s+1)$,
$(B-1)\mapsto(B+1)$ and one factor (\ref{est1}) by (\ref{est0}). Then
two factors (\ref{est0}) are merged with two volume factors
(\ref{volumefactor}) to give two factors (\ref{est2}).

Finally, we have to consider the case where $m,n$ are located on a
different loop of the amputated subgraph $\gamma_1$ than $k,l$. The
contraction joins these loops, giving $g=g_1+1$. If
the resulting loop carries at least one external leg we have
$B=B_1-1$, whereas we get $B=B_1-2$ if
the resulting loop does not carry any external legs. We first
consider the case that there is a further external leg on the
$n,m$-loop in $\gamma_1$. We take in the propagator the maximum over
$k,n$ and sum the subgraph for given $l,n$ over $k$ and possibly the
outgoing index of the $n$-trajectory. The result is independent of
$l,n$. Next, we sum the propagator for given $m$ over $l$ and finally
apply the remaining $\mathcal{E}^{s}$-summations and the summation
over $m$ to $\gamma_1$. We get the same estimates as in (\ref{ab8E}) with
$(B-1)$ replaced by $(B+1)$ and $g$ by $(g-1)$.

If there are no further external legs on the contracted loop we would
maximise the propagator over $k,n$, then sum $\gamma_1$ over $k$ for
given $l$, next sum the propagator over $l$ for given $m$. For each
resulting pair $k,l$ the remaining $m$-summation leaves $m-n$
constant. We thus have to use a volume factor in order to exhaust the
freedom of $m-n$, combining one factor (\ref{est0}) and the volume
factor (\ref{volumefactor}) to (\ref{est2}). We thus confirm
(\ref{ANnorm1}) for any contraction of the form (\ref{ab8}).

\bigskip

It is obvious that all examples not discussed in detail are treated in
the same manner. We conclude that (\ref{ANnorm1}) provides the correct
bounds for the interaction coefficients of $\phi^4$-matrix model with
cut-off propagator described by the three exponents
$\delta_0,\delta_1,\delta_2$.  \hfill $\square$

\section{Discussion}
\label{disc}

By solving the Polchinski equation perturbatively we have derived a
power-counting theorem for non-local matrix models with arbitrary
propagator.  

Our main motivation for the renormalisation group investigation of
non-local matrix models was to tackle the renormalisation problem of
field theories on noncommutative $\mathbb{R}^D$ from a different
perspective. The momentum integrals leading to the parametric integral
representation are not absolutely convergent; nevertheless
one exchanges the order of integration. In momentum space one can
therefore not exclude the possibility that the UV/IR-mixing is due to
the mathematically questionable exchange of the order of integration. 

The renormalisation group approach to noncommutative field theories in
matrix formulation avoids these problems. We work with cut-off
propagators leading to finite sums and take absolute values of the
interaction coefficients throughout. Oscillating phases never appear;
they are not required for convergence of certain graphs.

Our power-counting theorem provides a necessary condition for
renormalisability: The two scaling exponents $\delta_0,\delta_1$ of
the cut-off propagator have to be large enough relative to the
dimension of the underlying space. In
\cite{Grosse:2003nw,Grosse:2004yu} we determine these exponents for
$\phi^4$-theory on noncommutative $\mathbb{R}^D$, $D=2,4$:
\begin{prp}
The propagator for the real scalar field on noncommutative
$\mathbb{R}^D$, $D=2,4$, is characterised by the scaling exponents 
$\delta_0=1$ and $\delta_1=0$. Adding a harmonic oscillator potential
to the action one achieves $\delta_0=\delta_1=2$.  
\end{prp}
We thus conclude that scalar models on noncommutative $\mathbb{R}^D$
are anomalous unless one adds the regulating harmonic
oscillator potential. 

The weak decay $\sim \Lambda^{-1}$ of the propagator leads to
divergences in $\Lambda \sim \Lambda_0 \to \infty$ of arbitrarily high
degree.  The appearance of unbounded degrees of divergences in field
theories on noncommutative $\mathbb{R}^4$ is often related to the
so-called UV/IR-mixing \cite{Minwalla:1999px}. We learn from the
power-counting theorem (Theorem~\ref{thm1}) that similar effects will
show up in any matrix model in which the propagator decays too slowly
with $\Lambda$.  This means that the correlation between distant modes
is too strong, i.e.\ the model is too non-local.

\begin{acknowledgement}
 
We are grateful to Jos\'e Gracia-Bond\'{\i}a and Edwin Langmann for
discussions concerning the integral representation of the
$\star$-product and its matrix base. We would like to thank Thomas
Krajewski for advertising the Polchinski equation to us and Volkmar
Putz for accompanying the study of Polchinski's original proof. We are
grateful to Christoph \mbox{Kopper} for indicating to us a way to
reduce in our original power-counting estimation the polynomial in
$\big(\ln \frac{\Lambda_0}{\Lambda_R}\big)$ to a polynomial in
$\big(\ln \frac{\Lambda}{\Lambda_R}\big)$, thus permitting immediately
the limit $\Lambda_0 \to \infty$.  We would also like to thank Manfred
Schweda and his group for enjoyable collaboration.

We are indebted to the Erwin Schr\"odinger Institute
in Vienna, the Max-Planck-Institute for Mathematics in the Sciences
in Leipzig and the Institute for Theoretical Physics of the University
of Vienna for the generous support of our collaboration.
\end{acknowledgement}

\end{document}